\documentclass[aps,prd,twocolumn,nofootinbib,showpacs,preprintnumbers,floatfix]{revtex4}
\usepackage{color}
\usepackage{graphicx}
\usepackage{amssymb}
\usepackage{hyperref}
\usepackage{fancyhdr,lastpage} 
\usepackage[latin1]{inputenc}
\usepackage[english]{babel}
\usepackage{lineno}
\def \simgt{\,\rlap{\lower 7.5 pt\hbox{$\mathchar \sim$}}\raise 3 pt \hbox{$>$}\,}
\def \simlt{\,\rlap{\lower 7.5 pt\hbox{$\mathchar \sim$}}\raise 3 pt \hbox{$<$}\,}

\def\lsim{\raise0.3ex\hbox{$<$\kern-0.75em\raise-1.1ex\hbox{$\sim$}}}
\def\gsim{\raise0.3ex\hbox{$>$\kern-0.75em\raise-1.1ex\hbox{$\sim$}}}

\makeatletter
\ifx\@bibitemshut\undefined\let\@bibitemShut=\@empty\fi
\makeatother

\begin{document}
\title{The equation of state in (2+1)-flavor QCD}
\author{
A. Bazavov$^a$, 
Tanmoy Bhattacharya$^{\rm b}$, 
C. DeTar$^{\rm c}$, 
H.-T. Ding$^{\rm d}$, 
Steven Gottlieb$^{\rm e}$, 
Rajan Gupta$^{\rm b}$,
P. Hegde$^{\rm d}$, \\
U.M. Heller$^{\rm f}$, 
F. Karsch$^{\rm g,h}$,
E. Laermann$^{\rm h}$, 
L. Levkova$^{\rm c}$, 
Swagato Mukherjee$^{\rm g}$, 
P. Petreczky$^{\rm g}$, \\
C. Schmidt$^{\rm h}$, 
C. Schroeder$^{\rm i}$,
R.A. Soltz$^{\rm i}$, 
W. Soeldner$^{\rm j}$,
R. Sugar$^{\rm k}$, 
M. Wagner$^{\rm e}$,
P. Vranas$^{\rm i}$
\\[1mm]
{\bf (HotQCD Collaboration)}
}
\affiliation{
\vspace{0.5cm}
$^{\rm a}$ Department of Physics and Astronomy, University of Iowa, Iowa City, 
IA 52240, USA \\
$^{\rm b}$ Theoretical Division, Los Alamos National Laboratory, Los Alamos, NM 87545, USA\\
$^{\rm c}$ Department of Physics and Astronomy, University of Utah, Salt Lake City, UT 84112, USA \\
$^{\rm d}$ Key Laboratory of Quark \& Lepton Physics (MOE),
Institute of Particle Physics, Central China Normal University,
Wuhan, 430079, China \\
$^{\rm e}$ Physics Department, Indiana University, Bloomington, IN 47405, USA\\
$^{\rm f}$ American Physical Society, One Research Road, Ridge, NY 11961, USA\\
$^{\rm g}$ Physics Department, Brookhaven National Laboratory, Upton, NY 11973, USA \\
$^{\rm h}$ Fakult\"at f\"ur Physik, Universit\"at Bielefeld, D-33615 Bielefeld, Germany\\
$^{\rm i}$ Physics Division, Lawrence Livermore National Laboratory, Livermore CA 94550, USA\\
$^{\rm j}$ Institut f\"ur Theoretische Physik, Universit\"at Regensburg,
D-93040 Regensburg, Germany\\ 
$^{\rm k}$ Physics Department, University of California, Santa Barbara, CA 93106, USA\\
}

\begin{abstract}
We present results for the equation of state in
(2+1)-flavor QCD using the highly improved staggered quark action and
lattices with temporal extent $N_{\tau}=6,~8,~10$, and $12$.  We show
that these data can be reliably extrapolated to the continuum limit
and obtain a number of thermodynamic quantities and the speed of sound
in the temperature range (130--400)~MeV. We compare our results
with previous calculations, 
and provide an analytic parameterization of
the pressure, from which other thermodynamic quantities can be
calculated, for use in phenomenology.  We show that the energy density
in the crossover region, $145~{\rm MeV} \leq T \leq 163$~MeV, defined by the
chiral transition, is $\epsilon_c=(0.18-0.5)~{\rm GeV}/{\rm fm}^3$,
$i.e.$, $(1.2-3.1)\ \epsilon_{\rm nuclear}$. At high temperatures, we
compare our results with resummed and dimensionally reduced
perturbation theory calculations.  As a byproduct of our analyses, we
obtain the values of the scale parameters $r_0$ from the static quark
potential and $w_0$ from the gradient flow.
\vspace{0.2in}
\begin{center}
\bf{\today}
\end{center}
\end{abstract}

\vspace{0.2in}

\maketitle
\section{Introduction}

At high temperatures, matter governed by strong interactions (strong
interaction matter) undergoes a deconfining transition to a new state,
in which the thermodynamics can be described in terms of quark and gluon
degrees of freedom.  The equation of state (EoS) of such matter, just
as for many other thermodynamic systems, is of fundamental
importance for understanding its composition as well as its static and
dynamical properties. Studying the properties of this matter using
Quantum Chromo-Dynamics (QCD) was made possible by the formulation of
lattice-regularized QCD \cite{Wilson:1974sk} and the development of
numerical algorithms for its analysis \cite{Creutz:1980zw}. Lattice
calculations of the QCD EoS were first performed in 
1980~\cite{Engels:1980ty}, and, driven by the steady growth in
computing resources and the development of new simulation algorithms, there now
exist precise results for the transition temperature
\cite{Aoki:2009sc,Bazavov:2011nk}, fluctuations of conserved charges  
\cite{Borsanyi:2011sw,Bazavov:2012jq,Bazavov:2014xya} as well as the
EoS. For recent reviews see for instance Refs.
\cite{Petreczky:2012rq,Philipsen:2012nu,DeTar:2009ef}.

The EoS contains information on the relevant degrees of freedom in the
thermal medium in different temperature regimes and reflects the
transition between different states of matter.  A quantitative
description of the QCD EoS over a wide temperature range is needed to
understand the expansion and cooling of matter in the early universe, 
as well as of the hot dense nuclear matter created in heavy ion
collisions.

To study the QCD EoS across a transition between different states of
matter, at which the internal degrees of freedom are highly correlated, 
requires nonperturbative techniques.  However, in the case of strong
interaction matter, the need for nonperturbative methods is not
restricted to the strongly interacting region close to the QCD
transition temperature, but is also needed far above this deconfining
transition where well-known infrared problems \cite{Linde:1980ts}
prohibit a straightforward perturbative analysis of QCD
thermodynamics.  Also, at low temperatures, where the hadron resonance 
gas models (HRG) for the
description of the hadronic equation of state are quite
successful~\cite{BraunMunzinger:2003zd}, lattice QCD calculations are
important as they provide the benchmark estimates of thermal
properties of in-medium hadrons and the EoS of hadronic matter. In
summary, simulations of lattice QCD provide the best approach over the
full phenomenologically interesting temperature range in which all
sources of errors can be quantified and systematically improved.

The deconfining transition in QCD, with small but non-zero values of
the light quark masses, is a rapid crossover that coincides with the
restoration of chiral symmetry~\cite{Aoki:2009sc,Bazavov:2011nk}.  In
fact, it is the latter that characterizes the second order phase
transition that occurs in the chiral limit of QCD at finite
temperature. At this phase transition the spontaneously broken chiral
symmetry is restored. The universal scaling properties of this chiral
transition are used to determine the pseudo-critical temperature $T_c$
at which the rapid crossover with the physical light and strange quark
masses takes place~\cite{Bazavov:2011nk}.  Extensive simulations of
lattice QCD at zero net baryon number density have established that
this crossover transition occurs at $T_c\sim 155$~MeV for the physical
spectrum of two light and a heavier strange quark
\cite{Aoki:2009sc,Bazavov:2011nk}. Even though there is no
well-defined separation of phases because of the crossover nature of
the transition, it is well established that many thermodynamic
properties change rapidly in the vicinity of $T_c$. Along with the
analysis of fluctuations in the chiral condensate that are used to
probe the restoration of chiral symmetry and to determine $T_c$, the
study of fluctuations in conserved charges provides clear evidence for
deconfinement of light and strange quark degrees of freedom, {\it i.e.}, 
a transition from hadronic to quark-gluon degrees of freedom around
$T_c$~\cite{Ejiri:2005wq}.

In this paper, we present a detailed analysis of the EoS that captures
the crossover transition and the temperature range that is relevant to
the hydrodynamic evolution of heavy ion collisions at the Relativistic 
Heavy Ion Collider (RHIC) and the Large Hadron Collider 
(LHC).  We performed high statistics simulations of lattice QCD on
lattices of size $N_\sigma^3 N_\tau$ for four values of
$N_{\tau}\equiv (aT)^{-1} = 6$, $8$, $10$, and $12$ and a large
spatial size $N_\sigma = 4 N_{\tau}$.  We use these data to show that a
controlled extrapolation to the continuum can be performed in
the temperature range $130~{\rm MeV} \le T \le 400~{\rm MeV}$.  We
also show that the rapid change in the energy density signaling the
liberation of quark-gluon degrees of freedom leads to an estimate of
the pseudo-critical temperature that is consistent with that obtained
from the analysis of chiral symmetry restoration. Lastly, we provide
an accurate parametrization of this EoS that can be used for
hydrodynamic modeling of heavy ion collisions (see
Ref.~\cite{Gale:2013da} for a recent review) and other
phenomenological studies of the thermodynamics of strong interaction
matter.
 
Most of the lattice QCD calculations of the thermodynamics of strong
interaction matter use the staggered fermion discretization
scheme. The main reason for this is that staggered fermions preserve
an essential remnant of the continuum $SU(2)_L\times SU(2)_R$ chiral
symmetry of the light quark sector and are, at the same time, the 
least demanding
computationally. For an overview of EoS calculations using other
fermion discretization schemes
see Refs.~\cite{Philipsen:2012nu,Umeda:2012er}.  Furthermore, simulations
of QCD thermodynamics using staggered fermions have been
systematically improved by eliminating ${\cal O}(a^2)$ cutoff effects
\cite{Heller:1999xz} and reducing the effects of the so-called taste
symmetry breaking, specific to the staggered fermion formulation, by
using smeared gauge links~\cite{Blum:1996uf,Orginos:1999cr}. 

A number of improved staggered formulations have been developed and
used to study QCD at finite temperature. In the past, we have
simulated the p4 and asqtad actions
\cite{Karsch:2000ps,Bernard:2006nj,Cheng:2007jq,Bazavov:2009zn,Cheng:2009zi}.
These actions eliminate tree-level ${\cal O}(a^2)$ cutoff effects on
lattices with moderate size, $N_\tau > 8$, but have large taste
symmetry violations at low temperatures.  The Wuppertal-Budapest collaboration
has used the stout-smeared staggered action~\cite{Borsanyi:2010cj}
that very effectively reduces taste symmetry violation effects but still shows
large ${\cal O}(a^2)$ cutoff effects at high temperatures.  The first
reliable continuum extrapolated results for the QCD EoS have recently
been obtained with this action \cite{Borsanyi:2013bia}.

The calculations presented in this paper are carried out using the
highly improved staggered quark (HISQ) action introduced by the HPQCD
collaboration~\cite{Follana:2006rc}.  It was designed to improve both
the taste symmetry and the quark dispersion relation by including
smeared one-link terms as well as straight three-link terms that 
completely eliminate ${\cal O}(a^2)$ discretization errors at tree
level. The HISQ action has turned out to yield the smallest violations
of taste symmetry among the currently used staggered fermion actions
\cite{Bazavov:2011nk,Bazavov:2013yv,Cea:2014xva}.  We have used it
extensively to carry out high precision studies of the chiral and
deconfinement aspects of the QCD transition which lead to the estimate
$T_c = 154(9)$~MeV for the QCD transition temperature. It has also
been used to study the fluctuations of conserved charges
\cite{Bazavov:2012jq,Bazavov:2012vg,Bazavov:2013dta,Bazavov:2013uja}
and various spatial and temporal correlation functions
\cite{Bazavov:2013zha,Kim:2013seh}.  The study of fluctuations of
conserved charges at high temperatures demonstrates, in particular,
that the HISQ action is indeed very effective in reducing cutoff
effects~\cite{Bazavov:2013uja}.

In this paper, we show that
continuum extrapolated results for the EoS of (2+1)-flavor QCD
obtained with the HISQ action are in good agreement with those 
obtained with the stout action \cite{Borsanyi:2010cj,Borsanyi:2013bia}
\footnote{
There was an error in the preliminary analyses of the EoS
with the HISQ/tree action presented in  
conference proceedings before 2014 \cite{Petreczky:2012gi,Bazavov:2012bp,Bazavov:2010sb}
 due to an incorrect normalization of the fermion
contribution to the trace anomaly. This error gave a larger value
of the trace anomaly for $T<300$ MeV. Preliminary results for the EoS 
with the HISQ/tree action prior to 2014 are, therefore, superseded.}.
There are, however, systematic differences which may start to 
become of relevance in the analysis of the approach to the perturbative
limit at high temperatures. We will discuss these features in more detail
in Secs.~\ref{sec4} and \ref{sec5}. 

The rest of the paper is organized as follows. In Sec.~\ref{sec2} we
discuss the lattice setup and the simulation
parameters. Section~\ref{sec3} contains the results for the trace
anomaly, which is the basic thermodynamic quantity obtained from
lattice calculations, and from which the EoS is
obtained. Section~\ref{sec4} discusses the extraction of thermodynamic
quantities in the continuum limit. This section ends with an
analytical parameterization of the EoS that matches the HRG estimates
below $T = 130$~MeV and the lattice data between $130$ and
$400$~MeV. In Sec. \ref{sec5}, we present results on observables that
depend on second order derivatives of the pressure with respect to
temperature, {\it i.e.}, the specific heat and the speed of sound. We
discuss their phenomenological importance. Also in Sec. \ref{sec5}, we
discuss how our results for the EoS connect to high temperature
perturbative calculations. Finally, Sec.~\ref{sec6} contains our
conclusions. Technical details of the calculations are given in the
appendices.

\section{Lattice setup}
\label{sec2}

We performed simulations of (2+1)-flavor QCD using the HISQ action and
the tree-level improved gauge action.  This combination is referred to
as the HISQ/tree action. The (2+1)-flavor simulations are defined by
three bare parameters, the gauge coupling $\beta=10/g^2$, the
light-quark mass $m_l=m_u=m_d$, and the heavier strange quark mass
$m_s$. For a given value of the gauge coupling, we tune the strange
quark mass to its physical value by matching the mass of the
fictitious unmixed pseudoscalar $\eta_{s\bar s}$ meson to 695~MeV.
The light quark mass is fixed as a fraction of the strange quark mass,
$m_l=m_s/20$. This is slightly above the physical ratio $m_l=m_s/27.3$
and corresponds to a pion mass of about 160~MeV in the continuum
limit. This difference should, however, give rise to negligible
effects in the calculation of the
EoS~\cite{Cheng:2009zi,Borsanyi:2010cj}.  Having fixed $m_s$ and
$m_l$, the continuum limit is taken along a line of constant physics
(LCP) controlled by a single parameter, the gauge coupling $\beta$.

The LCP for the HISQ/tree action and $m_l=m_s/20$ has been
established, and reported in Ref.~\cite{Bazavov:2011nk}, based on a
set of zero-temperature ensembles that span the range of gauge
couplings $\beta=5.9-7.28$. In that study, the associated sets of 
finite temperature
ensembles were generated on lattices with temporal extent $N_\tau=6$,
$8$, and $12$, and a fixed aspect ratio $N_\sigma/N_\tau=4$ for
the spatial extent. In Appendix~\ref{app_ensembles}, we list
all the zero- and finite-temperature gauge field ensembles used in
this study along with the final statistics.  Here we briefly summarize
the additional simulations carried out, and the improvements made,
compared to those presented in Ref.~\cite{Bazavov:2011nk}.  
\begin{enumerate}
\item[(i)] 
Additional zero- and finite-temperature ensembles were generated
  at $\beta=7.373$, $7.596$, and $7.825$.
\item[(ii)] 
A new set of finite temperature lattices with $N_\tau=10$ were
  generated.
\item[(iii)] 
The statistics are substantially increased for all existing
  ensembles, in some cases by more than an order of magnitude compared
  to Ref.~\cite{Bazavov:2011nk}.
\item[(iv)] 
The determination of the LCP on finer lattices is improved by
  measurements of the static quark potential and the hadron spectrum
  on new zero-temperature ensembles.
\end{enumerate}

The lattice spacing $a$, corresponding to the coupling $\beta$, was determined by
calculating the scales $r_0$ \cite{Sommer:1993ce} and $r_1$
\cite{Bernard:2004je}, defined in terms of the static potential as
\begin{equation}
\left . r^2 \frac{dV}{dr} \right|_{r_i}=C_i\;\; ,\;\; i=0,1 \ ,
\label{eq:r0_r1}
\end{equation}
where $C_0=1.65$ and $C_1=1.0$.  At each $\beta$, these scales are
determined by first extracting the potential $V(r)$ by fitting the
lattice data to 
\begin{equation}
V(r) = C + \frac{B}{r} + \sigma r \ ,
\label{eq:potential}
\end{equation}
and then calculating its derivative in intervals around the values of
$r_1$ and $r_0$, as described in Ref.~\cite{Bazavov:2011nk}.  The
details of the determination of $r_0/a$ and $r_1/a$ and
the extrapolation of the ratio, $r_0/r_1$, to the continuum limit are
given in Appendix~\ref{app_r0r1}. The extrapolated result is
$r_0/r_1=1.5092(39)$, which gives $r_0=0.4688(41)$~fm using the
physical value $r_1=0.3106(14)(8)(4)$~fm \cite{Bazavov:2010hj}.  This
estimate of $r_0$ is in agreement with $r_0=0.48(1)(1)$~fm given in
Ref.~\cite{Aoki:2009sc}.

To crosscheck the precision of the determination of the lattice
spacing, we also calculated the scale $w_0$ first proposed in
Ref.~\cite{Borsanyi:2012zs}. The details of this calculation are also given
in Appendix~\ref{app_r0r1}, and we obtain $w_0/r_1 = 0.5619(21)$ in
the continuum limit. This translates to $w_0 = 0.1749(14)$~fm, in
agreement with $w_0 = 0.1755(18)(4)$~fm given in
Ref.~\cite{Borsanyi:2012zs}.

We have also measured the masses and decay constants of several light
hadrons.  These allow us to improve the determination of the LCP at
weaker coupling and provide further crosschecks on the scale
setting in the continuum limit. 
We find that the different ways to set the lattice scale
using hadronic observables agree with each other and the scale
determined using $r_1$ within the estimated errors.  The details of
these analyzes are presented in Appendix~\ref{hadronic}.

\section{The trace anomaly}
\label{sec3}

\begin{figure}
\includegraphics[width=8.5cm]{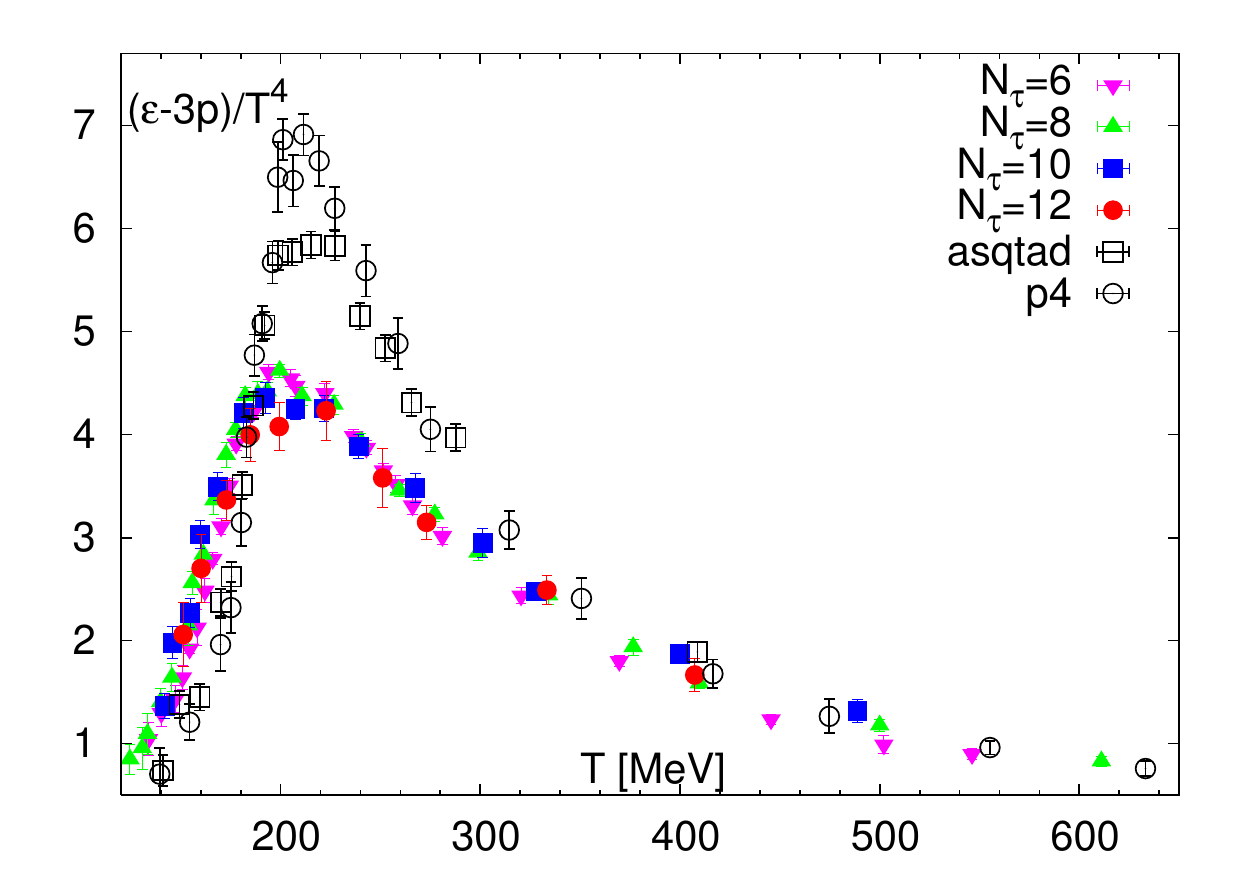}
\caption{The trace anomaly calculated with the HISQ/tree action at
  different $N_{\tau}$ and compared with results from previous calculations 
  with the
  p4 and asqtad actions on $N_{\tau}=8$ lattices~\cite{Bazavov:2009zn},
  except for the two highest temperatures, where we show the
  $N_{\tau}=6$ p4 data from Ref.~\cite{Petreczky:2009at} and
  Ref.~\cite{Cheng:2007jq}, respectively.  }
\label{fig:e-3p}
\end{figure}

\begin{figure*}
\includegraphics[width=0.45\textwidth]{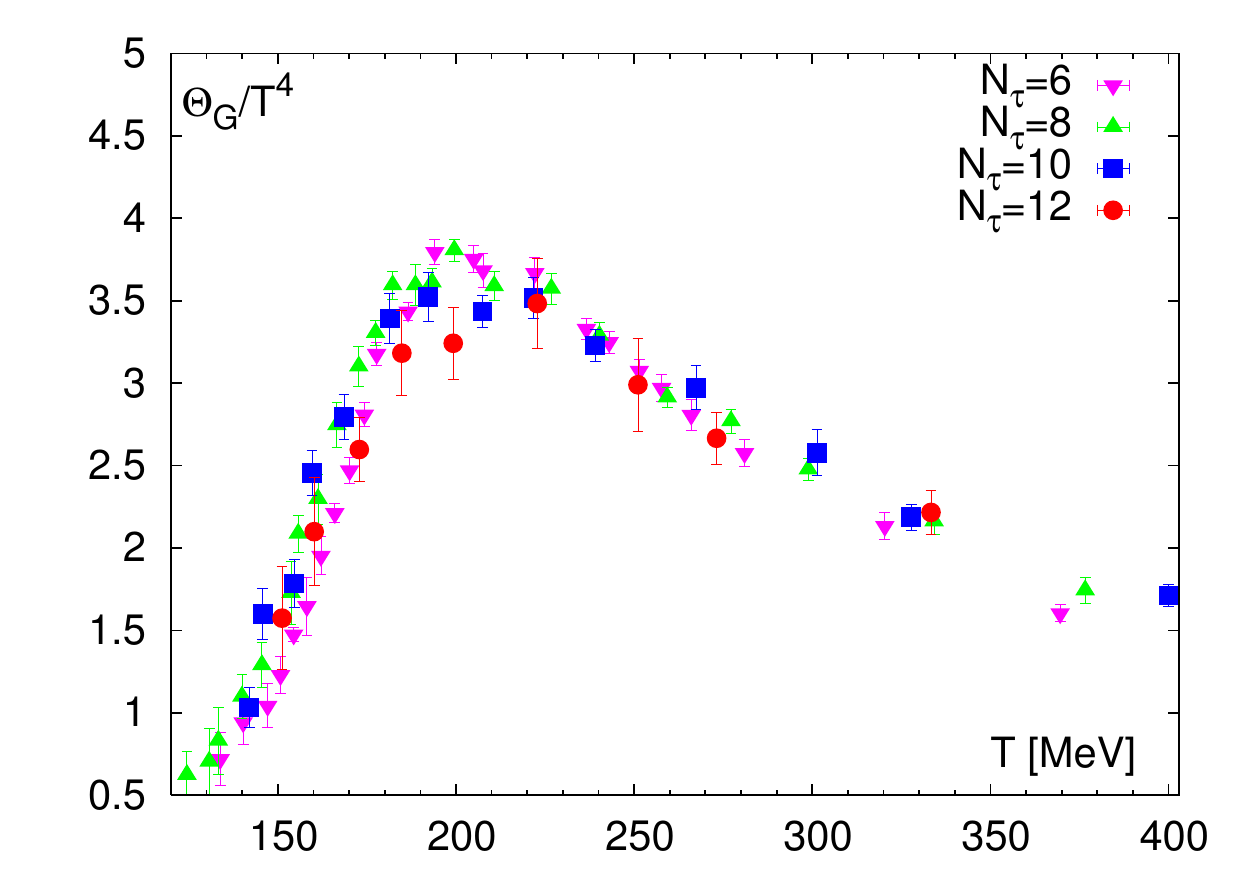}
\includegraphics[width=0.45\textwidth]{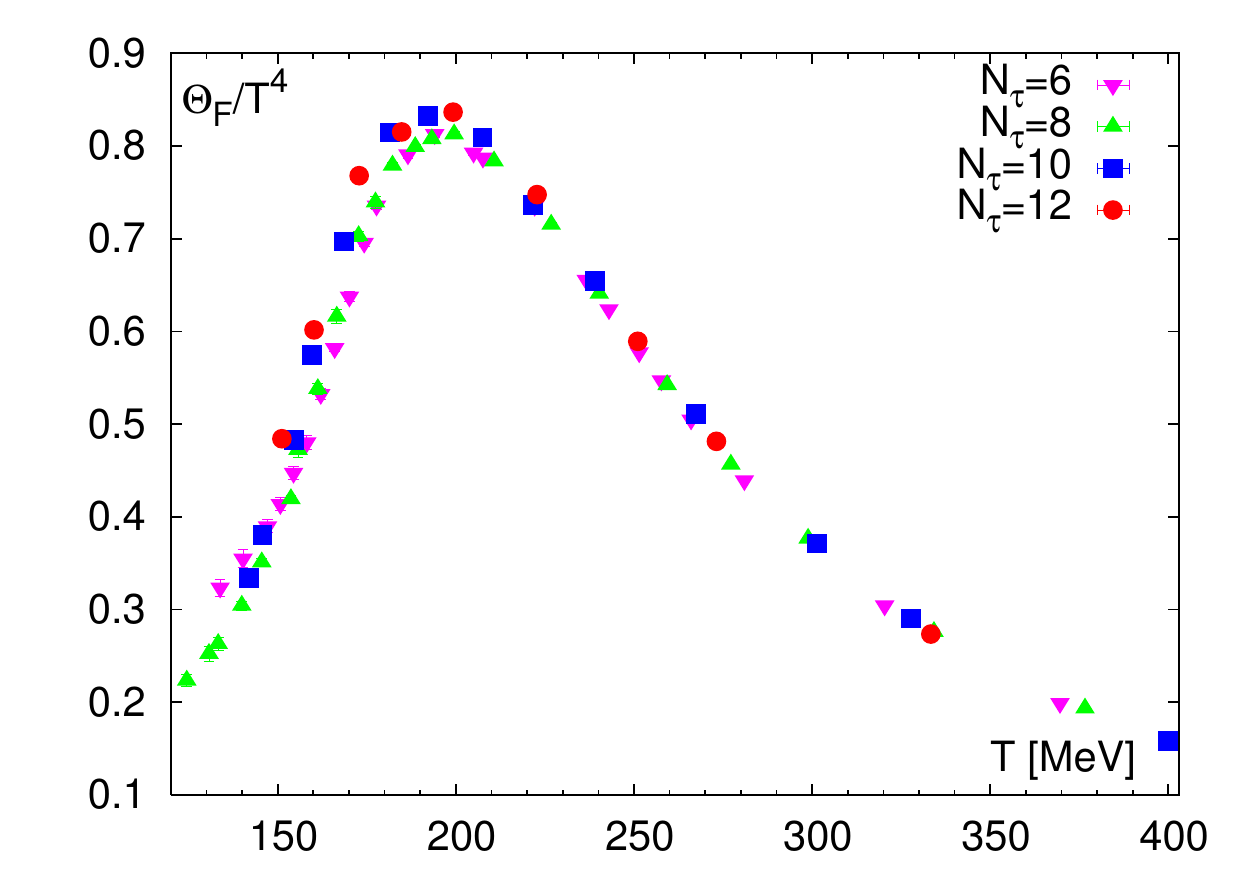}
\caption{The gluonic (left) and fermionic (right) parts of the trace
  anomaly for different $N_{\tau}$. See text for details.}
\label{fig:e-3p_SG}
\end{figure*}

The QCD partition function on a hypercubic lattice of size $N_\sigma^3N_\tau$, after integration over the 
fermion degrees of freedom, is given by
\begin{equation}
  Z(\beta,N_\sigma,N_\tau)=\int \prod_{x,\mu} dU_{x,\mu} e^{-S(U)}\; ,
\end{equation}
where $U_{x,\mu}\in SU(3)$ are the gauge field variables, labeled by
$x$ and $\mu$, defined on the links between lattice points and the Euclidean
action $S(U)$ is the sum of the gauge and fermionic parts:
\begin{equation}
  S(U) = \beta S_G(U) - S_F(U)\; .
\end{equation}
The temperature in physical units is set by the temporal extent $N_{\tau}$ 
of the lattice and 
related to the lattice spacing $a$ as $T=1/(aN_\tau)$. 

The trace of the energy-momentum tensor,  also called trace anomaly or
the interaction measure, is related to the pressure $p$ as 
(see Ref.~\cite{Cheng:2007jq})
\begin{equation}
\frac{\Theta^{\mu\mu}(T)}{T^4} =\frac{\epsilon -3p}{T^4}=T\frac{d}{d T}
\left(\frac{p}{T^4}\right) \; ,
\label{theta_p}
\end{equation}
with $\epsilon$ denoting the energy density.
$\Theta^{\mu\mu}(T)$ can be defined on the lattice as 
the total derivative of $\ln Z$ with respect to the lattice spacing $a$:
\begin{equation}
\Theta^{\mu\mu}=\epsilon -3p = -\frac{T}{V}\frac{d\ln Z}{d \ln a}\; .
\label{e3p_dlnZ}
\end{equation}
The right hand side of Eq.~(\ref{e3p_dlnZ}) is straightforward to evaluate on
the lattice and gives
\begin{eqnarray}
&
\displaystyle
\frac{\epsilon-3p}{T^4} \equiv
\frac{\Theta^{\mu\mu}_G(T)}{T^4} +
\frac{\Theta^{\mu\mu}_F(T)}{T^4} \; , \\[2mm]
&
\displaystyle
\frac{\Theta^{\mu\mu}_G(T)}{T^4}=
R_\beta
\left[ \langle s_G \rangle_0 - \langle s_G \rangle_\tau \right] N_\tau^4 \; , 
\label{e3pG}  \\[2mm]
&
\displaystyle
\frac{\Theta^{\mu\mu}_F(T)}{T^4}  
= - R_\beta R_{m} [
2 m_l\left( \langle\bar{\psi}\psi \rangle_{l,0}
- \langle\bar{\psi}\psi \rangle_{l,\tau}\right)  \nonumber \\[2mm] 
&
\displaystyle
\qquad\qquad + m_s \left(\langle\bar{\psi}\psi \rangle_{s,0}
- \langle\bar{\psi}\psi \rangle_{s,\tau} \right )
 ] N_\tau^4 \; .
\label{e3pF}
\end{eqnarray}
Here $\langle s_G\rangle_{\tau (0)}$ is the expectation value of
the action density for the gauge fields evaluated at
finite (zero) temperature and $\langle \bar{\psi}\psi \rangle_{l(s),\tau (0)}$ 
stands for the expectation values of light ($l$) and strange ($s$) quark 
chiral condensates evaluated at finite (zero) temperature. 
Subtracting the zero temperature values
in the above expressions ensures that all thermodynamic quantities are
finite in the continuum limit.  In Eq.~(\ref{e3pF}), we have used the
single flavor normalization for both the light and strange quark
condensates as in previous works
\cite{Cheng:2007jq,Bazavov:2009zn,Bazavov:2011nk}.  The
nonperturbative beta function and mass renormalization function are
defined as \cite{Cheng:2007jq,Bazavov:2009zn}
\begin{eqnarray}
&
\displaystyle
R_{\beta}(\beta) = \frac{r_1}{a} \left( {{\rm d} (r_1/a) \over {\rm d} \beta} \right)^{-1}\; ,\label{Rbeta}\\
&
\displaystyle
R_m(\beta) = \frac{1}{m_s(\beta)}
\frac{{\rm d} m_s(\beta)}{{\rm d}\beta}\; .
\label{Rm}
\end{eqnarray}
The determination of these functions is discussed in Appendices
\ref{app_r0r1}--\ref{app:observables}.  In the above equations, we explicitly separated the
contributions to the trace anomaly that come from purely gluonic
operators $\Theta^{\mu\mu}_G(T)$ and fermionic operators
$\Theta^{\mu\mu}_F(T)$.  Even though we will refer to them as the
gluonic and fermionic parts, it would be misleading to consider
$\Theta^{\mu\mu}_F(T)$ as the quark contribution to the trace
anomaly. For example, for massless quarks $\Theta^{\mu\mu}_F(T)/T^4$
is zero, while massless quarks certainly contribute to the trace
anomaly. At high temperatures, where the effect of nonzero quark
masses is expected to be small, the quark contribution almost
exclusively comes from $\Theta^{\mu\mu}_G(T)$. As we will see below,
this expectation is confirmed by our numerical data.  The above
separation of the trace anomaly into $\Theta^{\mu\mu}_G(T)$ and
$\Theta^{\mu\mu}_F(T)$ is, however, useful in the analysis of
lattice data as they are expected to be affected differently by the
taste symmetry breaking inherent in staggered fermions and because the
statistical errors are also different.

The pressure can be calculated using the integral method, {\it i.e.},
by inverting Eq.~(\ref{theta_p}):
\begin{equation}
\frac{p(T)}{T^4} = \frac{p_0}{T_0^4} + \int_{T_0}^T dT'\frac{\Theta^{\mu\mu}}{T^{\prime5}} \; .
\label{p_int}
\end{equation}
The choice of the reference temperature $T_0$ and pressure $p_0$ is discussed
in Sec.~\ref{sec4}.
All other thermodynamic quantities, defined as appropriate derivatives of the partition
function with respect to the temperature, can be calculated from Eqs.~(\ref{theta_p}) and
(\ref{p_int}) by using standard thermodynamic identities.

Since the trace anomaly is the central quantity in the lattice
calculations of the EoS, we discuss its properties in some detail.  In
Fig.~\ref{fig:e-3p}, we compare results for the trace anomaly obtained
with the HISQ/tree action on lattices with temporal extent $N_\tau=
6$, $8$, $10$, and $12$ with our previous findings using the p4 and
asqtad actions \cite{Cheng:2007jq,Bazavov:2009zn,Petreczky:2009at}.
The cutoff effects are much smaller in the HISQ/tree action and the
height of the peak is significantly reduced. Below the peak, the
HISQ/tree data are larger than the p4 and asqtad results, but
significantly smaller at temperatures around and higher than the
peak. These large deviations reflect the fact that the asqtad and the
p4 actions have much larger cutoff effects at low temperatures and in
the crossover region (see discussions in Ref.~\cite{Bazavov:2011nk}).
The smaller taste violations of the HISQ action lead to a smaller 
root-mean-square mass in the pseudoscalar sector \cite{Bazavov:2011nk},
{\it i.e.}, to a smaller average pion mass, which leads to a larger trace
anomaly as well as larger pressure and energy density in the low temperature,
hadronic region. 
For $T>350$~MeV, we find reasonably good agreement between the results
obtained with different actions. This is, to some extent, expected as
at such high temperatures, {\it i.e.}, at small $a$, all the above actions 
should have small cutoff effects. This expectation has been demonstrated 
in the calculations of quark-number susceptibilities
\cite{Bazavov:2013uja}. In Sec.~\ref{sec4}, we show that having
small cutoff effects in the data with the HISQ/tree action allows us
to make robust continuum extrapolations and obtain a precise EoS in
the temperature range 130--400~MeV.

\begin{figure}
\includegraphics[width=8cm]{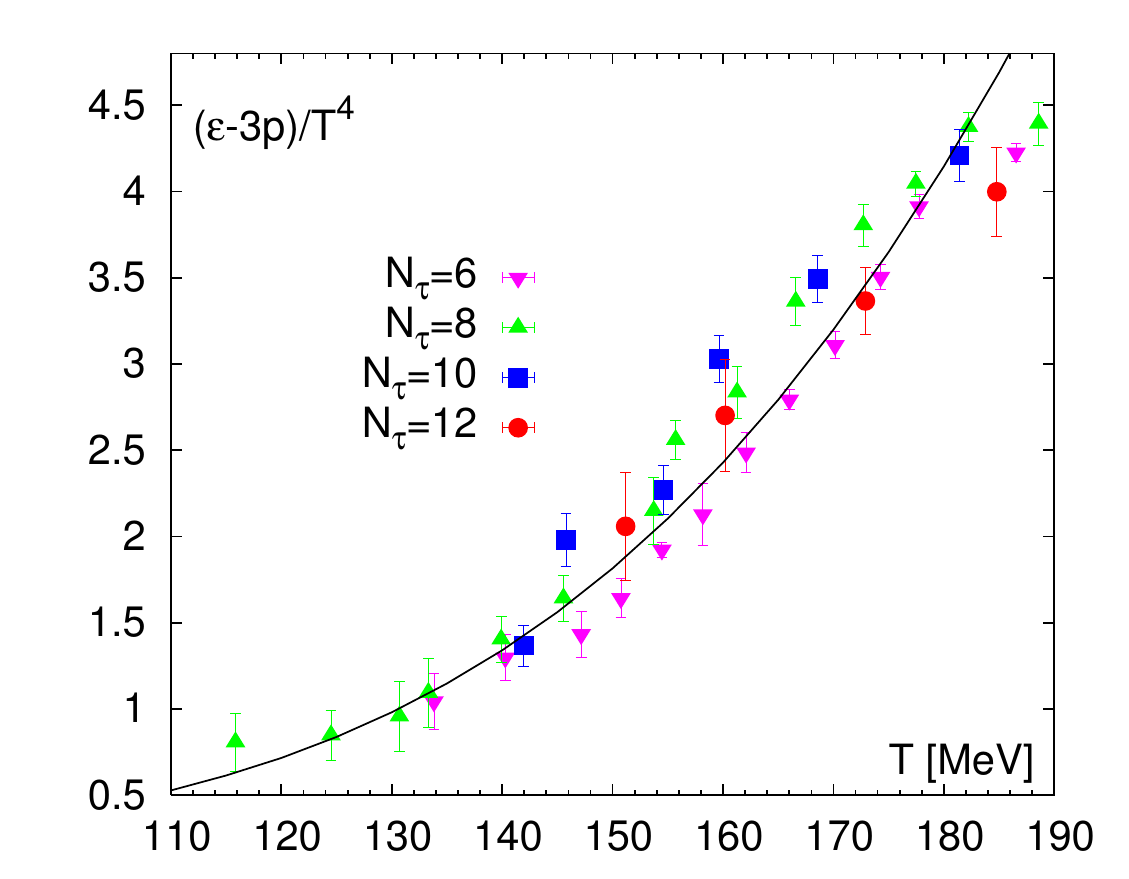}
\caption{The trace anomaly in the low temperature region compared
with the hadron resonance gas model (solid line).}
\label{fig_e3p_lowT}
\end{figure}

A closer look at the HISQ/tree action data shown in
Figs.~\ref{fig:e-3p}, \ref{fig:e-3p_SG}, and \ref{fig_e3p_lowT} reveals
some cutoff effects at low temperatures and in the peak region. It is
instructive to discuss these cutoff effects separately in terms of the
gluonic, $\Theta_G^{\mu\mu}$, and the fermionic, $\Theta_F^{\mu\mu}$,
contributions defined in Eqs. (\ref{e3pG}) and (\ref{e3pF}),
respectively, and shown in Fig. \ref{fig:e-3p_SG}.  We find that the
trace anomaly is dominated by the gluonic part. The fermionic
contribution is about $(20-25)\%$ of the gluonic contribution in the peak
region, rises to $\sim 35\%$ below it, and becomes much smaller at high
temperatures. At $T=400$~MeV, it is only about $10\%$.
Around the peak, $\Theta_G^{\mu\mu}$, and consequently
the trace anomaly, shows a decrease with increasing $N_{\tau}$, {\it i.e.},
the continuum limit is approached from above.  

The statistical errors and the lattice discretization effects in the
HISQ/tree data are smaller in the fermionic part compared to the
gluonic part. In $\Theta_F/T^4$, we observe significant cutoff
effects only at the lowest temperature $T\sim 133$~MeV, where the
$N_\tau=6$ and 8 data differ by about 30\%. This small size of cutoff
effects in $\Theta_F^{\mu\mu}$ with the HISQ/tree action in the low
temperature region is in contrast to results obtained using the
asqtad and the p4 actions, where the fermionic part showed
significantly larger cutoff effects.  We also note that cutoff
effects arising from taste symmetry violations have opposite effects
in $\Theta_F^{\mu\mu}$ and $\Theta_G^{\mu\mu}$. While a larger
root-mean-square (RMS) mass for the pions leads to smaller values of
$\Theta_G^{\mu\mu}$ at low temperatures, it leads to larger values in
$\Theta_F^{\mu\mu}$ as the chiral condensates are larger for larger
pion masses.

In the total trace anomaly, significant discretization effects are
observed only in the peak and low-temperature regions.  Within errors,
we find no cutoff effects on comparing $N_\tau=6,~8$, and $10$, data
for $T<145$~MeV.  In the interval $145~{\rm MeV} < T < 170~{\rm MeV}$,
we observe some cutoff dependence, with the largest difference
between the $N_\tau=6$ and $8$ data.

At low temperatures, all thermodynamic quantities are expected to be
well-described by the hadron resonance gas (HRG) model, in which all
the hadrons and hadron resonances are assumed to contribute to the
thermodynamics as non-interacting particles.  Many previous studies
have confirmed this expectation
\cite{Karsch:2003vd,Cheng:2008zh,Huovinen:2009yb,Borsanyi:2010bp,Borsanyi:2011sw,Bazavov:2012jq,Bazavov:2013dta}.
The trace anomaly in the HRG model is given by 
\begin{eqnarray}
&
\displaystyle
\left( \frac{\epsilon - 3p}{T^4}\right)^{HRG} =\nonumber\\
&
\displaystyle
\sum_{m_i\le m_{max}}
\frac{d_i}{2\pi^2}
\sum_{k=1}^\infty
\frac{(-\eta_i)^{k+1}}{k}
\left( \frac{m_i}{T}\right)^3 K_1(\frac{km_i}{T}) \; ,
\label{e3plow}
\end{eqnarray}
where different particle species of mass $m_i$ have degeneracy factors
$d_i$ and $\eta_i = -1 (+1)$ for bosons (fermions).  The particle
masses are taken from the Particle Data Book \cite{Beringer:1900zz},
including all known states up to the resonance mass of
$m_{max}=2.5$~GeV.  We compare the predictions of the HRG model with
our data for the trace anomaly in the low temperature region in
Fig.~\ref{fig_e3p_lowT}.  For $T<145$~MeV the lattice data do not show
any significant $N_{\tau}$ dependence and are in good agreement
with estimates from the HRG model.  This agreement will be used in
an important way for the continuum extrapolation and for the calculation
of the pressure described in the next section.  For temperatures in
the interval $145~{\rm MeV} < T < 170~{\rm MeV}$, the $N_{\tau}=8,~10$
and $12$ lattice data lie above the HRG curve, while the $N_{\tau}=6$
data lie systematically below.  In Sec.~\ref{sec4}, we show
that the cutoff effects in the $N_{\tau}=6$ data are large in this 
temperature interval.

\section{Thermodynamics in the continuum limit}
\label{sec4}

In this section, we describe the calculation of the pressure and the
energy and entropy densities in the continuum limit. The main step in 
this calculation is the extrapolation of the lattice data for the trace 
anomaly $\Theta^{\mu\mu}$ from $N_{\tau}=6$, $8$, $10$,
and $12$ lattices to the continuum limit. Noting that the leading
lattice discretization effects ($N_\tau$ dependence) for staggered
fermions are expected to be proportional to $(a T)^2 \sim
1/N_{\tau}^2$, we use the fit ansatz
\begin{equation}
\frac{\Theta^{\mu\mu}(T)}{T^4} = 
A + \sum_{i=1}^{n_k+3} B_i \times S_i(T) + 
\frac{C + \sum_{i=1}^{n_k+3} D_i \times S_i(T)}{N_\tau^2}  \; ,
\label{eq:SplineFit}
\end{equation}
where $n_k$ denotes the number of knots in the interior of the fit
interval and the $S_i$ are a set of basis cubic splines with
discontinuities only in the third derivative at the specified knots as
described below.\footnote{Note that when knots are coincident,
  successively lower derivatives are discontinuous. All the splines
  are defined to go to zero at the lower end of the fit interval as 
  we explicitly include the constants A and C in our fit ansatz.}
The positions of the knots and the constants $A,\ B_i, \ C,$
and $D_i$ are parameters that are determined by the fit. To test whether
it is sufficient to keep just the leading $1/N_\tau^2$ term, we also
considered the next, $O(1/N_\tau^4)$, correction, 
\begin{equation}
\frac{E + \sum_{i=1}^{n_k+3} F_i \times S_i(T)}{N_\tau^4} \; . 
\end{equation}
Adding these terms to the quadratic fit, given in Eq.~(\ref{eq:SplineFit}),
defines the quartic fit also discussed below.

The basic assumption underlying the proposed fit ansatz is that the
data and the variation with $N_\tau$ can be described by a set of
piecewise continuous splines of cubic order.  The temperature interval
to be fitted is divided into sub-regions by a finite number of
internal knots $n_k$, which we further assume are independent of
$N_\tau$. The number and position of these knots specify a set of
basis splines that forms a complete set over the full interval, {\it i.e.},
any piecewise continuous cubic function can be fitted by them.  The
number of knots needed depends, in general, on the complexity of the
data; and the total number of basis splines invoked by the fit depends
on the number of knots specified. The positions of the knots are
outputs of the least-square minimization procedure we use.

A number of choices need to be specified before we can discuss the fits. 
\begin{itemize}
\item[(i)] The errors in each data point for $\Theta^{\mu\mu}(T)/T^4$
  are assumed to be normally distributed and independent, 
  since these come from independent simulations.
\item[(ii)] The entire analysis is done within a bootstrap procedure
  using 20,001 samples.  This number was chosen to make the sampling
  error in the bootstrap estimate of the standard error $1\%$. The
  bootstrap samples were generated by selecting each data point from a
  normal distribution with its width given by the quoted error. The
  final error band for $\Theta^{\mu\mu}(T)/T^4$ is given by the $1
  \sigma$ spread of the bootstrap values at each temperature. The
  statistical package R~\cite{Rpackage,RpackageMASS,RpackageHmisc} was
  used to implement this analysis.
\item[(iii)] The values of temperature at which simulations have been
  done are not uniform, in particular we do not have much data on
  $N_\tau=10$ and $12$ lattices for $T < 130$~ MeV and
  $T>400$~MeV. Our results will, therefore, be restricted to the range
  $130~{\rm MeV} \leq T \leq 400$~MeV.
\item[(iv)] Our goal is to use the minimum number of knots, and thus,
  the minimum number of parameters. We studied the $\chi^2$ resulting
  from the least-square minimization procedure to settle on the number
  of knots.
\item[(v)] We analyze the data using both the quadratic and quartic
  ansatz and with and without the $N_\tau=6$ data. Our final results
  are obtained using the quadratic fit without the $N_\tau=6$ data.
\item[(vi)] The data on $N_\tau = 8$ lattices for $T \leq 130$~MeV are
  insufficient to constrain the fits at the lower end. We, therefore,
  use the estimate $\Theta^{\mu\mu}(T)/T^4 = 1.007$ with slope $0.032$
  at $T=130$~MeV, obtained from the HRG model, for the continuum
  extrapolated value. To justify this choice we note that the HRG
  model is a good approximation at this temperature and insensitive to
  possible higher resonances missed in the hadron spectrum
  \cite{Bazavov:2014xya}.  Indeed, we find that the lattice data and
  the HRG estimates agree for $T< 145$ MeV. To take into account the
  uncertainty in the HRG estimates, both the estimate and the slope
  were picked using a Gaussian distribution about their central values
  with a conservatively chosen width, $10\%$ of their respective
  values.We implemented this constrain by replacing the spline $B_1$
  by a term proportional to $T-130$~MeV with its coefficient given by
  the HRG value. This constraint, therefore, reduces the number of
  free parameters in Eq.~(\ref{eq:SplineFit}) by two.
\item[(vii)] The data on $N_\tau = 8$ lattices in the temperature
  range (400--610)/~MeV was used to stabilize the quadratic fits up to
  $400$~MeV.  For the quartic fits, both the $N_\tau = 6$ and $8$ data
  at $T > 400$~MeV were used.
\end{itemize}

\begin{figure}
\includegraphics[width=8cm]{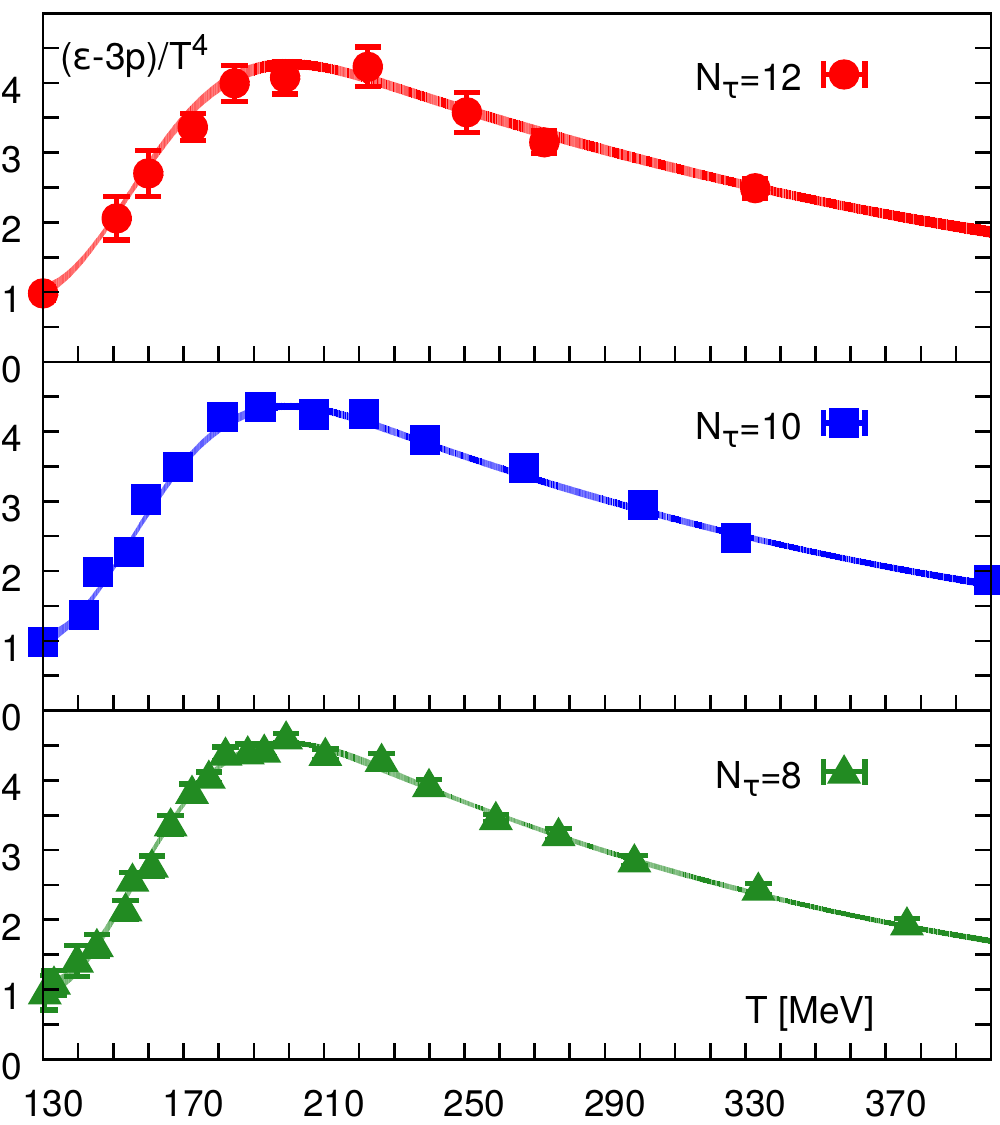}
\caption{The data for the trace anomaly and the result (thick lines
  showing the $1 \sigma$ bootstrap error bands) of applying
  Eq.~(\ref{eq:SplineFit}) with $N_\tau = 8$, $10$, and $12$. The
  parameters in Eq.~(\ref{eq:SplineFit}) and their errors, defining
  this final fit, were determined from these data as discussed in the
  text.  The error bands shown are generated by the same bootstrap
  process used to estimate the fit parameters and their errors. The
  additional $2\%$ error that is added to the final continuum result
  to account for the uncertainty in the determination of the
  temperature scale as discussed in the text is not included in these
  plots.  }
\label{fig:FitFinal3}
\end{figure}

To decide on the number of knots to use, we fit the $N_\tau=8, \ 10$,
and $12$ data with the quadratic ansatz with 2--4 internal knots. The
fit with two knots was the most stable and the $\chi^2$ did not
improve significantly with additional knots. The choice of two knots
is consistent with the observation that the data show three main
regions: the low temperature region $T \lsim 175$~MeV, the peak region
(175--225~MeV) and the high temperature region $T \gsim 225$~MeV. The
fit parameters and the location of the knots are outputs of the
$\chi^2$ minimization procedure. This fit has 53 data points, 5
basis splines and 12 free parameters, {\it i.e.}, the 10 parameters
remaining in Eq.~(\ref{eq:SplineFit}) after imposing the HRG value and
slope at $T=130$~MeV and the locations of the two knots. 
This ansatz fits all the
data, and the $\chi^2/{\rm dof}=0.9$ for ${\rm dof}=41$ was well
distributed, $i.e.$, it was not dominated by a few points nor by any
one of the three regions.  The distribution of the positions of the
knots over the 20,001 samples had central values of $170$~MeV and $229$~MeV
with a standard deviation of 8~MeV. This fit, called the final fit, is
used for our continuum results as the tests itemized below did not
improve upon it:
\begin{itemize}
\item[(i)]
Adding more knots to the final fit did not improve the fit. The
additional parameters were poorly determined, and in most bootstrap
samples two or more knots were coincident.
\item[(ii)]
We added the $N_\tau=6$ data to the final fit.  The $\chi^2/{\rm dof}$
increased and the fit became skewed.  It adjusted to preferentially
fit the low error $N_\tau = 6$ points and the $\chi^2$ became
dominated by the $N_\tau=12$ data below the peak. We concluded that
the quadratic ansatz is insufficient to fit the data at all four
$N_\tau$ values.
\item[(iii)]
We explored the quartic ansatz to fit the data at all four $N_\tau$
values.  In this case, the best fit required three knots. The
resulting error band overlaps with that of the the final fit except in
the peak region, where it is about $1 \sigma$ lower.  The position of
the knots are not as stable as in the final fit, and in many bootstrap
samples, two knots were coincident. To summarize, the final quadratic fit 
was preferred over the quartic fit as it is based on data closer to the
continuum limit, has the least number of parameters and fits the data
well as shown in Fig.~\ref{fig:FitFinal3}.
\end{itemize}

\begin{figure*}[t]
\includegraphics[width=8cm]{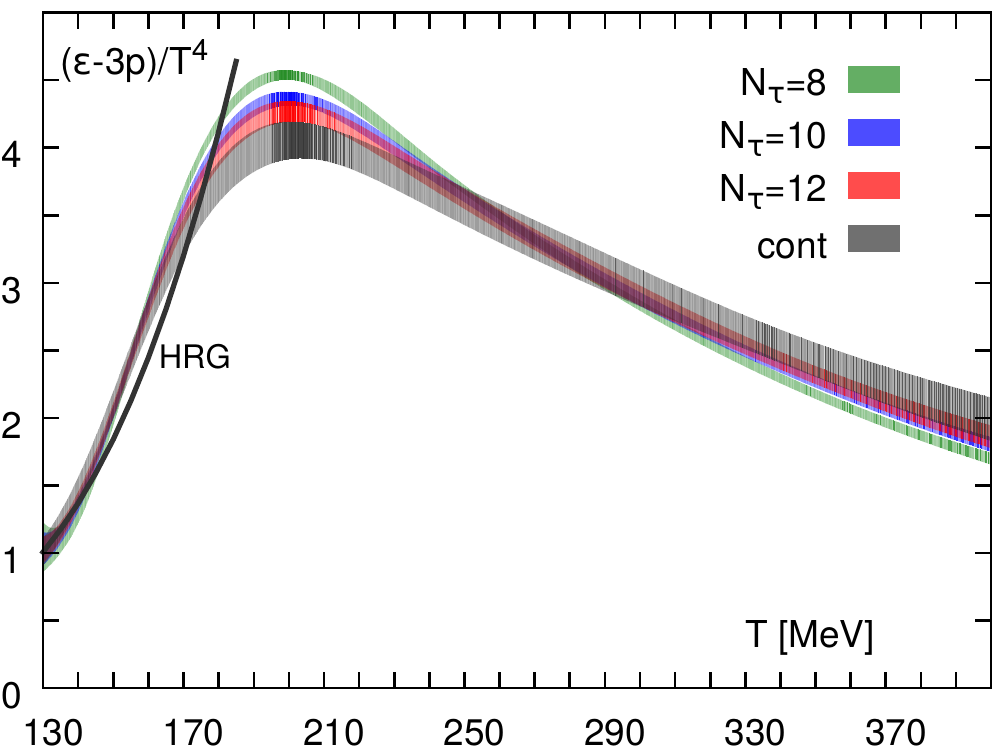}~~~
\includegraphics[width=8cm]{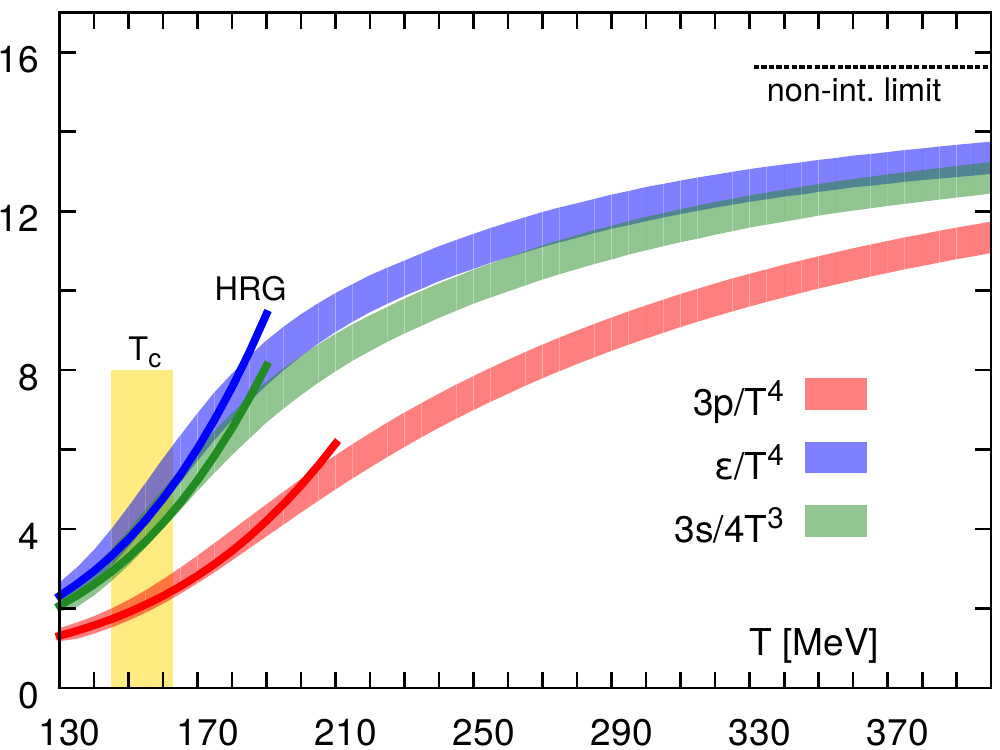}
\caption{Spline fits to the trace anomaly for several values of
the lattice spacing $aT=1/N_\tau$ and the result of our continuum 
extrapolation (left). Note that the error bands shown here do not
include the 2\% scale error. The right hand panel shows 
  suitably normalized pressure, energy density, and entropy density
  as a function of the temperature. In this case the 2\% scale error 
  is included in the error bands. The dark lines show the
  prediction of the HRG model. The horizontal line at $95 \pi^2/60$ 
  in the right panel corresponds to the ideal gas limit for the
  energy density and the vertical band marks the crossover 
  region, $T_c=(154\pm 9)$~MeV.}
\label{fig:cont}
\end{figure*}

The quality of the final fit using Eq.~(\ref{eq:SplineFit}) is
demonstrated in Fig.~\ref{fig:FitFinal3} where we show that the
bootstrap error bands of the final fit describe the $N_\tau = 8$,
$10$, and $12$ data very well.  On the other hand, as stated
previously, we find that the $N_\tau=6$ data lie outside the range of
applicability of the quadratic ansatz.  The same error bands are
compared with the final continuum extrapolated result (black band) in
Fig.~\ref{fig:cont}.

Having determined the final fit, we obtained the pressure $p/T^4$ by
numerically integrating the bootstrap samples for $\Theta^{\mu\mu}(T)$ between
$130$~MeV and $400$ MeV using Eq. (\ref{p_int}).  For the integration constant
$p_0$, the pressure at $T=130$~MeV, we picked a value from a normal
distribution with the mean value $p_0/T_0^4=0.4391$, again taken from the
HRG model, and width $0.0439$, a conservative $10\%$ error estimate on
this HRG value. Since the estimate of $p_0/T_0^4$ is independent of the
calculation of $\Theta^{\mu\mu}(T)$, this choice effectively adds a 
$ \delta p_0$ in
quadrature to the errors from integrating $\Theta^{\mu\mu}(T)/T^4$.  
Knowing
$\Theta^{\mu\mu}(T)/T^4 \equiv (\epsilon-3 p)/T^4$ and $p/T^4$, 
it is straightforward to derive the energy density, $\epsilon$, and the
entropy density $s=(\epsilon+p)/T$.

The final systematic error that is folded into the estimates of all
the thermodynamic quantities is the uncertainty in the determination
of the lattice scale $a$, and thus the values of the temperature $T$ used
in the fits.  Based on the uncertainty analyses in the determination
of the lattice scale $a$ ($\sim 1.3\%$) and tuning of the $m_s$ to
stay on the LCP presented in Appendices~\ref{app_r0r1} and
\ref{app_masses}, we assigned an overall conservative $2\%$
uncertainty in $T$, which we add linearly to the error estimates
already assigned by the bootstrap process. In practice, at each $T$
and for each observable, we picked the minimum and maximum values of
the $1 \sigma$ bootstrap envelope in the region $T\pm 2\%$. This new
envelope is then used as the final uncertainty band for all the
continuum results shown in the figures and discussed below.

Our continuum extrapolated results for the trace anomaly and other
thermodynamic observables are shown in Fig.~\ref{fig:cont} and the
data are given in Table~\ref{tab:fin}.  For $T<150$~MeV, the trace
anomaly is well approximated by the HRG estimate shown by
the solid line in Fig.~\ref{fig:cont} (left).  For $T>150$~MeV, the
$N_\tau \geq 8$ lattice results are systematically higher than the HRG
estimate as shown in Fig.~\ref{fig_e3p_lowT}, and the slopes of the
HRG and continuum extrapolated curves start to differ as shown in
Fig.~\ref{fig:cont}.  In the peak region, $(\epsilon -3 p)/T^4$ has a
maximum of about $4.05(15)$ at $T\sim 204$~MeV.  This maximal value
from simulations with the HISQ/tree action is significantly smaller
than our previous results with the p4 and asqtad actions which were
incorporated in the HotQCD parametrization \cite{Bazavov:2009zn} of
the EoS, as well as in the s95p parametrization of the EoS that is
frequently used in hydrodynamic models \cite{Huovinen:2009yb}.

\begin{table*}
\begin{tabular}{|c|c|l|r|r|c|l|}
\hline
$T [{\rm MeV}]$ & $\Theta^{\mu\mu}/T^4$ & \multicolumn{1}{c|}{$p/T^4$} & \multicolumn{1}{c|}{$\epsilon/T^4$}  & 
\multicolumn{1}{c|}{$s/T^3$} &  $C_V/T^3$ & \multicolumn{1}{c|}{$c_s^2$} \\
\hline
130 &  1.01(-10)(+19) &  0.439(-44)(+65) &   2.33(-16)(+33) &   2.77(-20)(+39)  & 16.5(-0.9)(+3.0)  & 0.168(-15)(+6)  \\
135 &  1.21(-21)(+23) &  0.481(-67)(+69) &   2.65(-35)(+38) &   3.13(-41)(+44)  & 20.4(-3.0)(+3.1)  & 0.153(-10)(+12)  \\
140 &  1.46(-24)(+25) &  0.529(-72)(+75) &   3.05(-41)(+43) &   3.58(-47)(+50)  & 24.5(-3.2)(+3.4)  & 0.146(-8)(+9)  \\
145 &  1.76(-27)(+28) &  0.586(-78)(+82) &   3.52(-46)(+48) &   4.11(-52)(+55)  & 28.6(-3.4)(+3.6)  & 0.144(-6)(+7)  \\
150 &  2.09(-29)(+30) &  0.651(-85)(+89) &   4.05(-50)(+52) &   4.70(-58)(+60)  & 32.6(-3.6)(+3.6)  & 0.144(-6)(+7)  \\
155 &  2.43(-31)(+32) &  0.726(-93)(+97) &   4.61(-54)(+56) &   5.34(-62)(+65)  & 36.2(-3.6)(+3.5)  & 0.148(-7)(+7)  \\
160 &  2.76(-32)(+32) &  0.808(-100)(+105) &   5.19(-57)(+59) &   6.00(-66)(+68)  & 39.3(-3.4)(+3.3)  & 0.153(-8)(+9)  \\
165 &  3.07(-32)(+31) &  0.898(-108)(+112) &   5.76(-59)(+60) &   6.66(-69)(+70)  & 41.8(-3.2)(+3.0)  & 0.159(-9)(+10)  \\
170 &  3.34(-31)(+30) &  0.994(-115)(+118) &   6.32(-60)(+60) &   7.32(-70)(+71)  & 43.8(-2.9)(+2.7)  & 0.167(-10)(+10) \\ 
175 &  3.56(-29)(+28) &  1.094(-121)(+124) &   6.85(-60)(+60) &   7.94(-71)(+71)  & 45.2(-2.6)(+2.4)  & 0.176(-11)(+10)  \\
180 &  3.74(-27)(+25) &  1.197(-126)(+129) &   7.33(-59)(+59) &   8.53(-71)(+70)  & 46.2(-2.4)(+2.1)  & 0.185(-11)(+10)  \\
185 &  3.88(-25)(+23) &  1.302(-130)(+133) &   7.78(-58)(+57) &   9.08(-71)(+71)  & 47.0(-2.2)(+1.9)  & 0.194(-11)(+10)  \\
190 &  3.97(-22)(+19) &  1.406(-134)(+136) &   8.19(-57)(+56) &   9.60(-69)(+68)  & 47.5(-1.9)(+1.7)  & 0.202(-10)(+10)  \\
195 &  4.03(-19)(+16) &  1.510(-137)(+139) &   8.56(-56)(+54) &  10.07(-68)(+67)  & 47.9(-1.7)(+1.6)  & 0.210(-10)(+10)  \\
200 &  4.05(-16)(+14) &  1.613(-140)(+141) &   8.89(-54)(+52) &  10.50(-67)(+65)  & 48.1(-1.6)(+1.5)  & 0.218(-10)(+10)  \\
205 &  4.05(-14)(+14) &  1.713(-142)(+143) &   9.19(-52)(+50) &  10.90(-65)(+63)  & 48.4(-1.5)(+1.6)  & 0.225(-10)(+10)  \\
210 &  4.03(-15)(+15) &  1.810(-143)(+143) &   9.46(-50)(+48) &  11.27(-64)(+62)  & 48.6(-1.6)(+1.6)  & 0.232(-10)(+10)  \\
215 &  3.99(-16)(+16) &  1.904(-144)(+144) &   9.70(-48)(+47) &  11.61(-62)(+60)  & 48.8(-1.6)(+1.7)  & 0.238(-10)(+9)  \\
220 &  3.94(-17)(+17) &  1.995(-144)(+144) &   9.93(-47)(+46) &  11.92(-61)(+59)  & 49.1(-1.7)(+1.8)  & 0.243(-9)(+9)  \\
225 &  3.88(-18)(+17) &  2.083(-145)(+144) &  10.13(-46)(+45) &  12.21(-59)(+58)  & 49.4(-1.8)(+1.8)  & 0.247(-9)(+8)  \\
230 &  3.82(-18)(+18) &  2.168(-145)(+144) &  10.32(-45)(+44) &  12.49(-59)(+58)  & 49.8(-1.8)(+1.9)  & 0.251(-8)(+8)  \\
235 &  3.76(-19)(+18) &  2.249(-144)(+143) &  10.50(-45)(+44) &  12.75(-59)(+58)  & 50.3(-1.9)(+1.9)  & 0.254(-8)(+7)  \\
240 &  3.69(-19)(+19) &  2.328(-144)(+143) &  10.68(-44)(+44) &  13.00(-58)(+58)  & 50.7(-1.9)(+1.9)  & 0.256(-8)(+7)  \\
245 &  3.63(-20)(+19) &  2.403(-144)(+143) &  10.84(-44)(+44) &  13.24(-58)(+57)  & 51.1(-1.9)(+1.9)  & 0.259(-7)(+7)  \\
250 &  3.57(-20)(+20) &  2.476(-143)(+142) &  10.99(-44)(+44) &  13.47(-57)(+57)  & 51.5(-1.9)(+1.9)  & 0.261(-7)(+6)  \\
255 &  3.50(-21)(+20) &  2.546(-143)(+142) &  11.14(-44)(+44) &  13.68(-57)(+57)  & 51.9(-1.9)(+1.9)  & 0.264(-7)(+6)  \\
260 &  3.44(-21)(+21) &  2.613(-143)(+142) &  11.28(-44)(+44) &  13.89(-58)(+57)  & 52.2(-1.9)(+1.9)  & 0.266(-7)(+6)  \\
265 &  3.38(-21)(+21) &  2.678(-142)(+141) &  11.41(-44)(+44) &  14.09(-58)(+57)  & 52.5(-1.9)(+1.8)  & 0.268(-6)(+6)  \\
270 &  3.32(-21)(+21) &  2.741(-142)(+141) &  11.54(-44)(+44) &  14.28(-57)(+57)  & 52.8(-1.8)(+1.8)  & 0.270(-6)(+6)  \\
275 &  3.26(-21)(+21) &  2.801(-141)(+141) &  11.66(-44)(+44) &  14.46(-57)(+57)  & 53.1(-1.8)(+1.8)  & 0.272(-6)(+5)  \\
280 &  3.20(-21)(+21) &  2.859(-141)(+140) &  11.77(-44)(+43) &  14.63(-57)(+57)  & 53.3(-1.8)(+1.7)  & 0.274(-6)(+5) \\
285 &  3.14(-21)(+21) &  2.915(-141)(+140) &  11.88(-43)(+43) &  14.80(-57)(+57)  & 53.6(-1.8)(+1.7)  & 0.276(-5)(+5)  \\
290 &  3.08(-21)(+21) &  2.969(-140)(+140) &  11.99(-43)(+43) &  14.95(-57)(+56)  & 53.8(-1.7)(+1.7)  & 0.278(-5)(+5)  \\
295 &  3.02(-20)(+21) &  3.021(-140)(+140) &  12.08(-43)(+43) &  15.11(-56)(+56)  & 54.0(-1.7)(+1.7)  & 0.280(-5)(+5)  \\
300 &  2.96(-20)(+21) &  3.072(-140)(+139) &  12.18(-43)(+43) &  15.25(-56)(+56)  & 54.2(-1.7)(+1.7)  & 0.282(-5)(+6)  \\
\hline
\end{tabular}
\caption{Continuum extrapolated results for the trace anomaly 
$\Theta^{\mu\mu}$, pressure $p$, energy density
$\epsilon$, entropy density $s$, specific heat $C_V$, and the square of the speed of sound $c_s^2$ in appropriate units of the temperature $T$. The 
asymmetry in the errors, given in the two brackets, arises from the 2\% 
systematic error in asigned to the temperature scale.}
\label{tab:fin}
\end{table*}

\begin{table*}[t]
\begin{tabular}{|c|c|l|r|r|c|l|}
\hline
$T [{\rm MeV}]$ & $\Theta^{\mu\mu}/T^4$ & \multicolumn{1}{c|}{$p/T^4$} & \multicolumn{1}{c|}{$\epsilon/T^4$}  &
\multicolumn{1}{c|}{$s/T^3$} &  $C_V/T^3$ & \multicolumn{1}{c|}{$c_s^2$} \\
\hline
305 &  2.91(-20)(+21) &  3.120(-139)(+139) &  12.27(-43)(+42) &  15.39(-56)(+55)  & 54.3(-1.7)(+1.7)  & 0.283(-5)(+6)  \\
310 &  2.85(-20)(+20) &  3.167(-139)(+139) &  12.35(-42)(+42) &  15.52(-56)(+55)  & 54.5(-1.7)(+1.7)  & 0.285(-6)(+6)  \\
315 &  2.79(-19)(+20) &  3.212(-139)(+138) &  12.43(-42)(+42) &  15.64(-56)(+55)  & 54.6(-1.7)(+1.7)  & 0.286(-6)(+6)  \\
320 &  2.74(-19)(+20) &  3.256(-139)(+138) &  12.51(-42)(+41) &  15.76(-55)(+55)  & 54.8(-1.7)(+1.7)  & 0.288(-6)(+6)  \\
325 &  2.69(-19)(+20) &  3.298(-138)(+138) &  12.58(-42)(+41) &  15.88(-55)(+54)  & 54.9(-1.7)(+1.7)  & 0.289(-6)(+7)  \\
330 &  2.63(-19)(+19) &  3.338(-138)(+137) &  12.65(-41)(+41) &  15.99(-54)(+54)  & 55.0(-1.7)(+1.7)  & 0.291(-6)(+7)  \\
335 &  2.58(-19)(+19) &  3.377(-138)(+137) &  12.71(-41)(+41) &  16.09(-54)(+54)  & 55.1(-1.7)(+1.8)  & 0.292(-6)(+7)  \\
340 &  2.53(-19)(+19) &  3.415(-137)(+137) &  12.78(-41)(+40) &  16.19(-54)(+53)  & 55.2(-1.7)(+1.8)  & 0.293(-7)(+7)  \\
345 &  2.48(-20)(+19) &  3.452(-137)(+136) &  12.83(-41)(+40) &  16.29(-54)(+53)  & 55.3(-1.7)(+1.8)  & 0.294(-7)(+7)  \\
350 &  2.43(-20)(+19) &  3.487(-136)(+136) &  12.89(-40)(+40) &  16.38(-53)(+53)  & 55.4(-1.8)(+1.9)  & 0.296(-7)(+7)  \\
355 &  2.38(-20)(+19) &  3.521(-136)(+135) &  12.94(-40)(+40) &  16.47(-53)(+53)  & 55.5(-1.8)(+1.9)  & 0.297(-7)(+7)  \\
360 &  2.33(-20)(+20) &  3.554(-136)(+135) &  13.00(-40)(+40) &  16.55(-53)(+53)  & 55.6(-1.8)(+1.9)  & 0.298(-7)(+7)  \\
365 &  2.29(-21)(+20) &  3.586(-135)(+134) &  13.04(-40)(+40) &  16.63(-53)(+53)  & 55.7(-1.9)(+1.9)  & 0.299(-7)(+7)  \\
370 &  2.24(-21)(+20) &  3.617(-135)(+134) &  13.09(-40)(+40) &  16.71(-53)(+53)  & 55.8(-1.9)(+2.0)  & 0.300(-7)(+7)  \\
375 &  2.20(-21)(+20) &  3.647(-134)(+134) &  13.14(-40)(+40) &  16.78(-53)(+53)  & 55.8(-1.9)(+2.0)  & 0.301(-7)(+7)  \\
380 &  2.15(-22)(+21) &  3.675(-134)(+133) &  13.18(-40)(+40) &  16.85(-53)(+53)  & 55.9(-2.0)(+2.0)  & 0.302(-7)(+7)  \\
385 &  2.11(-22)(+21) &  3.703(-134)(+133) &  13.22(-40)(+41) &  16.92(-53)(+53)  & 56.0(-2.0)(+2.0)  & 0.302(-7)(+7)  \\
390 &  2.07(-22)(+21) &  3.730(-133)(+132) &  13.26(-40)(+41) &  16.99(-53)(+53)  & 56.1(-2.0)(+2.1)  & 0.303(-7)(+7)  \\
395 &  2.03(-22)(+22) &  3.756(-133)(+132) &  13.30(-40)(+41) &  17.05(-53)(+53)  & 56.2(-2.0)(+2.1)  & 0.304(-7)(+7)  \\
400 &  1.99(-22)(+22) &  3.782(-132)(+132) &  13.34(-40)(+41) &  17.12(-53)(+53)  & 56.2(-2.1)(+2.1)  & 0.304(-7)(+7)  \\
\hline
\end{tabular}

\vspace{2mm}
\centerline{Table~\protect\ref{tab:fin} continued}
\end{table*}

The final continuum extrapolated estimates of the pressure, energy
density and entropy density are shown in Fig.~\ref{fig:cont} (right)
and compared with HRG predictions for $T<170$~MeV. Again, there is
reasonable agreement for $T<150$~MeV. Above $T=150$~MeV, HRG estimates
lie along the lower edge of the error-band of the lattice estimates.

\begin{figure}
\includegraphics[width=8cm]{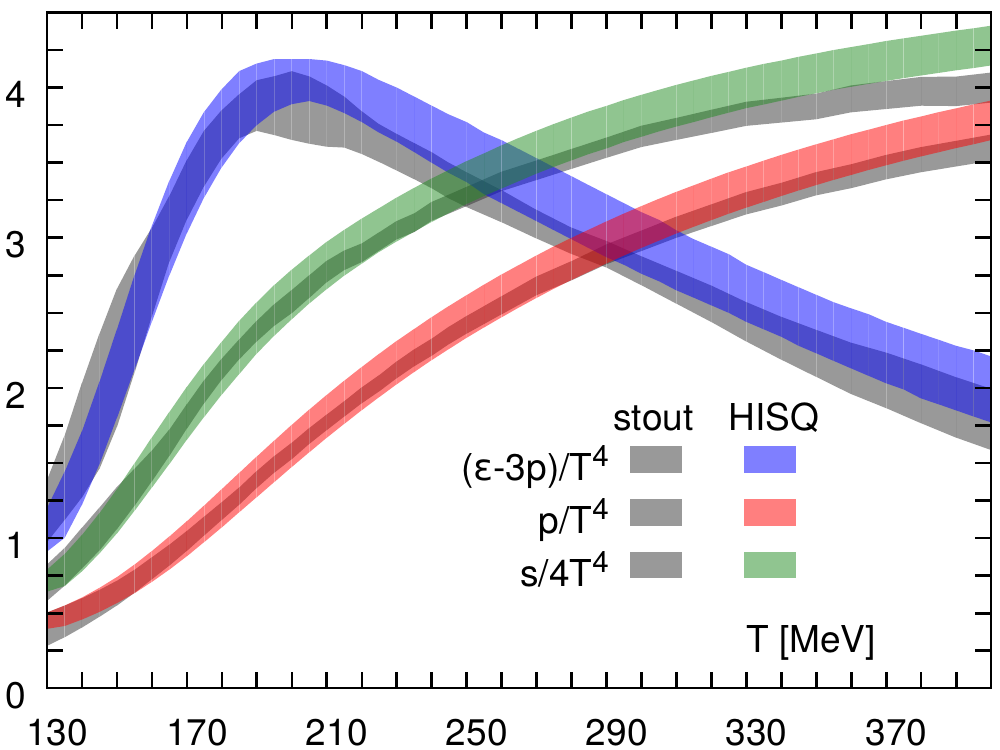}
\caption{The comparison of the HISQ/tree and stout results for the trace anomaly,
the pressure, and the entropy density. 
}
\label{fig:comp}
\end{figure}

We can now compare our results with the results obtained by the
Wuppertal-Budapest Collaboration using the stout action
\cite{Borsanyi:2013bia}. This comparison is shown in
Fig. \ref{fig:comp} for the trace anomaly, the pressure and the
entropy density. We find good agreement in the trace anomaly with the
stout results over the full temperature range ($130-400$)~MeV. Note,
however, that above the peak the central values with the stout action
lie systematically below ours.  As a result, our estimates of the
pressure become
systematically larger for $T>200$ MeV. By $T=400$~MeV, the difference
between the central values in the two calculations increases to about
$6\%$. The two results, however, still agree within errors.  The
difference in the entropy density reaches about $7\%$ by $T=400$~MeV,
and in this case the two estimates differ by about $2\sigma$.  These
differences suggest that more detailed calculations of the trace
anomaly at higher temperatures are needed. In particular, it would be
important to see if the differences persist at higher temperatures
where a comparison with resummed perturbative calculations should be
possible (see Sec.~\ref{sec5}.C).

\subsection{Parametrization of the equation of state}

We close this section by providing an analytical parametrization of
the pressure of (2+1)-flavor QCD, summarized in Table~\ref{tab:fin},
that can be used in phenomenological applications and hydrodynamic
modeling of strong interaction matter. We choose an ansatz that
incorporates basic features of the low and high temperature limits,
{\it i.e.}, it ensures that the pressure becomes exponentially small at low
temperatures and approaches the ideal gas limit at high
temperatures. We find that the following parametrization provides an
excellent description of all bulk thermodynamic observables discussed in
the previous sections, including the specific heat and speed of sound
that require second derivatives of $p/T^4$ with respect to the
temperature to be discussed in the next section,
\begin{eqnarray}
&&
\frac{p}{T^4} = \frac{1}{2}\hspace{-0.03cm}\left( 1+ \tanh (c_t (\bar{t} -t_0)) \right) \cdot  \nonumber \\
&&
\hspace{-0.3cm}\frac{p_{id}+ a_n/\bar{t}+b_n/\bar{t}^2+c_n/\bar{t}^3+d_n/\bar{t}^4}{1+a_d/\bar{t}+b_d/\bar{t}^2+c_d/\bar{t}^3+d_d/\bar{t}^4} \; ,
\label{eosfit}
\end{eqnarray}
where $\bar{t} = T/T_c$ and the QCD transition temperature
$T_c=154$~MeV is a conveniently chosen normalization.  In this
parametrization, $p_{id}= 95 \pi^2/180$ is the ideal gas value of
$p/T^4$ for massless 3-flavor QCD. It is also the appropriate infinite
temperature limiting value for QCD with light and strange quarks 
that could be refined to include additional perturbative
corrections. However, at present we do not see any need for this. We
also note that fixing $c_n=c_d=0$ gives an excellent
parametrization of all our numerical data and is in good
agreement with the HRG estimate, at least down to
$T=100$~MeV. Furthermore, this parametrization agrees with the 
$N_\tau=8$ data well beyond $T=400$~MeV.

\begin{table}[t]
\begin{center}
\begin{tabular}{|c|c|c|c|c|}
\hline
$c_t$&$a_n$&$b_n$&$c_n$&$d_n$ \\
3.8706&-8.7704&3.9200&0&0.3419 \\
\hline
\hline
$t_0$&$a_d$&$b_d$&$c_d$&$d_d$ \\
0.9761&-1.2600&0.8425&0&-0.0475 \\
\hline
\end{tabular}
\end{center}
\caption{Parameters used in the ansatz given in Eq.~(\ref{eosfit}) for 
the pressure of (2+1)-flavor QCD in the temperature interval
$T\in [100~{\rm MeV}, 400~{\rm MeV}]$.}
\label{tab:eosfit}
\end{table}

The values of the parameters in our ansatz for the pressure,
Eq.~(\ref{eosfit}), are summarized in Table~\ref{tab:eosfit}.
The results of this ansatz for the speed of sound, energy density, and specific
heat are compared with our continuum extrapolated error bands 
in Figs.~\ref{fig:cs2} and \ref{fig:Cvfit}.

\section{Specific heat, the speed of sound and deconfinement}
\label{sec5}

\begin{figure}
\includegraphics[width=8cm]{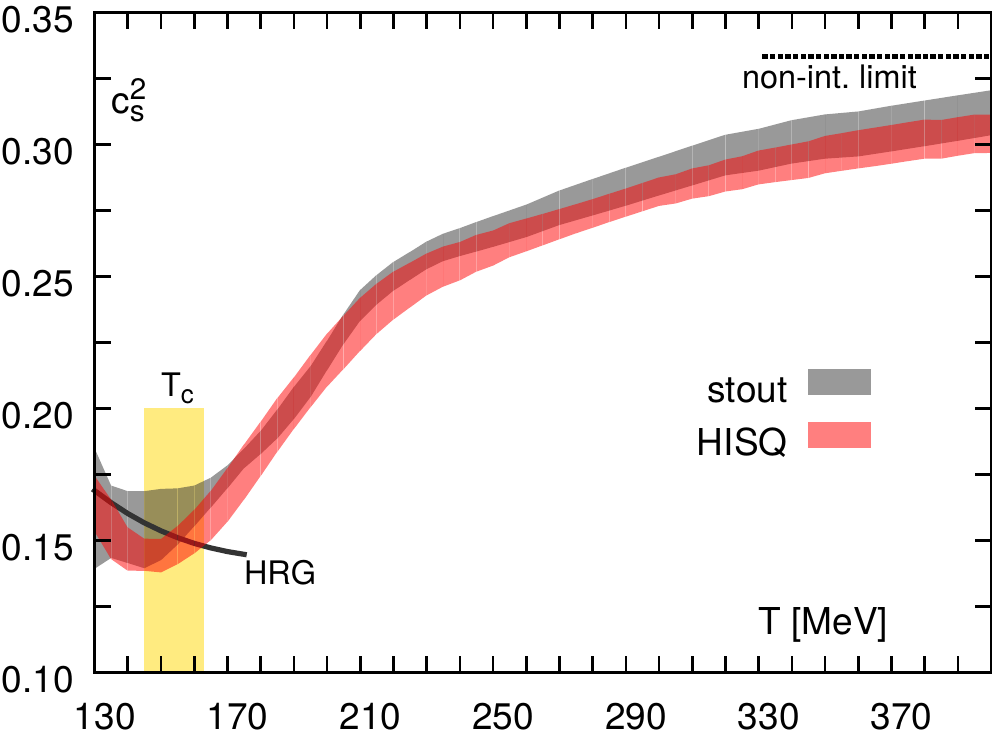}

\includegraphics[width=8cm]{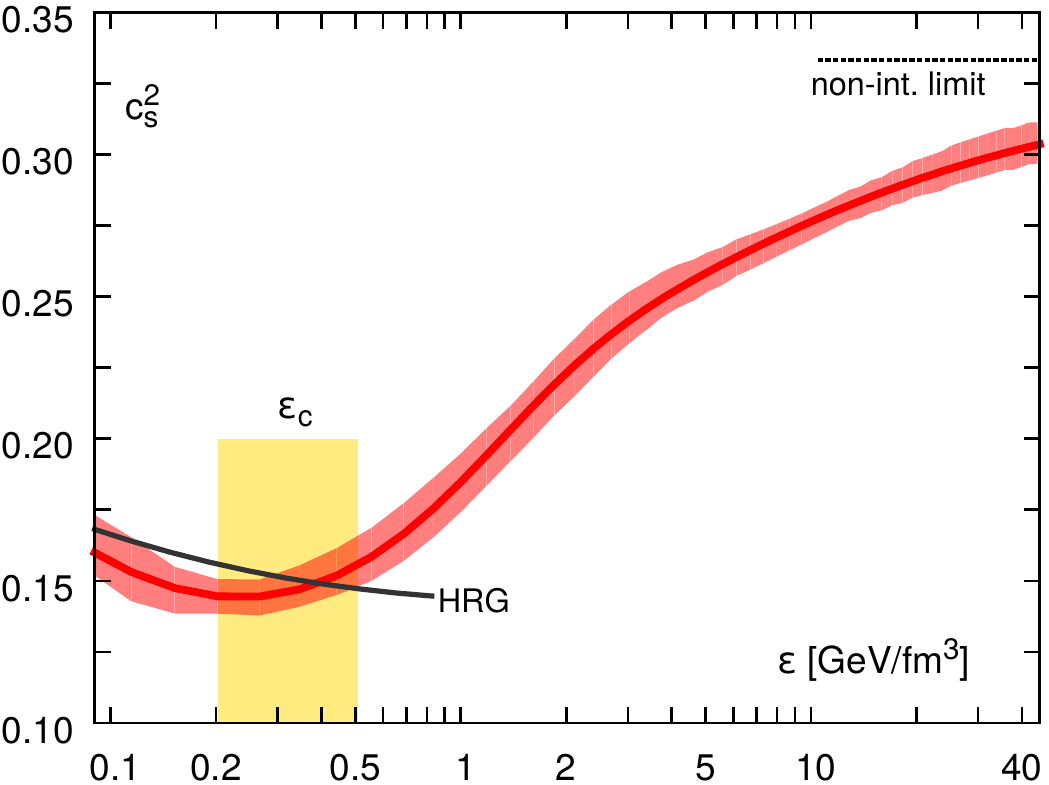}
\caption{The speed of sound squared from lattice QCD and the HRG model
versus temperature (top) and energy density (bottom). In the upper figure,  
our results (HISQ) are compared with those obtained with the stout 
action \cite{Borsanyi:2013bia}. 
The vertical band marks the location of the crossover region
$T_c=(154\pm 9)$~MeV in the upper figure and the corresponding range in 
energy density , $\epsilon_c=(0.18-0.5)~{\rm GeV/fm}^3$,
in the lower figure. The dark line within each error band is the prediction of 
the analytical parameterization given in Eq.~(\protect\ref{eosfit}).}
\label{fig:cs2}
\end{figure}

All thermodynamic quantities, for fixed light and strange quark
masses, depend on a single parameter---the temperature.  In
Section~\ref{sec4}, we derived the basic thermodynamic observables
($\epsilon,~p,~s$) from the continuum extrapolated trace anomaly
$\Theta^{\mu\mu}(T)$.  We now discuss two closely related observables
that involve second order derivatives of the QCD partition function
with respect to the temperature, {\it i.e.}, the specific heat,
\begin{equation}
C_V=\left. \frac{\partial \epsilon}{\partial T}\right|_V \equiv \left( 
4 \frac{\epsilon}{T^4} + T 
\left. \frac{{\partial} (\epsilon/T^4)}{\partial T}\right|_V \right) T^3 
\, ,
\label{CV}
\end{equation}
and the speed of sound,
\begin{equation}
c_s^2 = \frac{\partial p}{\partial \epsilon}  = 
\frac{\partial p/\partial T}{\partial \epsilon/\partial T} 
= \frac{s}{C_V} \; .
\label{cs}
\end{equation}
The quantity $T {{\rm d} (\epsilon/T^4)}/{{\rm d} T}$ can be calculated
directly from the trace anomaly and its derivative with respect to
temperature,
\begin{equation}
T \frac{{\rm d} \epsilon/T^4}{{\rm d} T} = 3 \frac{\Theta^{\mu\mu}}{T^4}
+T \frac{{\rm d} \Theta^{\mu\mu}/T^4}{{\rm d} T} \; .
\label{dedT}
\end{equation}
These identities show that the estimates for the specific heat and the
speed of sound should be of a quality similar to $\epsilon/T^4$ or
$p/T^4$.  In Figs.~\ref{fig:cs2} and \ref{fig:Cvfit}, we show the agreement 
between the bootstrap error bands for these quantities and the estimates
obtained by taking second order derivatives of the analytic
parameterization for $p/T^4$ given in Eq.~\ref{eosfit}.  The latter are 
shown as dark lines inside the bootstrap error bands. 

\subsection{Speed of sound, the softest point of the EoS and the 
critical energy density}

In Fig.~\ref{fig:cs2} (top), we show the speed of sound as a function
of the temperature and compare our results with those obtained by
using the stout action \cite{Borsanyi:2013bia}.  We find that the HISQ/tree 
and the stout results
agree within the estimated errors. The softest point of the EoS
\cite{Hung:1994eq} at $T\simeq (145-150)$~MeV, {\it i.e.}, at the
minimum of the speed of sound, lies on the low temperature side of the
crossover region.  At this point, the speed of sound is
only slightly below the corresponding HRG value. This follows from the
good agreement between HRG estimates and our lattice QCD results for
the energy density and the pressure. Furthermore, the value
$c_s^2\simeq 0.15$ is roughly half way between zero, the value
expected at a second order phase transition with diverging specific
heat\footnote{In the case of QCD the 
specific heat and therefore also the speed of sound stays finite even 
at a second order phase transition in the chiral limit.}, and the value 
for an ideal massless gas, $c_s^2=1/3$. At the
high temperature end, $T\sim350$ MeV, it reaches within $10\%$ of the
ideal gas value.

The softest point of the EoS is of interest in the phenomenology of
heavy ion collisions as it characterizes the temperature and energy
density range in which the expansion and cooling of matter slows
down. The system spends a longer time in this temperature range, and
one expects to observe characteristic signatures from this regime.  To
facilitate a more direct comparison with experiments, we show $c_s^2$ as a function of
the energy density in physical units in Fig.~\ref{fig:cs2} (bottom)
using the parametrization given in Eq.~\ref{eosfit} to convert
temperature to energy density. At the softest point, the energy density
is only slightly above that of normal nuclear matter,
$\epsilon_{\rm nuclear}=150~{\rm MeV}/{\rm fm}^3$.  In the crossover
region, $T_c=(154 \pm 9)$~MeV~\cite{Bazavov:2011nk}, 
the energy density varies from $180~{\rm MeV}/{\rm fm}^3$ at the
lower edge to $500~{\rm MeV}/{\rm fm}^3$ at the upper edge, slightly above
the energy density inside the proton $\epsilon_{\rm proton}=~450~{\rm
MeV}/{\rm fm}^3$. 

The QCD crossover region, thus, starts at or close to the softest
point of the EoS and the entire crossover region corresponds to
relatively small values of the energy density, $(1.2-3.1)
\epsilon_{\rm nuclear}$. This value is about a factor of four smaller than that
of an ideal quark-gluon gas in this temperature range.  In the next
subsection, we will discuss to what extent this has consequences for
the size of fluctuations in the energy density, {\it i.e.}, the specific
heat.

\subsection{Specific heat and deconfinement}

\begin{figure}
\includegraphics[width=8cm]{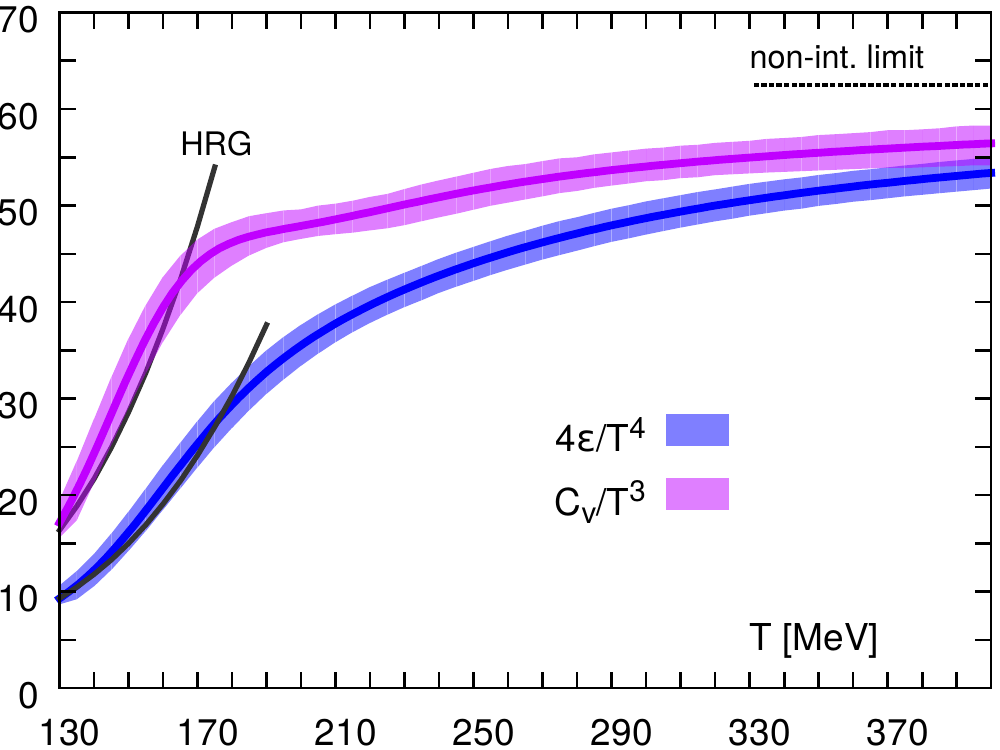}
\caption{Error bands showing the continuum extrapolation of the
  specific heat and energy density and solid lines obtained from 
  the parametrization given in Eq.~(\ref{eosfit}). Also shown are
  the HRG estimates at low temperatures and
  the ideal gas limit at high temperatures.  }
\label{fig:Cvfit}
\end{figure}

The intuitive characterization of deconfinement at the 
QCD phase transition is that the liberation of many new degrees
of freedom give rise to a rapid increase in the energy density, ideally
with an infinite slope at $T_c$ 
as in a conventional second order phase transition.
This rapid rise would then show up as a peak (or even a divergence) 
in the specific heat,
which could serve as an indicator for the pseudo-critical (or critical)
temperature. However, the specific heat of (2+1)-flavor QCD, shown in 
Fig.~\ref{fig:Cvfit}, exhibits a rapid increase but no peak. In 
the crossover region, $C_V/\epsilon\simeq 8/T$ 
is a factor of two larger than for an ideal quark-gluon gas; 
the specific heat reaches about half of its ideal gas value,
$(C_V/T^3)_{ideal} = 4 (\epsilon/T^4)_{ideal}= 95\pi^2/15$; and 
the energy density reaches only about one quarter of its
limiting high temperature, ideal gas value. 

The analysis of the quark-mass dependence of the QCD transition
temperature, the chiral condensate and, in particular, the peak in the
chiral susceptibility suggest that for physical values of the quark
masses QCD is sufficiently close to the chiral limit to be sensitive
to the chiral phase transition~\cite{Bazavov:2011nk} and exhibit an
almost universal, pseudo-critical behavior controlled by it. The peak
observed in the chiral susceptibility is dominated by the second
derivative of the singular part of the free energy with respect to the
quark mass~\cite{Bazavov:2011nk}. Extending that generic scaling
analysis, one may have expected that, for physical quark masses, the
pseudo-critical behavior would also lead to large fluctuations in the
energy density and that the specific heat would exhibit a peak in the
crossover region controlled by the second derivative with respect to
temperature of the same singular part of the free energy.

There may be at least two reasons for the difference in behavior
between the chiral susceptibility and the specific heat, which are
second derivatives of the partition function with respect to the quark
mass and the temperature, respectively. First, thermal fluctuations
are controlled by the thermal critical exponent $\alpha$, {\it i.e.}, $C_V
/T^3 \sim |T-T_c|^{-\alpha}$.  In the 3-d $O(4)$ universality class,
which is relevant for the chiral phase transition, the exponent
$\alpha \simeq -0.21$ is negative~\cite{Engels:2011km}. Consequently, 
unlike the chiral susceptibility, the specific heat stays
finite at $T_c$ even in the chiral limit. The singular part of the
free energy~\cite{Engels:2011km},  which gives the leading temperature
dependence in the vicinity of $T_c$, contributes only a cusp in $C_V$. 
This can be seen by examining the energy density near $T_c$, 
\begin{equation}
\frac{\epsilon}{T^4} = e_0 + e_1 \left(\frac{T-T_c}{T_c}\right) +
{\cal O}\left( |T-T_c|^{1-\alpha} \right) \; ,
\end{equation}
where the dominant contribution, $e_0$, comes from the regular part
and the singular contributions, $\sim |T-T_c|^{1-\alpha}$, are
sub-dominant.  From Eq.~(\ref{CV}), we get 
\begin{equation}
\frac{C_V}{T^3} = c_0 + \frac{A^{\pm}}{\alpha} \left| \frac{T-T_c}{T_c} \right|^{-\alpha}  +
{\cal O}\left( T-T_c \right) \; ,
\label{Cvbar}
\end{equation}
with $c_0=4e_0+e_1$ and $A^+$ ($A^-$) are the amplitudes above (below)
$T_c$. The ratio of these amplitudes is universal and positive;
$A^+/A^- = 1.842(43)$ in the 3-d O(4) universality
class~\cite{Engels:2011km}. Since $\alpha$ is negative, the singular
part gives only a cusp, which should persist in the chiral limit
but may not be easy to detect if the regular contributions are large.

\begin{figure}
\includegraphics[width=8cm]{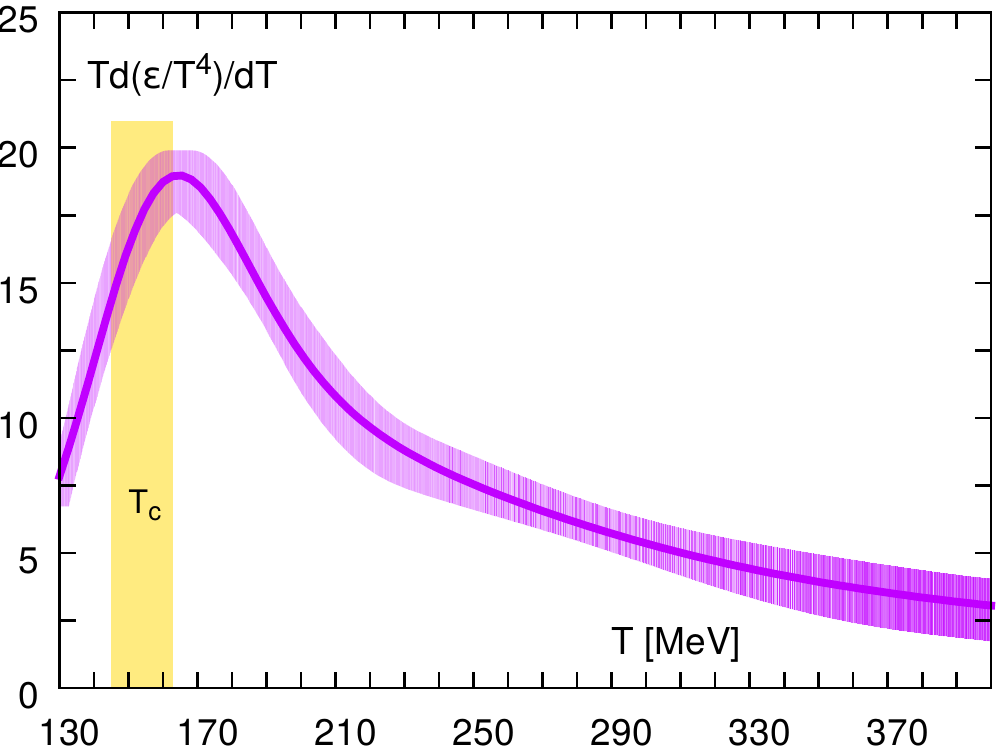}
\caption{Derivative of $\epsilon/T^4$ with
respect to temperature. The vertical band gives the
chiral crossover temperature determined from the location of the peak
in the disconnected chiral susceptibility.}
\label{fig:CVtilde}
\end{figure}

The second reason for the lack of a peak in $C_V/T^3$ is that the 
contributions from the regular
part of the free energy are large in the high temperature phase~\cite{Engels:2011km}, 
and are ${\cal O}(g^0)$ at infinite temperature. 
Furthermore, as discussed above, the regular terms dominate 
even in the crossover region. To make this observation more explicit, we note 
from Eq.~(\ref{CV}) that $C_V/T^3$ can be written 
in terms of the energy density, $\epsilon/T^4$, and its derivative,
\begin{equation}
T \frac{{\rm d} \epsilon/T^4}{{\rm d} T}  \equiv \frac{\overline{C}_V}{T^3} \; .
\label{Cbar}
\end{equation} 
The dominant singular terms are contained in the second term
($\overline{C}_V/T^3$) or, more specifically, in the temperature
derivative of the trace anomaly, {\it i.e.}, the second term in
Eq.~(\ref{dedT}).  The contribution of the regular terms to
$\overline{C}_V/T^3$ is strongly suppressed at high temperatures; it
is zero in the infinite temperature ideal gas limit and receives
contributions starting at ${\cal O}(g^4)$ in perturbation
theory. Thus, while $C_V$ and $\overline{C}_V$ have identical leading
contributions from the singular part near $T_c$, the contribution from
the regular part is much smaller in $\overline{C}_V$. Consequently,
the singular behavior is not masked and $\overline{C}_V$ has a
pronounced peak close to the chiral crossover region as shown in
Fig.~\ref{fig:CVtilde}.  To summarize, the location of the peak in the
temperature derivative of $\epsilon/T^4$ is a good indicator of
deconfinement, {\it i.e.}, the liberation of quark-gluon degrees of freedom,
and occurs close to the chiral transition in QCD as shown in
Fig.~\ref{fig:CVtilde}.

\subsection{Approach to the perturbative limit}

In this subsection, we discuss how our results for the (2+1)-flavor EoS
connect to analytic calculations at high temperatures.

At sufficiently high temperatures, thermodynamics should be
describable in terms of a weakly interacting quark-gluon gas, and at
infinite temperature all thermodynamic quantities will converge to the
ideal gas limit. Plots in Fig.~\ref{fig:cont} show that at our highest
temperature value, $T=400$~MeV, the entropy and energy density and
pressure are still $13\%$, $18\%$ and $27\%$, respectively, below the
ideal gas limit.  In contrast to other quantities, {\it e.g.},
susceptibilities of conserved charge fluctuations, these deviations
from the ideal gas limit are still quite large. This is probably due
to large nonperturbative contributions in the gluonic sector of QCD
which are present in bulk thermodynamic observables but are suppressed
in observables that, at tree level, only depend on the quark sector of
QCD.

Although, for some observables, resummation \cite{Haque:2014rua} or
dimensional reduction \cite{Laine:2006cp} based perturbative
calculations show good agreement with lattice QCD calculations already
at temperatures $T\sim 400$~MeV, for others this is not the case.  In
particular, their functional dependence on temperature is still
significantly different in this temperature range, which can lead to
larger differences in higher order derivatives between perturbative
and lattice QCD calculations. As our current continuum-extrapolated
EoS is limited to $T<400$~MeV, we cannot perform a detailed comparison
with perturbation theory but point out a few qualitative features.

\begin{figure}
\includegraphics[width=8cm]{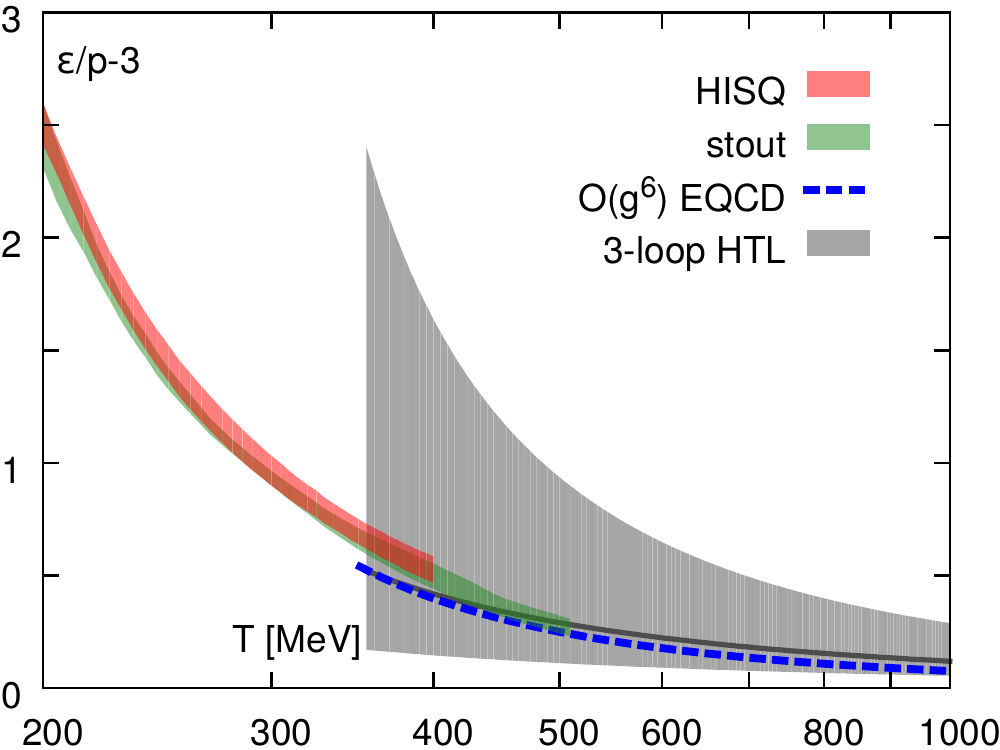}
\caption{The ratio of the trace anomaly and the pressure
from (2+1)-flavor QCD calculations with the HISQ and stout actions,
respectively. These results are compared to HTL and EQCD (dashed line) 
calculations. 
The black line corresponds to the HTL calculation with renormalization 
scale $\mu=2 \pi T$.}
\label{fig:ep}
\end{figure}

We have shown in Fig.~\ref{fig:comp} that continuum-extrapolated
results for the trace anomaly obtained with the stout and the HISQ
discretization schemes agree within errors. At high temperatures,
however, the HISQ results are systematically above the stout
results. This propagates into other thermodynamic observables, {\it e.g.},
the pressure. The systematic differences, however, cancel to a large
extent in ratios. For example, the ratio of the trace anomaly and the 
pressure,
\begin{equation}
\frac{\Theta^{\mu\mu}}{p} = \frac{\epsilon}{p} -3 \; ,
\label{ratio}
\end{equation}
is in excellent agreement between the two calculations and thus
provides a good starting point for a comparison with high temperature
perturbative calculations.  In Fig.~\ref{fig:ep}, we show results for
the ratio $\Theta^{\mu\mu} / p$ and compare with perturbative
calculations performed in the Hard Thermal Loop (HTL)
\cite{Haque:2014rua} and Electrostatic QCD (EQCD) \cite{Laine:2006cp}
schemes. The broad band for the three-loop HTL calculation corresponds
to varying the renormalization scale in the interval $\mu= (1-4)\pi T$
and the black line in this band corresponds to $\mu = 2 \pi T$.  The
EQCD and HTL results for $\mu =2 \pi T$ are in good agreement, and the
lattice QCD results approach these estimates for $T\gsim 500$~MeV.

\begin{figure}
\includegraphics[width=8cm]{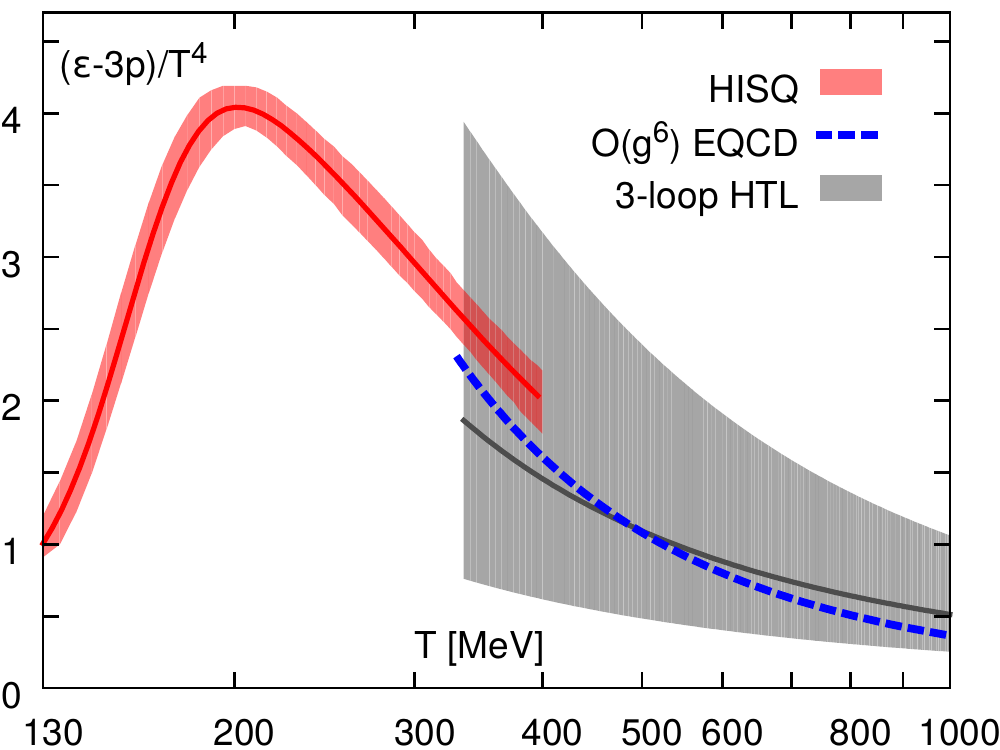}
\includegraphics[width=8cm]{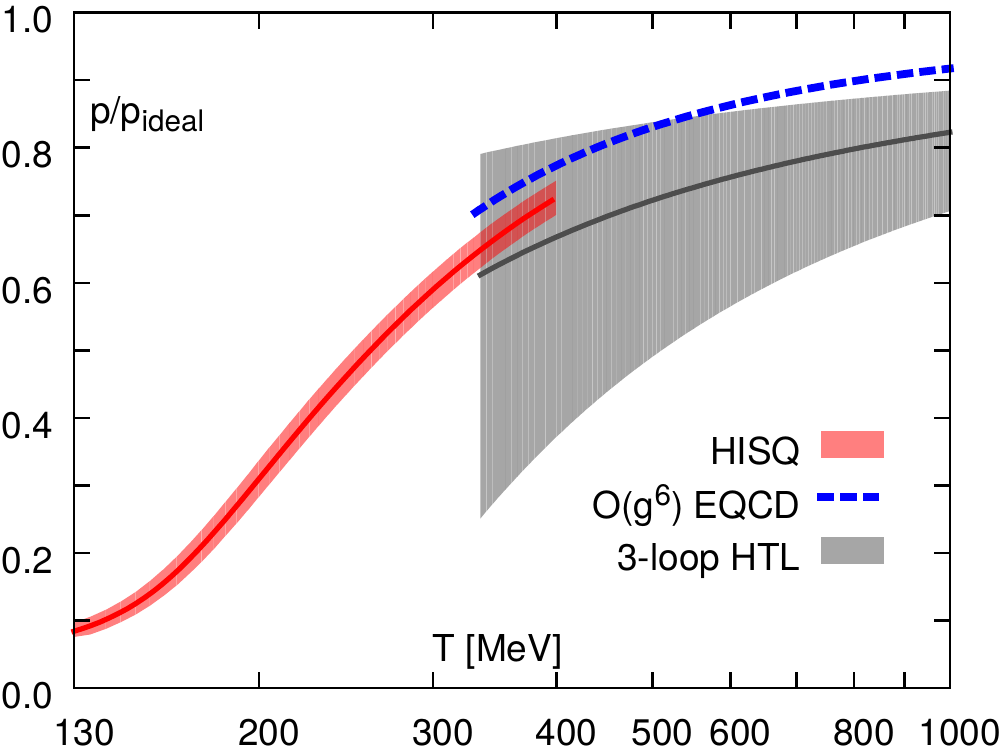}
\caption{Comparison of the (2+1)-flavor calculation of the trace
  anomaly (top) and pressure (bottom) with HTL and EQCD (dashed line)
  calculations.  The black line corresponds to the HTL calculation
  with renormalization scale $\mu=2 \pi T$.  Note that this solid line
  would move up for the trace anomaly and move down for the pressure
  if the scale $\mu$ in HTL is reduced.  }
\label{fig:pert}
\end{figure}

In Fig.~\ref{fig:pert}, we compare the three-loop HTL estimates 
with lattice QCD calculations of the trace anomaly (top) and the
pressure (bottom). Also shown, with a dashed line in Figs.~\ref{fig:ep}
and~\ref{fig:pert}, is the result of an ${\cal O}(g^6)$ calculation 
performed in the dimensional reduction scheme (EQCD). The lattice QCD
results are in qualitative agreement with these perturbative
calculations, with the ${\cal O}(g^6)$ EQCD estimate lying below the
lattice QCD results for the trace anomaly and above for the pressure
at $T=400$~MeV. 

One could try fixing the scale uncertainty in the HTL calculation by
matching one of the observables to the lattice QCD result, {\it e.g.}, the
pressure.  Results for other observables, {\it e.g.}, the trace anomaly
would then be parameter free predictions. It is clear from
Fig.~\ref{fig:pert} that such a simultaneously agreement between HTL
and lattice QCD calculations of $p/T^4$ and $(\epsilon -3p)/T^4$ is
not forthcoming. Making the HTL and the lattice QCD estimates agree
for the trace anomaly by reducing the value for the renormalization
scale $\mu$ would decrease the HTL results for $p/T^4$ even further
and, thereby, increase the deviation from the lattice QCD results.

Lastly, the EQCD result for the pressure in the temperature range
$(400-1000)$~MeV is about 10\% larger than the HTL result with $\mu =
2 \pi T$. To resolve the open question whether at these high
temperatures the pressure obtained from lattice QCD calculations is
better described by the HTL or the EQCD calculations requires lattice
simulations at higher temperatures.

\section{Conclusions}
\label{sec6}

We have calculated the trace anomaly and the equation of state in
(2+1)-flavor QCD with almost physical quark masses using the HISQ/tree
action on lattices with temporal extent $N_{\tau}=6,~8,~10$, and $12$.
We find that the lattice discretization errors in the HISQ/tree action
are small, and we obtain reliable continuum extrapolated results for a
number of thermodynamic quantities for $130~{\rm MeV} < T < 400~{\rm MeV}$.  
In fact, the trace anomaly calculated on the $N_{\tau}=12$ lattices
agrees with the continuum-extrapolated results within errors.  Our
main results are summarized in Figs.~\ref{fig:e-3p}, \ref{fig:cont},
and \ref{fig:comp}.  Based on these results, we propose in
Eq.~\ref{eosfit} an analytical parameterization of the pressure for
use in phenomenological studies that matches the HRG estimates below
$T = 130$~MeV and the lattice data between $(130-400)$~MeV.

We have compared our new results obtained using the HISQ/tree action
with our previous calculations performed using the asqtad and the p4
actions~\cite{Bazavov:2009zn}, and with the recent continuum
extrapolated stout results~\cite{Borsanyi:2013bia}.  For $T<300$~MeV,
the HISQ/tree results are very different from the results obtained
using the p4 and the asqtad actions on $N_\tau=8$ lattices, $i.e.$, without
an extrapolation to the continuum limit. At higher temperatures, the
results show reasonable agreement as expected since all three actions 
have small lattice artifacts.

Results for our continuum extrapolated trace anomaly presented in
Sec.~\ref{sec4} agree well with those from the stout
action~\cite{Borsanyi:2013bia}.  The discrepancy between the HotQCD
results and the stout results discussed
in~\cite{Bazavov:2009zn,Borsanyi:2013bia} was due to the large cutoff
effects in the previous estimates with the p4 and the asqtad actions and because
our  earlier results had not been extrapolated to the continuum limit.

We find reasonably good agreement for the pressure obtained using the
stout and the HISQ/tree actions for $T<300$~MeV as shown in
Fig.~\ref{fig:comp}. At higher temperatures, there is some tension
between the two estimates because the results for the trace anomaly,
$\Theta^{\mu\mu}(T)/T^4$, obtained 
with the HISQ/tree action lie systematically above those
from the stout action. Consequently, the pressure, which is the
integral of $\Theta^{\mu\mu}(T)/T^5$, will start to differ significantly at high
temperatures if the observed trends persist.  In this paper, we
focused on the temperature region $130~{\rm MeV} < T < 400~{\rm MeV}$,
which is the most relevant for phenomenological applications. Over
this temperature range, the difference is unlikely to have a significant
effect on the modeling of the hydrodynamic evolution of the system
produced in heavy ion collisions (see the discussion in
Ref.~\cite{Huovinen:2009yb}).  It is important to check, however, if
this tension persists at higher temperatures, especially if one wants
to determine to what extent the quark-gluon plasma is strongly or
weakly coupled by comparing lattice and resummed perturbation theory
results for the pressure or for the entropy density. Such calculations
are left for future studies.

\section*{Acknowledgments}
\label{ackn}
This work has been supported in part by contracts DE-AC02-98CH10886,
DE-AC52-07NA27344, DE-FC02-12ER41879, DE-FG02-92ER40699, DE-FG02-91ER-40628,
DE-FG02-91ER-40661, DE-FG02-04ER-41298, DE-KA-14-01-02, DE-SC0010120 with the
U.S. Department of Energy, and NSF grants PHY07-03296, PHY07-57333, PHY10-67881, 
PHY08-57333, PHY-1212389, and PHY13-16748, 
the Bundesministerium f\"ur Bildung und
Forschung under grant 06BI9001  and 05P12PBCTA, and the EU Integrated
Infrastructure Initiative HadronPhysics3.  The numerical simulations
have been performed on BlueGene/L computers GPU cluster (Edge) at Lawrence Livermore
National Laboratory (LLNL), the New York Center for Computational
Sciences (NYCCS) at Brookhaven National Laboratory, on BlueGene/P and BlueGene/Q computers at
Argonne Leadership Computing facility, on BlueGene/P
computers at NIC, Juelich, US Teragrid (Texas Advanced Computing
Center), at NERSC, GPU clusters at University of Bielefeld, the OCuLUS cluster at University of Paderborn, and on clusters of the USQCD collaboration in JLab and FNAL.
This research was supported in part by Lilly Endowment, Inc.,
        through its support for the Indiana University Pervasive
        Technology Institute, and in part by the Indiana METACyt
        Initiative. The Indiana METACyt Initiative at IU is also
        supported in part by Lilly Endowment, Inc.
We thank Nathan Brown for help with the $w_0$ scale calculations
and Michael Strickland for providing us with data from the HTL 
resummed perturbative calculations.
\appendix
\section{HISQ ensembles and topological charge history}\label{app_ensembles}

\subsection{HISQ ensembles}

To simulate the HISQ/tree action, we use the same Rational Hybrid
Monte Carlo algorithm~\cite{Clark:2004cp} with mass preconditioning
\cite{Hasenbusch:2001ne} as in the previous study
Ref.~\cite{Bazavov:2011nk}. Details of these simulations are given in
Ref.~\cite{Bazavov:2010ru} and in Table \ref{tab:hisq_0.05ms_runs} we
present the key lattice parameters of our simulation, namely the gauge
coupling $\beta=10/g^2$, the quark masses, the lattice dimensions, the
accumulated statistics in terms of molecular dynamics time units (TU),
and the length of the trajectories.  The zero temperature lattices
were saved every 5 TUs (or 6 TU for the fine lattices), and the finite
temperature lattices were saved every 10 TUs.

\begin{table*}[p]
\begin{tabular}{|c|c|c||c|r|c||c|r|c|c|}
\hline
\multicolumn{3}{|c||}{ } & \multicolumn{3}{c||}{$T=0$} & $N_\tau=6$ & $N_\tau=8$ & $N_\tau=10$ & $N_\tau=12$ \\
\hline
$\beta$ &  $m_l$  &  $m_s$  & $N_s^3\times N_\tau$ & $TU$~  & length & $TU$ & \multicolumn{1}{c|}{$TU$} & $TU$ & $TU$    \\
\hline
5.900   & 0.00660 & 0.1320  & $24^3 \times 32$     & 3700     &    1/4 & 30290 & ---    & ---    & ---  \\       
5.950   & 0.00615 & 0.1230  & $24^3 \times 32$     & 4715     &    1/4 & 30990 & ---    &        & ---  \\
6.000   & 0.00569 & 0.1138  & $24^3 \times 32$     & 4890     &    1/3 & 31730 & ---    & ---    & ---  \\
6.025   & 0.00550 & 0.1100  & $24^3 \times 32$     & 5250     &    1/3 & 33990 & ---    & ---    & ---  \\
6.050   & 0.00532 & 0.1064  & $24^3 \times 32$     & 4655     &    1/3 & 32100 & ~74210  & ---    & ---  \\
6.075   & 0.00518 & 0.1036  & $24^3 \times 32$     & 4085     &    1/3 & 32990 & ---    & ---    & ---  \\  
6.100   & 0.00499 & 0.0998  & $28^3 \times 32$     & 4190     &    1/3 & 39900 & ---    & ---    & ---  \\
6.125   & 0.00483 & 0.0966  & $32^4$               & 8645     &    1/3 & 32990 & ~67720  & ---    & ---  \\  
6.150   & 0.00468 & 0.0936  & $24^3 \times 32$     & 7795     &    1/3 & 31130 & ---    & ---    & ---  \\
6.175   & 0.00453 & 0.0906  & $32^4$               & 9080     &    1/3 & 30990 & ~60480  & ---    & ---  \\
6.195   & 0.00440 & 0.0880  & $32^4$               & 8445     &    1/2 & 33150 & ~25790  & ---    & ---  \\
6.245   & 0.00415 & 0.0830  & $32^4$               & 8505     &    1/2 & 30990 & ~28070  & ---    & ---  \\   
6.285   & 0.00395 & 0.0790  & $32^4$               & 7350     &    1/2 & 30990 & ~40250  & ---    & ---  \\
6.341   & 0.00370 & 0.0740  & $32^4$               & 6705     &      1 & 30990 & ~33310  & ---    & ---  \\
6.354   & 0.00364 & 0.0728  & $32^4$               & 8000     &      1 & 30990 & 220312 & ---    & ---  \\
6.390   & 0.00347 & 0.0694  & $32^4$               & 4602     &      1 & ---   & 269636 & ---    & ---  \\
6.423   & 0.00335 & 0.0670  & $32^4$               & 7970     &      1 & 30990 & 113315 & ---    & ---  \\
6.460   & 0.00320 & 0.0640  & $32^3 \times 64$     & 2900     &      1 & ---   & ~84841  & ---    & ---  \\
6.488   & 0.00310 & 0.0620  & $32^4$               & 19465    &      1 & 30990 & ~65281  & 103060 & ---  \\
6.515   & 0.00302 & 0.0604  & $32^4$               & 17385    &      1 & 30990 & 140212 & 108530 & ---  \\
6.550   & 0.00291 & 0.0582  & $32^4$               & 8805     &      1 & 30990 & 136781 & ---    & ---  \\
6.575   & 0.00282 & 0.0564  & $32^4$               & 21455    &      1 & 30990 & 144241 & 106750 & ---  \\
6.608   & 0.00271 & 0.0542  & $32^4$               & 21195    &      1 & 30990 & 171977 & 113920 & ---  \\
6.664   & 0.00257 & 0.0514  & $32^4$               & 21200    &      1 & 30990 & ~94440  & 175500 & ---  \\
6.740   & 0.00238 & 0.0476  & $48^4$               & 8005     &      1 & ---   & ~88520  & 217740 & ~48230   \\
6.800   & 0.00224 & 0.0448  & $32^4$               & 39077    &      1 & 30990 & 110200 & 299550 & ~57136   \\
6.880   & 0.00206 & 0.0412  & $48^4$               & 8095     &      1 & ---   & 110020 & 360690 & ~65678   \\
6.950   & 0.00193 & 0.0386  & $32^4$               & 39670    &      1 & 30990 & 117780 & 318700 & ~76080   \\
7.030   & 0.00178 & 0.0356  & $48^4$               & 16390    &      1 & ---   & ~96991  & 152330 & ~97801   \\ 
7.150   & 0.00160 & 0.0320  & $48^3 \times 64$     & 8094     &      2 & 29620 & ~96342  & 163900 & 106150  \\
7.280   & 0.00142 & 0.0284  & $48^3 \times 64$     & 7956     &      2 & 37340 & 103748 & 118460 & 110330  \\
7.373   & 0.00125 & 0.0250  & $48^3 \times 64$     & 9246     &      2 & 20780 & 116390 & 108100 & 164450  \\
7.596   & 0.00101 & 0.0202  & $64^4$               & 9514     &      2 & 36650 & 120000 & 113510 & 171020  \\
7.825   & 0.00082 & 0.0164  & $64^4$               & 9536     &      2 & 44390 & 119200 & 116070 & 105970  \\
\hline
\end{tabular}
\caption{Parameters used in simulations with the HISQ/tree action on
  $N_\tau=6$, $8$ $10$, and $12$ lattices and the LCP defined by
  $m_l/m_s = 0.05$.  The quark masses are given in units of the
  lattice spacing $a$. The statistics in molecular dynamics time units
  $TU$ are given for both the zero and finite temperature runs. The
  column ``length'' lists the length of the trajectory in TU before
  the Metropolis accept-reject step for zero-temperature runs.  All
  the finite temperature lattices have trajectories of unit length
  except for $\beta=5.90$ and $5.95$, where it was 0.5 TU. The lattice
  sizes used for the finite temperature simulations were $24^3 \times
  6$, $32^3 \times 8$, $40^3 \times 10$, and $48^3 \times
  12$. Measurements were performed after 1 TU on all ensembles, except
  for the large zero-temperature lattices with length=2 TU, where they
  were performed every 2 TUs. }
\label{tab:hisq_0.05ms_runs}
\end{table*}

\subsection{Topological charge history}\label{app_topology}

The topological charge history gives an indication of the ergodicity
of the molecular dynamics evolution.  Ideally, we want a reasonably
good coverage of the most probable topological charge sectors.  This
occurs when tunneling between the topological charge sectors is
reasonably frequent.  It is expected that the tunneling rate decreases
as the lattice spacing is decreased.  Therefore, to test ergodicity in our
molecular dynamics evolution, we look at the least favorable case, namely
our finest lattices.

\begin{figure*}
\includegraphics[width=0.4\textwidth]{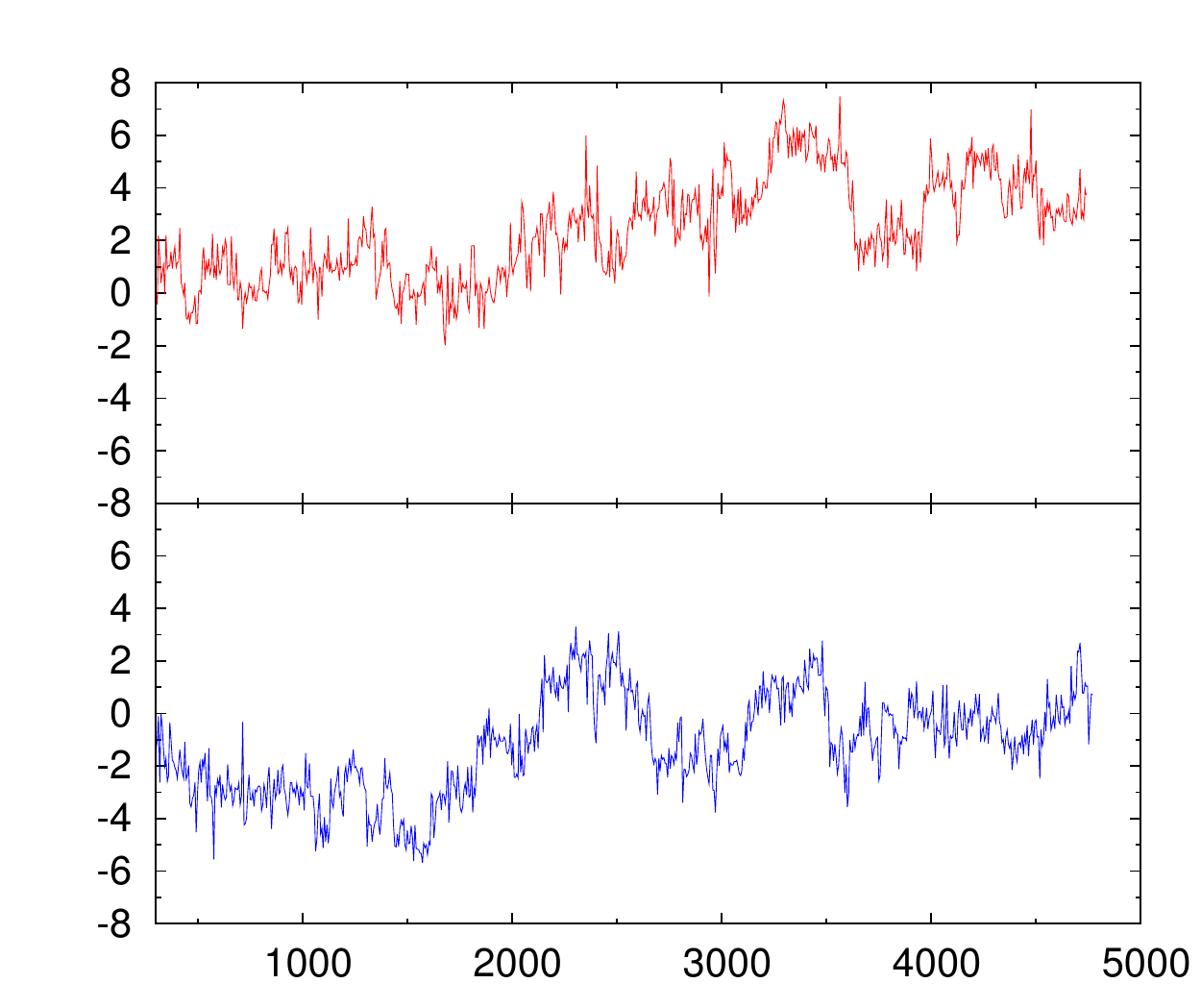}
\includegraphics[width=0.4\textwidth]{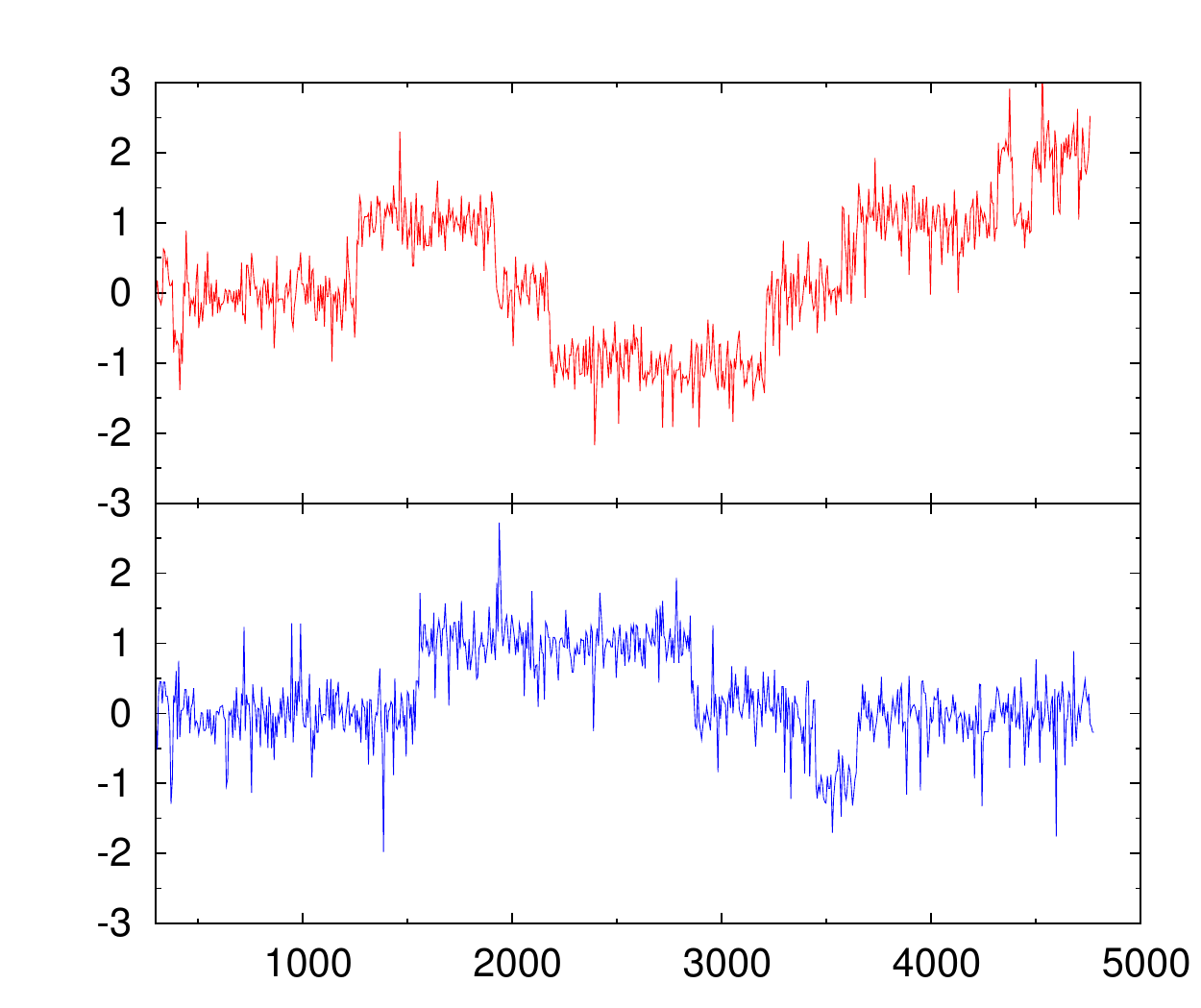}
\caption{The evolution of topological charge in Monte-Carlo
time for $\beta=7.596$ (left) and $\beta=7.825$ (right).
The top and bottom panels correspond to two different streams for hybrid
Monte-Carlo evolution.}
\label{fig:topology}
\end{figure*}

In our previous study we checked the evolution of the topological
charge in our simulations down to lattice spacings $a=0.066$ fm, 
and found that it fluctuated quite rapidly \cite{Bazavov:2011nk}. 
In the present study the lattice spacing for our two finest lattices 
corresponding to $\beta=7.596$ and $\beta=7.825$ 
is smaller still, namely, $a=0.049$~fm and $0.041$~fm, respectively. 
In Fig. \ref{fig:topology} we show the evolution of the topological
charge for those two ensembles.  The figures show a slower
tunneling rate than in our previous study, but we still see a
reasonable coverage of the topological charge sectors.

\section{Lattice scale}\label{app_r0r1}

To translate lattice observables to physical dimensionfull quantities,
we need to measure the lattice spacing.  We consider three methods for
determining the scale: the static-quark-potential parameters $r_1$ or $r_0$, the
gradient flow parameter $w_0$ \cite{Borsanyi:2012zs}, and the kaon
decay constant $f_K$.  Our preferred method uses the static quark
potential.  The other methods are used as a cross check.  We discuss here
the the former three ways to set the scale and defer the discussion
of $f_K$ to Appendix \ref{app_masses}.

\subsection{Static quark potential}

The static quark potential is used indirectly to set the scale.  In
brief, a standard radius $r_0$ \cite{Sommer:1993ce} or $r_1$
\cite{Bernard:2004je} is calculated from the measured heavy quark
potential using Eqs.~(\ref{eq:r0_r1}) and (\ref{eq:potential}).  On
any ensemble, the standard radii are first determined in lattice units:
$r_0/a$ and $r_1/a$.  The values of $r_0$ and $r_1$ in the continuum
limit are known in physical units from other lattice studies, based
on, for example, the experimental value of $f_\pi$.  From them and
$r_0/a$ one can infer the value of $a$.

The static potential for the HISQ/tree action has been studied in
Ref.~\cite{Bazavov:2011nk} for a large range of gauge couplings
$\beta$. We extended these studies in the following ways. We improved
significantly the statistical accuracy of the calculation of the
static potential at the highest beta values considered in
Ref.~\cite{Bazavov:2011nk}, namely for $\beta = 6.88,~6.95,~7.03,~7.15$ and
$7.28$.  To them we added calculations of the potential at
$\beta=6.740,~7.373,~7.596$, and $7.825$. As in our previous study, we
used Coulomb gauge fixing and calculated the potential from the
correlation of two Wilson lines of length $\tau$ at distance $R$. 
The potential is then obtained from the logarithm of the ratio
of two such correlators at neighboring $\tau$ values. 
We fit this ratio to a constant plus a term that decays
exponentially in Euclidean time $\tau$ in the interval
$[\tau_{min}:\tau_{max}]$.  We also studied the variation of the
potential due to different choices 
$[\tau_{min}:\tau_{max}]$ to estimate possible systematic errors.
To be specific,  we
used finally [2:4], [3:7], [4:9], [3:9], [4:9], [4:10], [4:10], [5:8],
and [6:11] for 
$\beta=6.740,~6.88,~6.95,~7.03,~7.15,~7.28,~7.373,~7.596$, and $7.825$,
respectively.

The scales $r_0/a$ or $r_1/a$ are then determined by separately
fitting the resulting potential to a Coulomb-plus-linear-plus-constant
form in the $r$-intervals around the values of $r_1$ and $r_0$, respectively.
We vary the fit intervals, and the variations in the extracted values of $r_1/a$ and $r_0/a$
are used as estimates of systematic errors. 
In most cases, the systematic errors are larger
than the statistical ones. The statistical and systematic errors are added in quadrature 
to estimate the total error for $r_0$ and $r_1$.
The values $r_0/a$, $r_1/a$ and their ratios $r_0/r_1$ determined in this study
as well as from Ref.~\cite{Bazavov:2011nk} are given in Table \ref{tab:r0}.
As in our previous study, the ratio $r_0/r_1$ appears to be independent of $\beta$ (lattice
spacing) within the estimated errors \cite{Bazavov:2011nk}. Accordingly, as before, we fit the values of $r_0/r_1$ 
given in Table~\ref{tab:r0} to a constant for $\beta \ge 6.423$ and
obtain
\begin{equation}
(r_0/r_1)_{cont}=1.5092 \pm 0.0039,~\chi^2/{\rm dof}=0.22 \; .
\label{r0pr1}
\end{equation}
This value agrees well with our previous estimate $r_0/r_1=1.508(5)$ \cite{Bazavov:2011nk}.
We also fit the ratio $r_0/r_1$ using only the data for $\beta \ge 6.664$ and $\beta \ge 6.608$,
obtaining $r_0/r_1=1.5083(44)$ and $r_0/r_1=1.5075(43)$ with similar $\chi^2/{\rm dof}$. These values agree well with the one given
in Eq. (\ref{r0pr1}).  Therefore we use Eq. (\ref{r0pr1}) as our final estimate for $r_0/r_1$.

\begin{table}
\begin{tabular}{|c|c|c|c|}
\hline
$\beta$ & $r_0/a$  & $r_1/a$    & $r_0/r_1$  \\
\hline
5.900 & 1.909(11)  & 1.230(133) & 1.552(168) \\
6.000 & 2.094(21)  & 1.386(80)  & 1.511(89)  \\
6.050 & 2.194(22)  & 1.440(31)  & 1.524(36)  \\
6.100 & 2.289(21)  & 1.522(30)  & 1.504(33)  \\
6.195 & 2.531(24)  & 1.670(30)  & 1.516(31)  \\
6.285 & 2.750(30)  & 1.822(30)  & 1.509(30)  \\
6.341 & 2.939(11)  & 1.935(30)  & 1.519(24)  \\
6.354 & 2.986(41)  & 1.959(30)  & 1.524(31)  \\
6.423 & 3.189(22)  & 2.096(21)  & 1.522(18)  \\
6.460 & 3.282(32)  & 2.165(20)  & 1.516(20)  \\
6.488 & 3.395(31)  & 2.235(21)  & 1.519(20)  \\
6.550 & 3.585(14)  & 2.369(21)  & 1.513(15)  \\
6.608 & 3.774(20)  & 2.518(21)  & 1.499(15)  \\
6.664 & 3.994(14)  & 2.644(23)  & 1.511(14)  \\
6.740 & 4.293(32)  & 2.856(11)  & 1.503(13)  \\
6.800 & 4.541(30)  & 3.025(22)  & 1.501(15)  \\
6.880 & 4.959(28)  & 3.265(23)  & 1.519(14)  \\
6.950 & 5.249(20)  & 3.485(22)  & 1.506(11)  \\
7.030 & 5.691(32)  & 3.763(13)  & 1.512(10)  \\
7.150 & 6.299(59)  & 4.212(42)  & 1.495(20)  \\
7.280 & 7.140(53)  & 4.720(33)  & 1.513(15)  \\
7.373 & 7.801(79)  & 5.172(34)  & 1.508(18)  \\
7.596 & 9.443(237) & 6.336(56)  & 1.490(40)  \\
7.825 & 11.51(378) & 7.690(58)  & 1.497(50)  \\
\hline
\end{tabular}
\caption{Values of $r_1$ and $r_0$ in lattice units for different $\beta$}
\label{tab:r0}
\end{table}

To determine the lattice spacing as function
of $\beta$, we fit $a/r_1$ to the Allton-type ansatz \cite{Allton:1996dn},
\begin{eqnarray}
& 
\displaystyle\label{asq_r0_fit}
\frac{a}{r_1}=\frac{c_0 f(\beta)+c_2 (10/\beta) f^3(\beta)}{
1+d_2 (10/\beta) f^2(\beta)}\; ,\label{ar1_fit} \\[3mm]
&
\displaystyle
f(\beta)=\left( \frac{10 b_0}{\beta} \right)^{-b_1/(2 b_0^2)} \exp(-\beta/(20 b_0))\; .\label{fbeta}
\end{eqnarray}
Here $b_0$ and $b_1$ are the well-known coefficients of the two-loop beta function,
which for the three-flavor case are $b_0=9/(16 \pi^2)$, $b_1=1/(4 \pi^4)$. 
At small $\beta$, the parameter $r_1$ is small in lattice units.  Therefore, to avoid
possibly large discretization effects, for $\beta < 6.423$, where $r_0/a$ is more
reliably determined, we use $r_1/a = r_0/a/(r_0/r_1)_{cont}$ with $(r_0/r_1)_{cont}$ from Eq. (\ref{r0pr1})
(see discussions in Ref.~\cite{Bazavov:2011nk}). The fit gives $\chi^2/{\rm dof}=0.25$ and
\begin{eqnarray}
&
c_0=43.1 \pm 0.3\; ,\\
&
c_2=343236 \pm 41191\; ,\\
&
d_2=5514 \pm 755\; .
\end{eqnarray}
The errors on the above fit parameters have also been estimated using the bootstrap method which
gives very similar results. The differences between the above parametrization of $r_1/a$ and the previous
one from Ref.~\cite{Bazavov:2011nk} are less than $0.2\%$ for $\beta<6.8$. For larger beta values the differences
are larger but do not exceed $1.3\%$. To convert all quantities to physical units, 
as in Ref.~\cite{Bazavov:2011nk}, we use the value $r_1=0.3106$~fm from \cite{Bazavov:2010hj}.

\begin{figure}
\includegraphics[width=7cm]{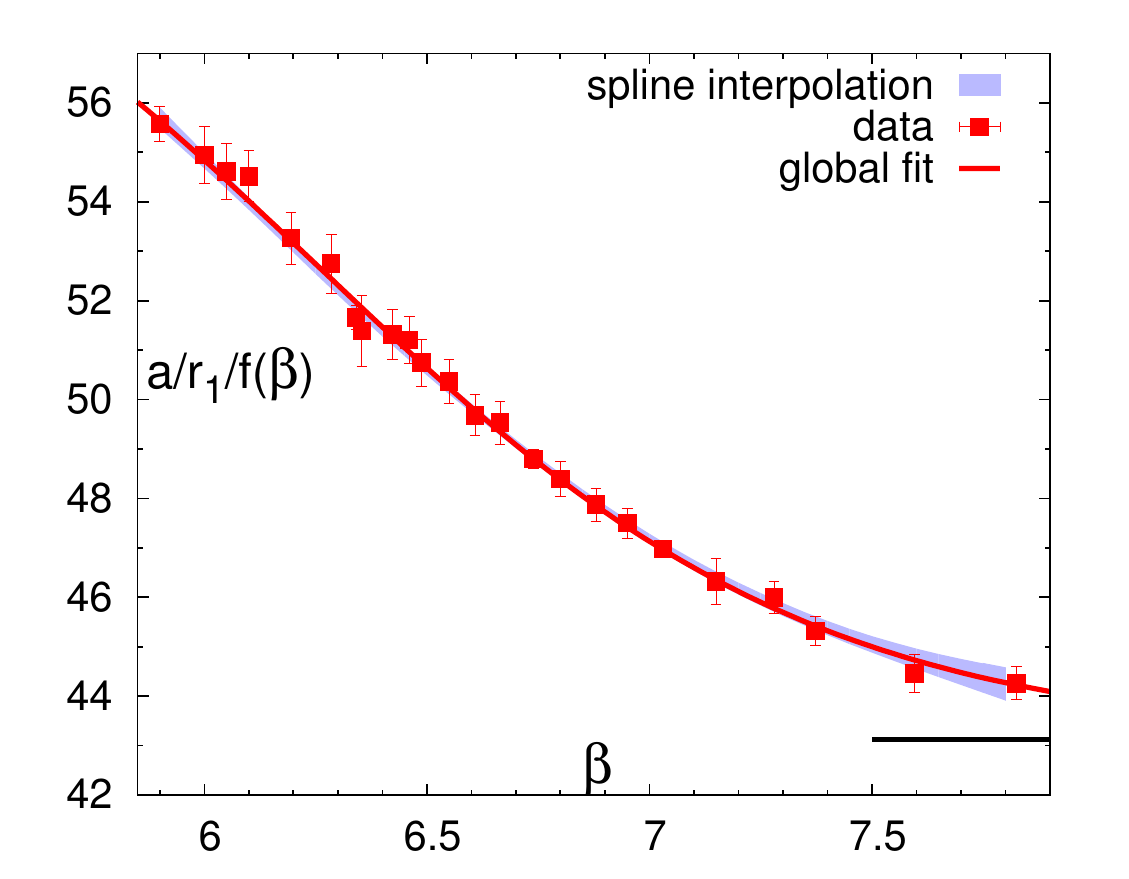}
\includegraphics[width=7cm]{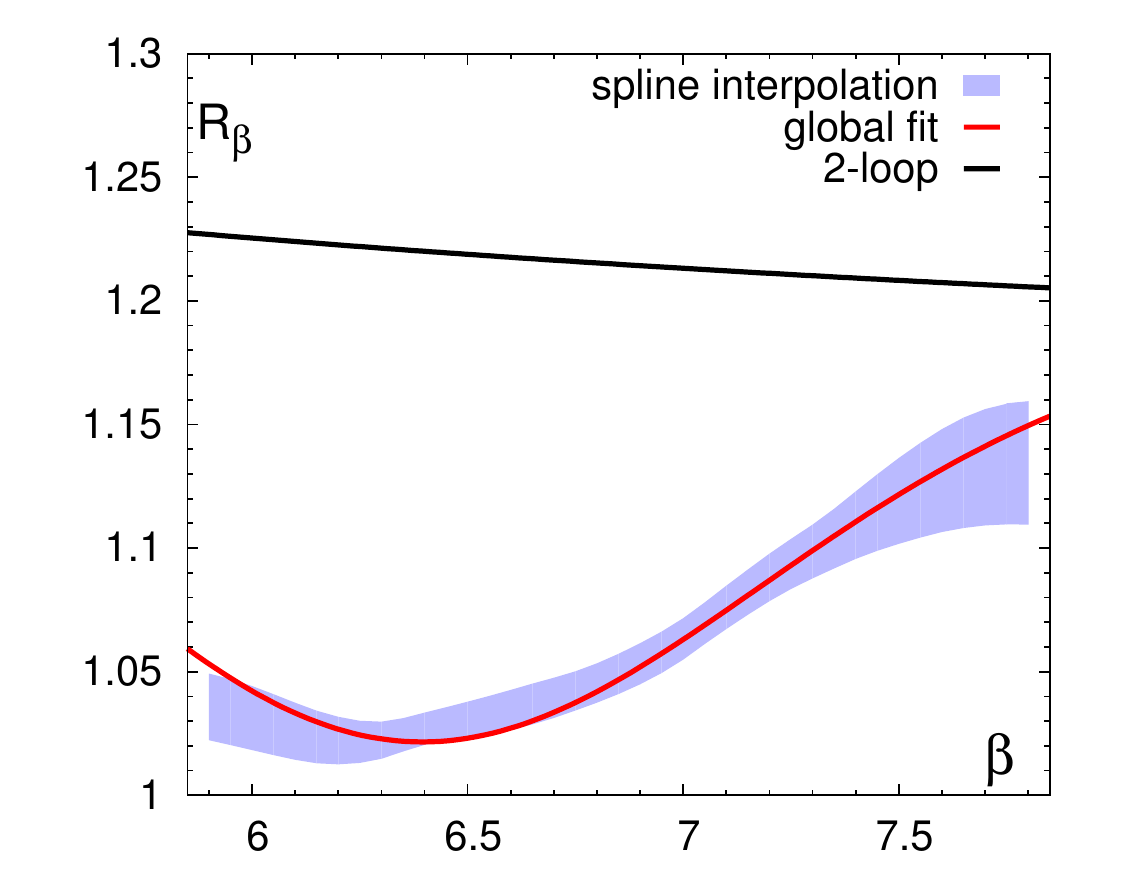}
\caption{The scale $a/r_1$ normalized by the asymptotic two-loop beta function 
(top) and the nonperturbative beta function, $R_{\beta}$, 
(bottom) as a function of $\beta$, which has been
derived from this using Eq.~(\ref{rbeta}); the fit and spline interpolations
are also shown.}
\label{fig:scale}
\end{figure}

To test the uncertainty in the scale parametrization, we also fit the
data for $a/r_1$ to the asymptotic form $f(\beta)$ times a smoothing
spline.  The smoothing spline is determined by minimizing the $\chi^2$
plus the integral of the square of the second derivative of the fit
function in the considered interval times a real parameter $sm$.  We
chose the largest possible value of the smoothing parameter $sm=0.7$
that still gives an acceptable $\chi^2/{\rm dof}=1.13$. To estimate the
uncertainties of the spline, we performed a bootstrap analysis. In
Fig.~\ref{fig:scale}, we show the $r_1$ scale as a function of $\beta$,
normalized by the asymptotic two-loop beta function $f(\beta)$. The
errors are bootstrap errors. The Allton-type fit and the smoothing
spline fits give very similar results as well as uncertainties. 

To calculate the EoS, we also need the nonperturbative beta
function
\begin{equation}
R_{\beta}=-a \frac{d \beta}{da}= \frac{r_1}{a} \left( \frac{d (r_1/a)}{d \beta} \right)^{-1} \; .
\label{rbeta}
\end{equation}
Figure~\ref{fig:scale} shows $R_\beta$ obtained from both the Allton-type and smoothing-spline fits,
together with bootstrap errors. The fit and the splines agree within the errors. 
The largest error in $R_{\beta}$ is about 3\%.
At sufficiently
large $\beta$, {\it i.e.}, close to the continuum limit, $R_{\beta}$ is expected to be given by its asymptotic two-loop form
\begin{equation}
R_{\beta}^{2-{\rm loop}}=20 b_0+ 200 b_1/\beta.
\end{equation}
The asymptotic limit is approached from below~\cite{Cheng:2007jq}, as
with the p4 action.  However, for the HISQ action, we see that the
deviations are at most $20 \%$ over the range considered, compared
with a factor of two deviation in the case of the p4
action~\cite{Cheng:2007jq}. In our calculations of the EoS we use
$R_\beta$ obtained from the fit with the Allton-type ansatz.

\begin{figure}
\includegraphics[width=7cm]{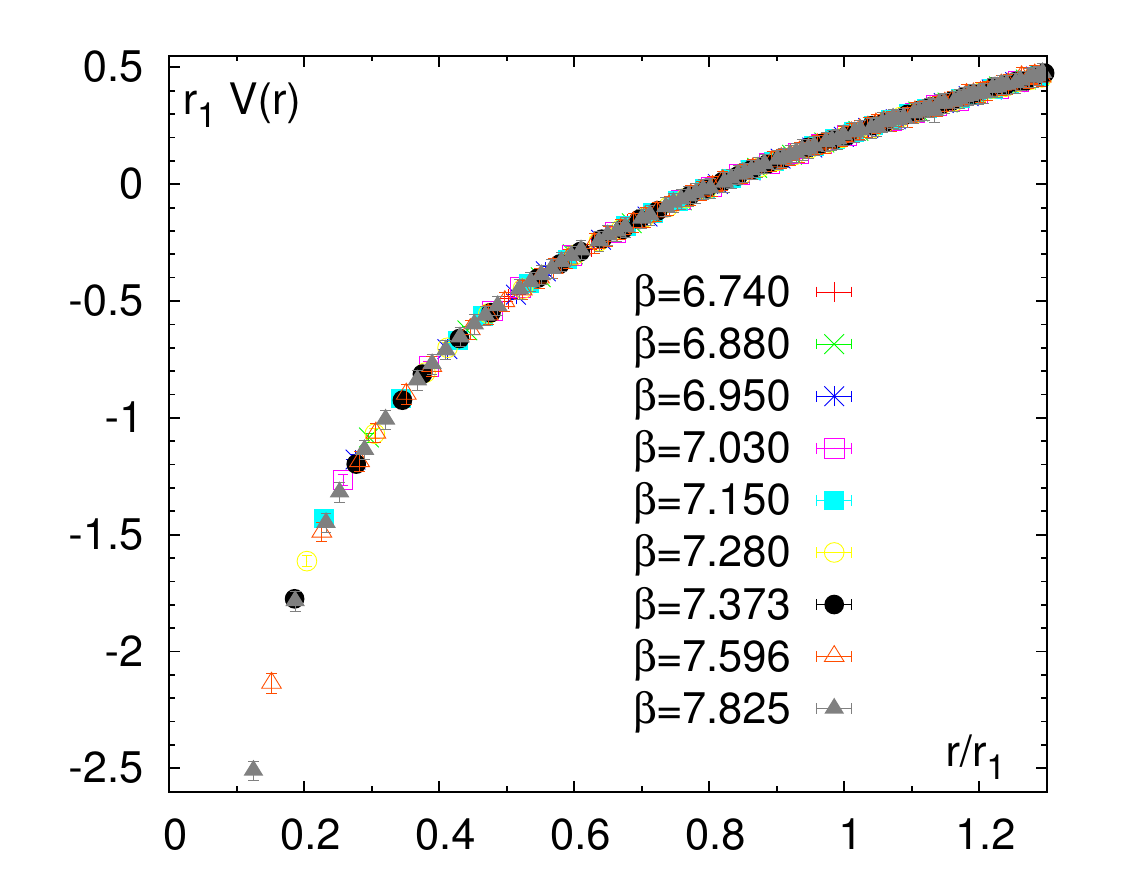}
\caption{The static potential versus distance in units of $r_1$ for 
different $\beta$. Here we use the improved estimator $r_I$ to define 
the distance $r$.}
\label{fig:potential}
\end{figure}

Finally, we compare the potential calculated at different
$\beta$. To do so, we normalize it with an additive
constant. We do this by requiring that the potential $V(r_1) =
0.2060/r_1$. This normalization condition is equivalent to the one
used in Ref.~\cite{Bazavov:2011nk}.  Here we choose $r_1$, because it
has smaller errors on fine lattices. The normalized potential in units
of $r_1$ is plotted in Fig.~\ref{fig:potential} against the tree-level
improvement radius $r \rightarrow r_I$, where $r_I$ is the improved
distance defined from the free lattice gluon propagator
\cite{Bazavov:2011nk}.  Down to distances $r=0.2 r_1$ or
$r=0.062$fm, we find no significant dependence on the lattice spacing
within the estimated errors.

\begin{figure}
\includegraphics[width=7cm]{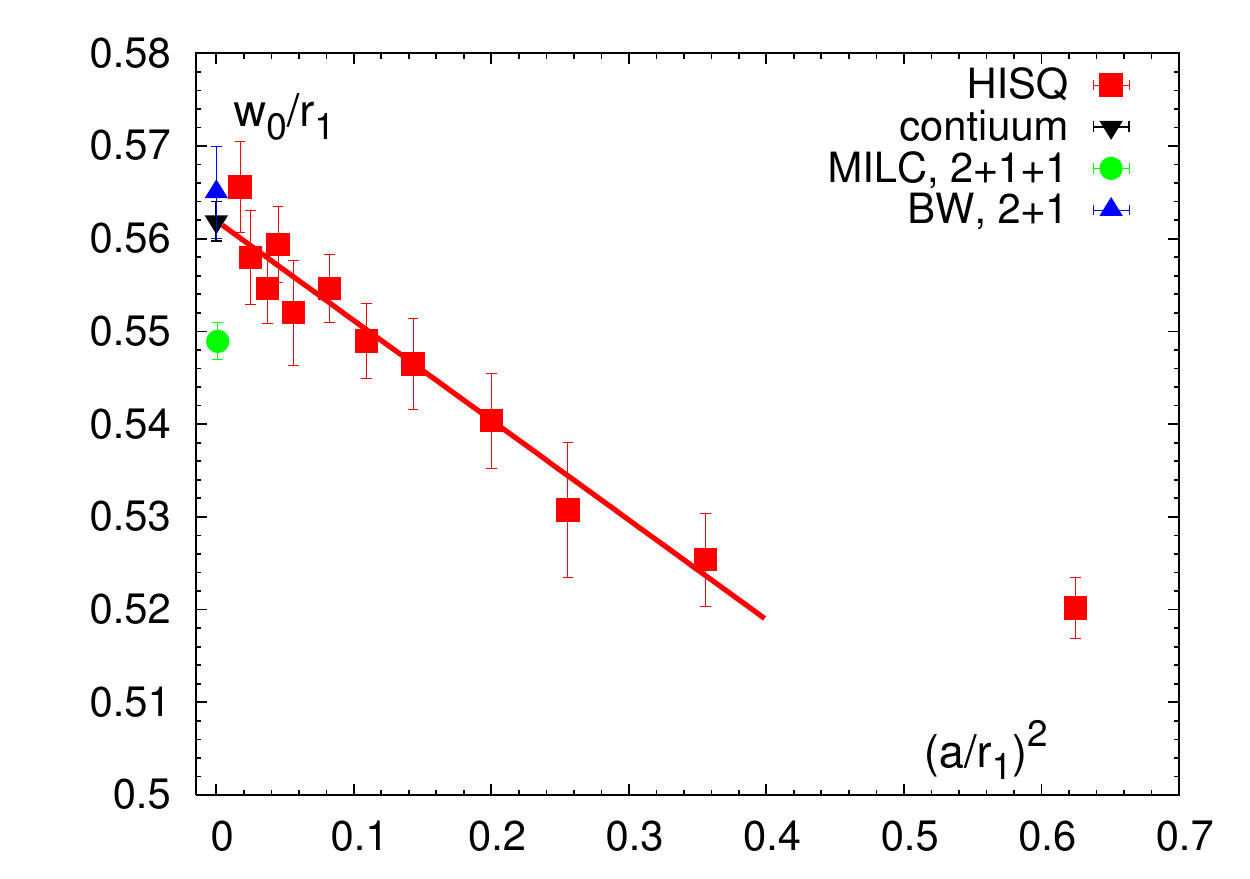}
\caption{The $w_0$ scale as function of the lattice spacing together with the continuum extrapolation.
Also shown are results from 2+1 flavor QCD \cite{Borsanyi:2012zs} and 2+1+1 flavor QCD \cite{Bazavov:2013gca}.}
\label{fig:w0r1}
\end{figure}

To cross check our determination of the lattice spacing, we also
calculated the scale $w_0$, defined from the gradient flow
\cite{Borsanyi:2012zs}. Our results for the $w_0$ scale are shown in
Fig.~\ref{fig:w0r1} in units of $r_1$.  As above, for $\beta<6.423$
the value of $r_1$ was estimated as $r_0/(r_0/r_1)_{cont}$.  As one
can see from the figure, this ratio appears to scale as $a^2$ for
$(a/r_1)^2 < 0.4$, {\it i.e.}, for $\beta \ge 6.195$.  We perform a
continuum extrapolation of the ratio $w_0/r_1$ using a simple form
$(w_0/r_1)_{cont}+h_w (a/r_1)^2$.  In the continuum limit, we obtain
$(w_0/r_1)_{cont}=0.5619(21)$ or $w_0=0.1749(14)$ fm. This value
agrees with the value quoted in Ref.~\cite{Borsanyi:2012zs},
$w_0=0.1755(18)(4)$ fm, within the estimated errors.
Our value of $w_0$ is higher than the preliminary value reported
by MILC $w_0=0.1711(2)(8)(2)(3)$fm for 2+1+1 flavor QCD \cite{Bazavov:2013gca}.
For the slope parameter we get $h_w=-0.1076(149)$ with $\chi^2/{\rm dof}=0.38$.
If we use the $w_0$ scale instead of the $r_1$ scale, the temperature values 
for $N_{\tau}=8$ lattices for $T<150$ MeV would be lower by $6\%$, and
for $N_{\tau}=10$ and $12$ calculations the differences in the
temperature scale would be only $4 \%$ or less.

\section{Hadronic observables}\label{app_masses}
\label{hadronic}

\subsection{Line of constant physics}

It is standard practice to present results for thermodynamic
quantities as a function of temperature at fixed, renormalized quark
masses.  We start by setting a constant value of the strange quark
mass $m_s$, preferably, its physical value, and then set the mass of
the light quarks to $m_s/20$.  
We determine the strange quark mass by requiring that the mass of the 
un-mixed pseudoscalar $s\bar s$ meson, $\eta_{s \bar s}$ is
equal to a prescribed value expressed in units of $r_1$.
We aim at the value suggested by leading order chiral perturbation theory, 
where the mass of $\eta_{s \bar s}$ meson in terms of the kaon and pion 
masses is, $M_{\eta_{s \bar s}}=\sqrt{2 m_K^2-m_{\pi}^2}=686$ MeV. In practice,
this requires some tuning and, as discussed later, it turns out
that our LCP is best described by the value
$M_{\eta_{s \bar s}}=695$~MeV.
Setting the line of constant physics (LCP) in this way
requires a combination of determining the lattice spacing in physical
units (see Appendix~\ref{app_r0r1}) and the hadron spectrum
at zero temperature, including, at least, the mass of the
$\eta_{s\bar s}$, a calculation with costs that mount as the lattice
spacing decreases.  Thus, some retuning is usually needed to correct for an
imprecise determination.

For the present study, we extend the LCP of our previous work~\cite{Bazavov:2011nk}
to weaker coupling in order to cover the range needed for $N_\tau=12$.
In our previous work, the hadron spectrum was measured
along the LCP up to $\beta=6.8$.  The masses of the pseudoscalar mesons were also
measured at $\beta=7.28$ with the relatively low statistics of about 1,400 equilibrated
time units.  Here, we added or extended nine
$T=0$ ensembles with $6.8 < \beta \le 7.825$ as described in Appendix~B.  On the
extended ensembles, in addition to measuring thermodynamic quantities needed
for the zero temperature subtraction, we 
measured the masses and decay constants of the pseudoscalar mesons 
and the masses of the vector mesons. These quantities allow us to 
quantify the lattice artifacts due to taste breaking and can be used as an
alternative
means to set the lattice spacing, thus providing additional validation of our
calculations.

\begin{table}
\begin{tabular}{|c|c|c|c|}
\hline
$\beta$ & $a M_{\pi}$   & $a M_K$       & $a M_{\eta_{s \bar s}}$ \\
\hline
5.900   & 0.20162(09) & 0.63407(17) & 0.86972(11)  \\
6.000   & 0.18381(37) & 0.57532(51) & 0.79046(27)  \\
6.195   & 0.15143(14) & 0.47596(16) & 0.65506(11)  \\
6.285   & 0.13823(50) & 0.43501(47) & 0.59951(28)  \\
6.354   & 0.12923(15) & 0.40628(20) & 0.55982(17)  \\
6.423   & 0.12022(12) & 0.37829(19) & 0.52161(17)  \\
6.460   & 0.11528(21) & 0.36272(34) & 0.50137(32)  \\
6.488   & 0.11245(15) & 0.35313(27) & 0.48716(17)  \\
6.515   & 0.10975(12) & 0.34453(29) & 0.47516(29)  \\
6.550   & 0.10629(16) & 0.33322(38) & 0.45989(24)  \\
6.575   & 0.10469(68) & 0.32521(55) & 0.44869(50)  \\
6.608   & 0.10001(17) & 0.31333(28) & 0.43286(29)  \\
6.664   & 0.09572(18) & 0.29837(37) & 0.41178(32)  \\
6.740   & 0.087991(64) & 0.27735(12) & 0.38342(10) \\
6.800   & 0.0849(18)  & 0.26387(99) & 0.36257(68)  \\
6.880   & 0.07714(16) & 0.24314(16) & 0.33630(11)  \\
7.030   & 0.06744(15) & 0.21202(19) & 0.29381(20)  \\
7.150   & 0.06126(18) & 0.19231(20) & 0.26631(16)  \\
7.280   & 0.05516(17) & 0.17209(19) & 0.23824(18)  \\
7.373   & 0.04990(22) & 0.15530(16) & 0.21531(12)  \\
7.596   & 0.04106(44) & 0.12896(30) & 0.17810(12)  \\
7.825   & 0.03425(23) & 0.10695(46) & 0.14731(15)  \\
\hline
\end{tabular}
\caption{
The pseudoscalar meson masses for the HISQ/tree action along the 
$m_l=0.05m_s$ LCP.}
\label{tab:mps}
\end{table}

\begin{figure}
\includegraphics[width=0.48\textwidth]{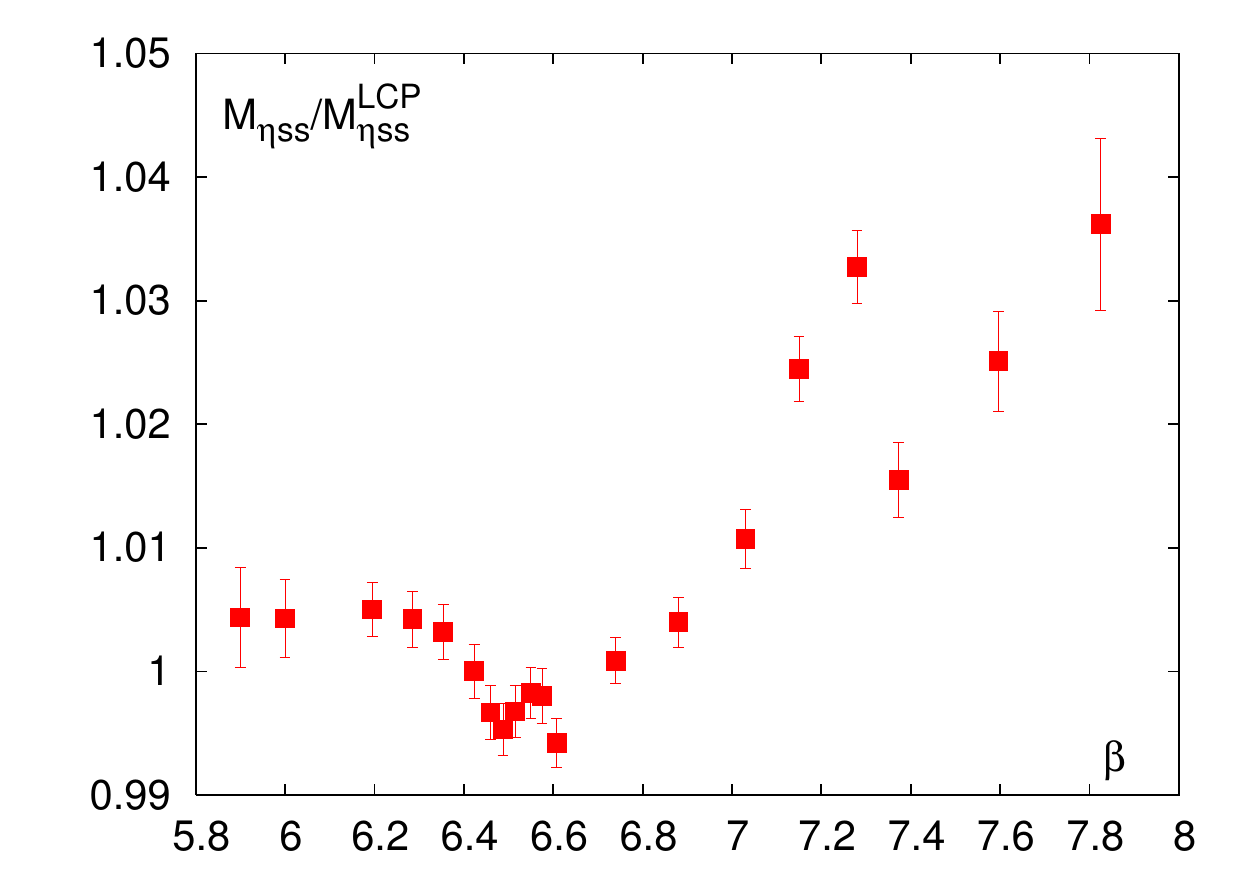}
\caption{The calculated masses of the $\eta_{s\bar s}$ meson normalized by
the chosen LCP value $M_{\eta_{s\bar s}}=695$~MeV as a function of $\beta$.}
\label{fig_m_eta}
\end{figure}

The masses of the three pseudoscalar mesons are given in lattice units in Table~\ref{tab:mps}.
The mass of the $\eta_{s\bar s}$ normalized by
the value $M_{\eta_{s\bar s}}=695$~MeV used to define the LCP 
is plotted in Fig.~\ref{fig_m_eta} as a function of $\beta$. 
As one can see, the
central values are systematically above the nominal value of $685.8$~MeV quoted
in Ref.~\cite{Davies:2009tsa}. The average value 
from ensembles with $\beta<7.03$ is about $695$~MeV. Therefore,
we define our LCP using this value.  That is, we choose a strange quark mass that gives
$M_{\eta_{s\bar s}}=695$~MeV.  The resulting strange quark mass is then about $2.6\%$ 
larger than its physical value.
For $\beta \le 7.03$  we find that
$M_{\eta_{s\bar s}}$ (and, in turn,
the strange quark mass) along the LCP agrees with the physical
values within (1--2)$\sigma$.
For the finest ensembles, $\beta>7.03$, we see a systematic deviation of $M_{\eta_{s\bar s}}$
towards higher values --- by as much as about 3.5\%.

\begin{figure}
\includegraphics[width=8cm]{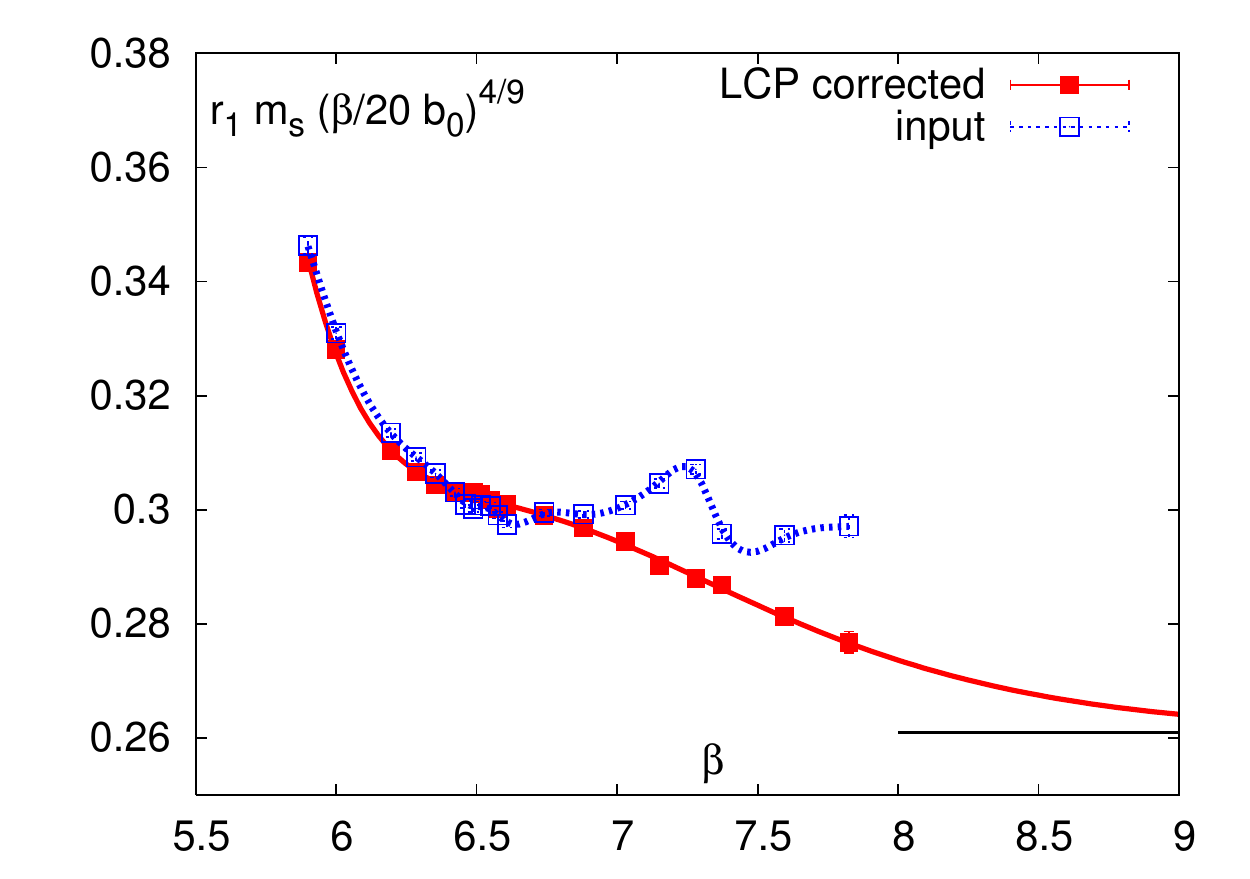}
\caption{The input strange quark mass and the strange quark mass along LCP together
with the respective fits shown as lines.
The input quark masses have been fitted with a smooth spline.
The horizontal line shows the assymptotic value.
}
\label{fig:ms}
\end{figure}

For the calculation of the trace anomaly, we need the strange quark
mass $m_s$ and its derivative as a function of $\beta$ along the LCP.
As we have seen, the strange-quark mass input into the simulation
drifts slightly above the LCP.  We correct for this drift using lowest
order chiral perturbation theory, $i.e.$, we assume that
$M_{\eta_{s\bar s}}^2$ is proportional to $m_s$ and calculate the
strange quark mass that gives $M_{\eta_{s\bar s}}=695$~MeV. This
corrected value is compared with the value used in the simulations in
Fig.~\ref{fig:ms}.  For the worst case, $\beta=7.825$, this amounts to
lowering $m_s$ used by about 7\% from the simulated value.  The pion
and kaon masses follow a pattern similar to that of the 
$\eta_{s\bar s}$ meson, i.e., they are roughly constant for $\beta< 7.03$ and
increase for larger beta values by approximately the same fractional
amount.

We then fit the product $r_1 m_s^{LCP}$ using a renormalization-group-inspired form
\begin{widetext}
\begin{equation}
r_1 m_s^{LCP} \equiv \tilde m_s= r_1 m^{RGI} \left( \frac{20 b_0}{\beta}\right )^{4/9} \frac{1+m_1 \frac{10}{\beta} f^2(\beta)+m_2 (\frac{10}{\beta})^2 f^2(\beta)
+m_3 \frac{10}{\beta} f^4(\beta) }{ 1+dm_1 \frac{10}{\beta} f^2(\beta)}\; ,
\end{equation} 
\end{widetext}
where $f(\beta)$ is the 2-loop beta function given by Eq.~\ref{fbeta}. For the fit parameters we get
\begin{eqnarray}
&
m^{RGI} = 0.2609 \pm 0.0030,\\
&
m_1=35600 \pm 6097,\\
&
m_2=-21760 \pm 3202,\\
&
m_3=(2.67 \pm 0.50)\cdot 10^7,\\
&
dm_1=2420 \pm 1346\\
&
\chi^2/{\rm dof}=0.51.
\end{eqnarray}
The resulting fit is shown in Fig.~\ref{fig:ms} together with a smoothing spline fit
to the input strange-quark masses.

\subsection{Pseudoscalar decay constants}

The decay constants of pseudoscalar mesons can be used to check the lattice scale
and to estimate the cutoff effects in the $T=0$ calculations. 
Results for the decay constants are shown in Table~\ref{tab:fpi}.

\begin{table}
\begin{tabular}{|c|c|c|c|c|}
\hline
$\beta$ & $a f_{\pi}$   & $a f_K$       & $a f_{\eta}$  & \# sources\\
\hline
6.000   & 0.11243(21) & 0.13224(31) & 0.15290(22)     &  1\\
6.195   & 0.09179(21) & 0.10835(13) & 0.12525(13)     &  1\\
6.285   & 0.08366(22) & 0.09826(13) & 0.11390(13)     &  1\\
6.354   & 0.07825(40) & 0.09146(19) & 0.10598(11)     &  1\\
6.423   & 0.07241(18) & 0.08515(11) & 0.09854(07)     &  2\\
6.460   & 0.06885(11) & 0.08185(09) & 0.09454(08)     &  4\\
6.515   & 0.06534(18) & 0.07707(15) & 0.08946(09)     &  4\\
6.575   & 0.06104(49) & 0.07265(19) & 0.08405(14)     &  2\\
6.740   & 0.052190(50) & 0.061731(41) & 0.071354(27)     &  2\\
6.800   & 0.04883(83) & 0.05774(18) & 0.06717(14)     &  1\\
6.880   & 0.045544(67) & 0.053749(42) & 0.062236(28)     &  2\\
7.030   & 0.03951(12) & 0.046566(68) & 0.054148(43)     &  2\\
7.150   & 0.03486(10) & 0.041636(68) & 0.048654(42)     &  2\\
7.280   & 0.03067(13) & 0.036894(64) & 0.043237(38)     &  2\\
7.373   & 0.02787(20) & 0.033825(82) & 0.039609(43)     &  2\\
7.596   & 0.02224(20) & 0.02741(21) & 0.032344(59)     &  2\\
7.825   & 0.01756(32) & 0.022526(85) & 0.026808(51)     &  2\\

\hline
\end{tabular}
\caption{Results for decay constants of the pseudoscalar mesons in
  lattice units for the HISQ/tree action along the $m_l=0.05 m_s$
  LCP. We use the normalization in which $f_\pi\sim 90$ MeV. In the
  last column, we list the number of source points used on each configuration
  to increase the statistics.}
\label{tab:fpi}
\end{table}

\begin{figure*}
\includegraphics[width=8cm]{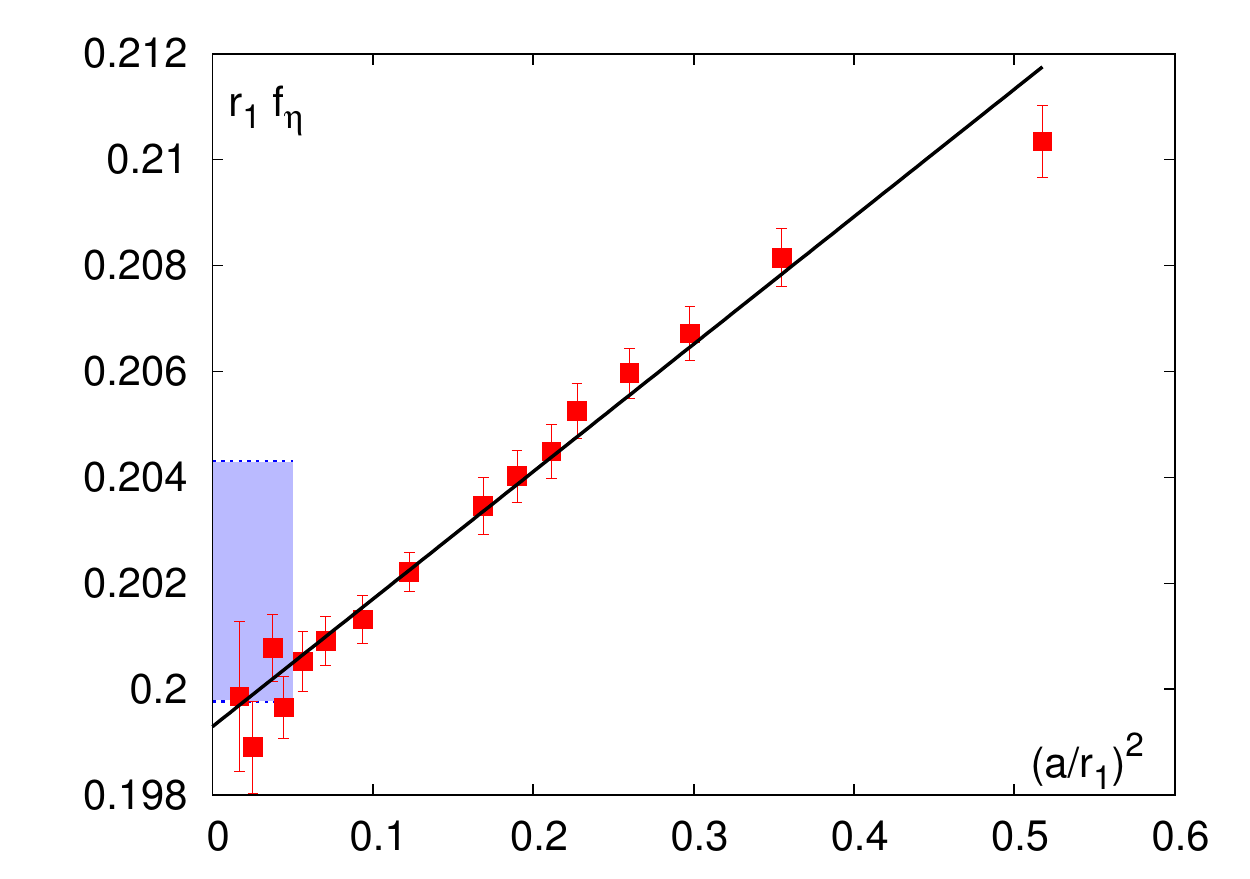} 
\includegraphics[width=8cm]{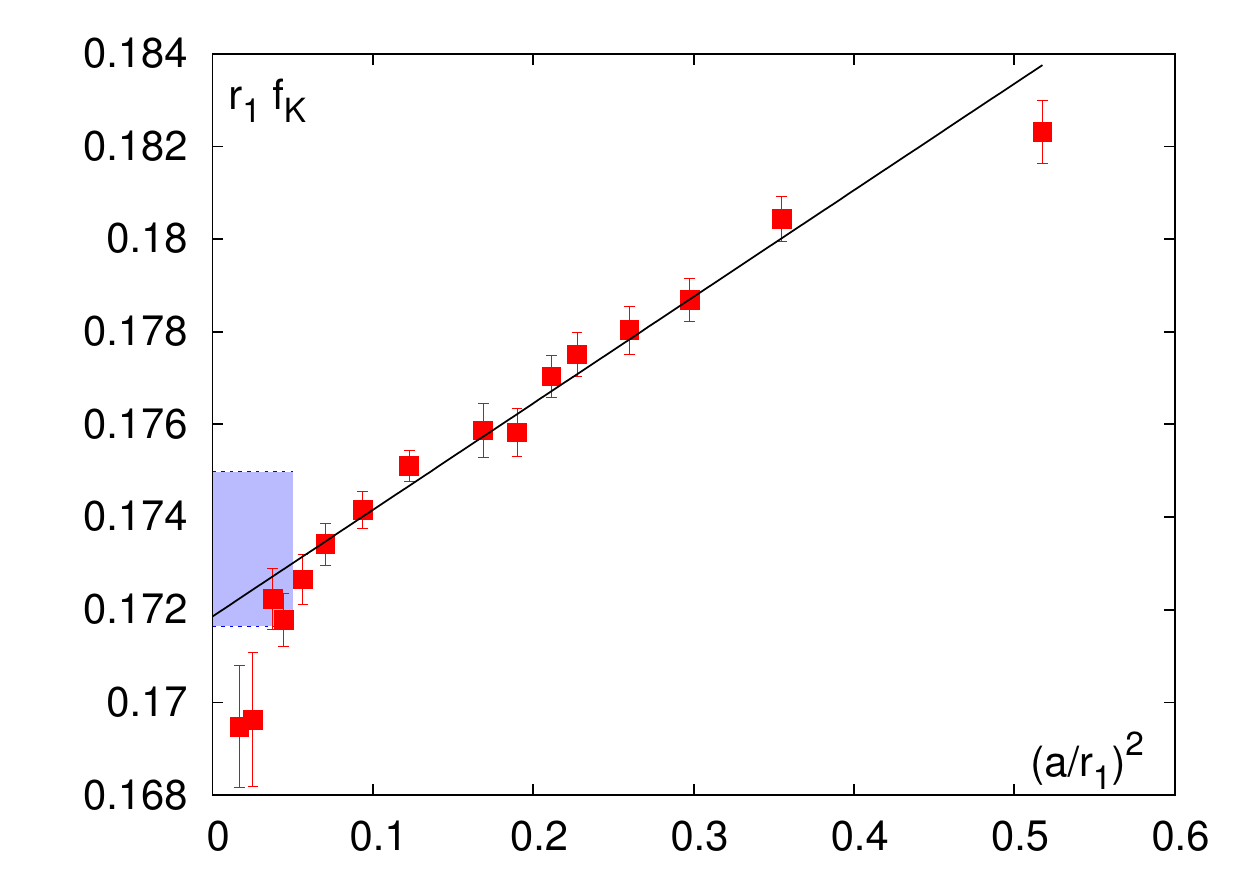}
\caption{The decay constants of $\eta_{s\bar s}$ meson (left) and kaon (right)
together with
the continuum extrapolations. 
The data are corrected for deviations from the LCP as described in the text. 
The bands show the value of these decay constants
from Ref.~\cite{Davies:2009tsa} and PDG. Here we use the value 
$r_1=0.3106(14)(8)(4)$ fm.
The errors on $r_1$ have been added linearly and combined with the errors on $f_{\eta}$ and
$f_K$ in quadratures.
}
\label{fig:fps}
\end{figure*}

Since they are quite sensitive to the values of the quark masses, we need
to take into account the deviations from the LCP, as well as the fact that even
on the LCP our quark masses are slightly heavier than the physical ones.
Thus, we need to interpolate/extrapolate
in the quark masses. To do this we assume that the pseudoscalar decay constants
depend linearly on the sum of the quark masses and use the numerical results given
in Table \ref{tab:fpi} to determine the slope for each value of $\beta$. 
The values of the $\eta_{s \bar s}$ 
meson decay constant $f_{\eta}$ and the kaon decay constant $f_K$ have been interpolated to the physical quark masses using this slope. The results
are shown in Fig.~\ref{fig:fps} in units of $r_1$. The kaon decay constant has large
finite size errors at the two smallest lattices spacings. Therefore, we do not include
the corresponding data in the fit.
We extrapolate the values
of $r_1 f_{\eta}$ and $r_1 f_K$ to zero lattice spacing assuming a simple
form 
\begin{equation}
f_i r_1 = (f_i r_1)^{cont}+e_i (a/r_1)^2,~~i=K, \eta.
\end{equation}
For the kaon decay constant we get $(r_1 f_K)^{cont}=0.17186(24)$ 
and $e_K=0.0230(11)$ with $\chi^2/{\rm dof}=1.20$.
For the $\eta_{s \bar s}$ decay constant we get 
$(r_1 f_{\eta})^{cont}=0.19930(24)$ and $e_{\eta}=0.024(12)$ with $\chi^2/{\rm dof}=0.77$.
The continuum extrapolation is also shown in Fig.~\ref{fig:fps}, where we compare it
with the value of $r_1 f_{\eta}$ quoted in Ref.~\cite{Davies:2009tsa},
and find reasonable agreement. 
The ``PDG'' value plotted there is based on the PDG value of $f_{\pi}$ and the value of
$f_K/f_{\pi}=1.194(5)$ from the recent FLAG review \cite{Aoki:2013ldr}, which gives 
$f_K=155.7(9)/\sqrt{2}$ MeV. We find agreement within estimated errors.

\subsection{Vector meson masses}

The masses of the three vector mesons $\rho$, $K^*$, and $\phi$ are listed in Table~\ref{tab:vec}. 
All three masses are adjusted to the LCP in a manner similar to the decay constants.
For $\rho$ and
$K^*$ the contribution of an excited state of the same parity is significant in the available
temporal range; however, fits with an excited state typically yield low confidence levels. 
Therefore, in reporting masses on our finest lattices we take, as a systematic error, 
the difference in the fitted masses with and without an excited state of the same parity.
This error is combined linearly with the statistical error in the table.

\begin{table}
\begin{tabular}{|c|c|c|c|}
\hline
$\beta$  & $a M_{\rho}$  &  $a M_{K^*}$  &  $a M_{\phi}$ \\
\hline
6.195    & 0.7562(36)	 &  0.8842(18)   &  1.0050(93)   \\
6.354    & 0.6375(35)	 &  0.7499(26)   &  0.8523(08)   \\
6.423    & 0.6047(43)	 &  0.6950(22)   &  0.7925(08)   \\
6.460    & 0.5784251)	 &  0.6709(43)   &  0.7644(22)   \\
6.488    & 0.5647(24)	 &  0.6478(22)   &  0.7363(07)   \\
6.550    & 0.5324(24)	 &  0.6118(20)   &  0.6929(14)   \\
6.608    & 0.5072(39)	 &  0.5757(08)   &  0.6523(10)   \\
6.664    & 0.4732(43)	 &  0.5501(26)   &  0.6180(10)   \\
6.740    & 0.4286(31)	 &  0.4996(22)   &  0.5732(05)   \\
6.880    & 0.2828(489)	 &  0.4359(17)   &  0.5000(04)   \\
7.030    & 0.2937(326)	 &  0.3750(126)  &  0.4333(09)   \\ 
7.150    & 0.2866(108)	 &  0.3387(107)  &  0.3901(15)   \\ 
7.280    & 0.2535(96)	 &  0.3026(20)   &  0.3467(25)   \\ 
7.373    & 0.2363(119)	 &  0.2774(33)   &  0.3165(06)   \\
7.596    & 0.1923(61)	 &  0.2272(25)   &  0.2593(15)   \\
7.825    & 0.1543(120)	 &  0.1884(57)   &  0.2140(19)   \\
\hline
\end{tabular}
\caption{
Masses of the vector mesons in lattice units.
}
\label{tab:vec}
\end{table}

\begin{figure}
\includegraphics[width=7cm]{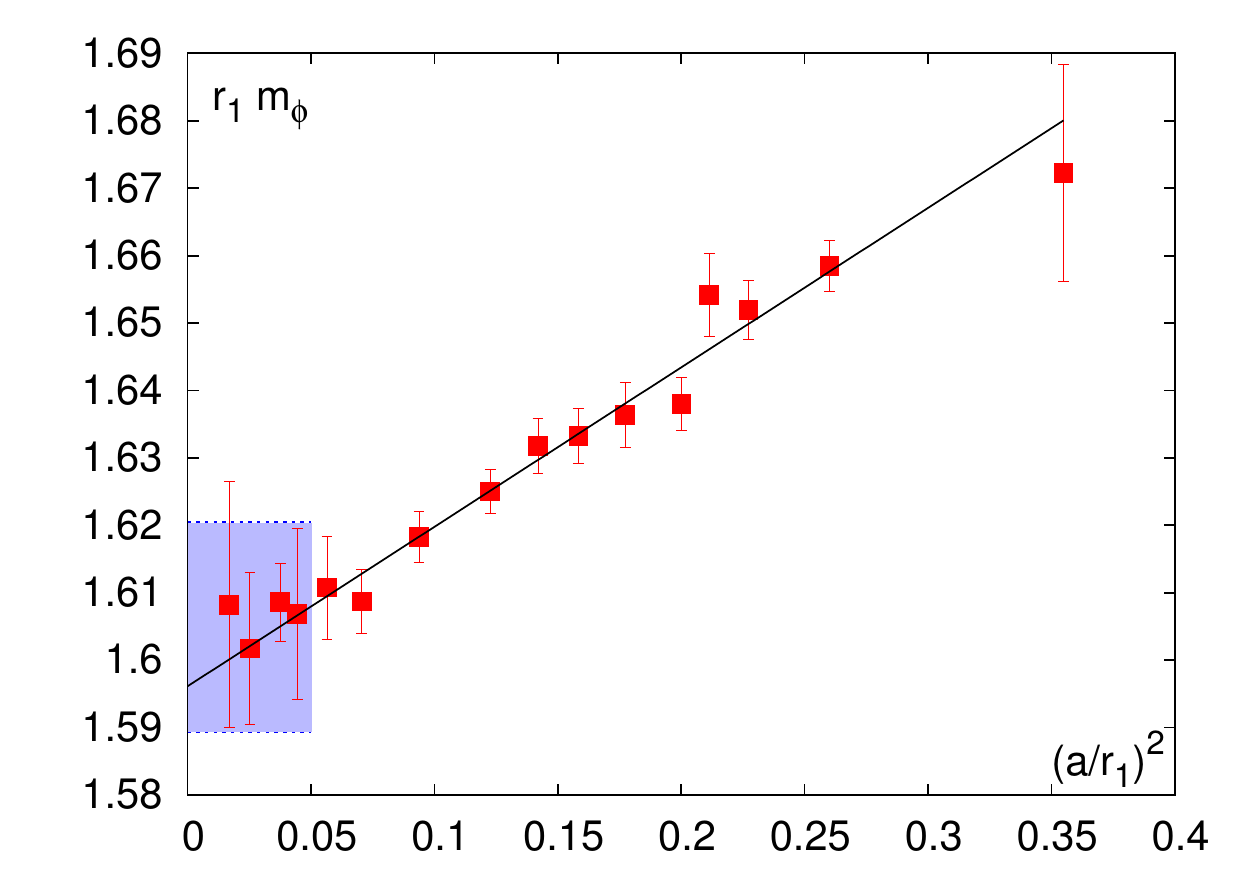}
\caption{The $\phi$ meson mass as function of the lattice spacing together
with the continuum extrapolation. 
The data are corrected for deviations from the LCP as described in the text.
The band shows the experimental value.}
\label{fig:mphi}
\end{figure}

In Fig.~\ref{fig:mphi}, we show the $\phi$ meson mass $m_{\phi}$ in $r_1$ 
units as a function of the lattice
spacing together with the continuum extrapolation. 
Again, we use the simple form $(r_1 m_{\phi})_{cont}+ g_{\phi} (a/r_1)^2$ 
to do the continuum extrapolations,
and get $(r_1 m_{\phi})=1.5961(30)$ and $g_{\phi}=0.236(18)$ 
with $\chi^2/{\rm dof}=0.42$.
Our continuum extrapolation agrees with the
experimental result (shown as the band). 

In summary, as one can see from Figs.~\ref{fig:fps} and
\ref{fig:mphi}, the hadronic observables provide additional valuable
cross-checks for the determination of the lattice spacing. The cutoff
dependence of $f_K$ and $m_{\phi}$ is very similar to the cutoff
dependence of $w_0$. Therefore, the change in the lattice spacing and
the temperature scale will be similar to the case when $w_0$ is used
to set the lattice spacing.
\section{Observables for EoS}
\label{app:observables}

In this appendix, we summarize the quantities we used to evaluate the trace
anomaly. They include the expectation values of local observables such as the 
gauge action density $s_G$ and the light and strange quark condensates. At nonzero
temperature, we also report results for the disconnected light and strange chiral 
susceptibilities ($\chi_l^{disc}$ and $\chi_s^{disc}$), {\it i.e.},
the fluctuations of the light and strange quark condensates, as well as the 
bare Polyakov loop $L_{bare}$.
We use the same definitions of these quantities as in Ref.~\cite{Bazavov:2011nk}.
In particular, the quark condensates are normalized per single flavor, the disconnected
chiral susceptibility for light quarks is normalized for two flavors, and the disconnected
chiral susceptibility for strange quarks is normalized for a single flavor.

The gauge action density and the quark condensates for zero temperature are given in 
Table \ref{tab0}. 
The observables at nonzero temperature are summarized in Tables \ref{tab6}, 
\ref{tab8}, \ref{tab10}, and \ref{tab12} for $N_{\tau}=6,~8,~10$, and $12$ 
lattices.

For the calculation of the trace anomaly, one also needs the lattice
spacing in units of $r_1$, the nonperturbative beta function $R_{\beta}$, the strange
quark mass as function of $\beta$ along the LCP and the mass renormalization function $R_m$ calculated along the LCP.
The parametrization of $r_1/a$ and $R_{\beta}$ and the calculation of their errors have been discussed
in Appendix \ref{app_r0r1}. In Table \ref{tab0}, we give these quantities with their errors.
The value of $m_s$ along the LCP has been discussed in Appendix \ref{app_masses}, where an explicit parametrization of
$\tilde m_s=r_1 m_s^{LCP}$ has been given. The mass renormalization function $R_m$ can be written
as 
\begin{equation}
R_m=-R_{\beta}^{-1} (1-\tilde R_m R_{\beta}),\quad\tilde R_m=\frac{1}{\tilde m_s}\frac{d \tilde m_s}{d \beta}.
\end{equation}
The values of $\tilde m_s$ and $\tilde R_m R_{\beta}$ are also given in 
Table \ref{tab0}. As one can see from Figs.~\ref{fig_m_eta} and
\ref{fig:ms}, the deviations of the input strange quark masses from
the LCP for $\beta \ge 7.03$ are at most $7\%$ and are below $1\%$ for
$\beta < 7.03$. 
To include the errors arising from the deviations from the LCP, we
assign a 1\% error to $m_s$ for  $\beta < 7.03$ and
10\% errors to $m_s$ for $\beta \ge 7.03$ in Eq. (9). All the
errors discussed above are added in quadratures to get the 
the total error estimate of the trace anomaly presented in Sec. III.
Our estimate of the systematic errors on the trace anomaly due to
the deviations from LCP includes only the difference between 
the input $m_s$ and the value of $m_s$ along the LCP. Since $m_l/m_s$ is kept
constant, 
there is no additional uncertainty due to $m_l$. The value
of $m_s$, however, will affect the expectation value of the gluon action
and the quark condensates shown in Eqs.(8,9). 
We did not estimate these effects, but 
based on the past experience \cite{Bernard:2006nj,Cheng:2009zi} we expect 
that these will be small.
\begin{table*}
\begin{tabular}{|l|l|l|l|l|l|l|l|}
\hline
\multicolumn{1}{|c|}{$\beta$} & \multicolumn{1}{c|}{$\langle s_G \rangle $} & 
\multicolumn{1}{c|}{$\langle \bar \psi \psi \rangle_l $} & 
\multicolumn{1}{c|}{$\langle \bar \psi \psi \rangle_s $}  & 
\multicolumn{1}{c|}{$r_1/a$} &  \multicolumn{1}{c|}{$R_{\beta}$} & 
\multicolumn{1}{c|}{$\tilde m_s$} & \multicolumn{1}{c|}{$R_{\beta} R_m$} \\
\hline
5.900  & 2.632783(101) &  0.049104(40)  &  0.105812(26) & 1.264( 5) &  1.048(17) & 0.16563(166) & -1.6999(114)   \\
5.950  & 2.597143(81)  &  0.043881(54)  &  0.098036(32) & 1.325( 5) &  1.043(17) & 0.16060(161) & -1.5925(97)    \\
6.000  & 2.561602(72)  &  0.038908(37)  &  0.090108(26) & 1.390( 4) &  1.038(16) & 0.15646(156) & -1.4972(77)    \\
6.025  & 2.544156(82)  &  0.036579(45)  &  0.086515(29) & 1.424( 4) &  1.036(15) & 0.15468(155) & -1.4545(66)    \\
6.050  & 2.526767(19)  &  0.034355(45)  &  0.083052(30) & 1.459( 4) &  1.034(15) & 0.15306(153) & -1.4149(60)    \\
6.075  & 2.509536(128) &  0.032248(36)  &  0.079962(31) & 1.494( 4) &  1.032(14) & 0.15160(152) & -1.3783(51)    \\
6.100  & 2.492425(78)  &  0.030236(36)  &  0.076478(23) & 1.531( 4) &  1.030(14) & 0.15028(150) & -1.3448(47)    \\
6.125  & 2.475464(21)  &  0.028340(18)  &  0.073315(10) & 1.568( 4) &  1.028(13) & 0.14908(149) & -1.3141(40)    \\
6.150  & 2.458696(42)  &  0.026433(38)  &  0.070257(16) & 1.607( 4) &  1.027(13) & 0.14799(148) & -1.2866(36)    \\
6.175  & 2.442094(39)  &  0.024793(20)  &  0.067305(13) & 1.646( 4) &  1.025(12) & 0.14701(147) & -1.2614(31)    \\
6.195  & 2.429035(30)  &  0.023470(19)  &  0.064892(12) & 1.679( 4) &  1.025(12) & 0.14629(146) & -1.2435(29)    \\
6.245  & 2.396804(11)  &  0.020449(14)  &  0.059664(10) & 1.762( 4) &  1.023(10) & 0.14470(145) & -1.2049(20)    \\
6.285  & 2.3716826(153)  &  0.0183176(150)  &  0.0557043(69) & 1.833( 4) &  1.022( 9) & 0.14362(144) & -1.1803(16) \\
6.341  & 2.3374387(197)  &  0.0156426(187)  &  0.0506613(102) & 1.936( 4) &  1.021( 8) & 0.14231(142) & -1.1538(12) \\ 
6.354  & 2.3297339(226)  &  0.0151142(157)  &  0.0495513(82) & 1.961( 4) &  1.021( 8) & 0.14204(142) & -1.1489(12)   \\
6.390  & 2.3084526(343)  &  0.0135987(151)  &  0.0464432(90) & 2.031( 4) &  1.021( 8) & 0.14133(141) & -1.1373(11)  \\
6.423  & 2.2894362(269)  &  0.0124345(110)  &  0.0440691(80) & 2.098( 5) &  1.022( 8) & 0.14072(141) & -1.1292(10)  \\
6.460  & 2.2685484(184)  &  0.0111926(157)  &  0.0413543(72) & 2.175( 5) &  1.022( 8) & 0.14008(140) & -1.1223(10)  \\
6.488  & 2.2530949(69)   &  0.0103368(80)  &  0.0395263(45) & 2.235( 5) &  1.023( 9) & 0.13962(140) & -1.1186(10)   \\
6.515  & 2.2384913(119)  &  0.0096409(162) &  0.0379912(95) & 2.295( 5) &  1.024( 9) & 0.13919(139) & -1.1160(10)   \\
6.550  & 2.2198533(261)  &  0.0087186(96)  &  0.0359905(69) & 2.375( 5) &  1.026(10) & 0.13864(139) & -1.1141(11)   \\
6.575  & 2.2068224(107)  &  0.0081675(80)  &  0.0345257(68) & 2.434( 5) &  1.027(10) & 0.13826(138) & -1.1134(11)   \\
6.608  & 2.1899477(85)   &  0.0074639(60)  &  0.0327412(57) & 2.513( 5) &  1.029(10) & 0.13776(138) & -1.1134(11)   \\
6.664  & 2.1620782(110)  &  0.0064352(101) &  0.0302931(50) & 2.654( 5) &  1.033(10) & 0.13691(137) & -1.1150(11)   \\
6.740  & 2.1257817(133)  &  0.0053945(96)  &  0.0272312(48) & 2.856( 5) &  1.039(10) & 0.13574(136) & -1.1196(12)   \\
6.800  & 2.0982834(130)  &  0.0045273(53)  &  0.0250900(34) & 3.026( 6) &  1.045(10) & 0.13479(135) & -1.1244(12)   \\
6.880  & 2.0630924(76)   &  0.0038178(57)  &  0.0224907(35) & 3.266( 7) &  1.053(10) & 0.13348(133) & -1.1313(12)   \\
6.950  & 2.0336080(100)  &  0.0030671(68)  &  0.0206297(36) & 3.491( 7) &  1.061(10) & 0.13230(132) & -1.1373(13)   \\
7.030  & 2.0012582(67)   &  0.0027057(50)  &  0.0186364(23) & 3.764( 8) &  1.071(10) & 0.13091(1309) & -1.1435(13)  \\
7.150  & 1.9552310(112)  &  0.0021015(39)  &  0.0162827(22) & 4.209(11) &  1.085(11) & 0.12878(1288) & -1.1507(15) \\
7.280  & 1.9083452(117)  &  0.0015991(68)  &  0.0140717(28) & 4.743(13) &  1.101(11) & 0.12646(1265) & -1.1552(16)  \\
7.373  & 1.8765238(46)   &  0.0012932(102) &  0.0122227(24) & 5.160(15) &  1.112(12) & 0.12481(1248) & -1.1562(17) \\
7.596  & 1.8053831(35)   &  0.0008701(71)  &  0.0095488(14) & 6.297(24) &  1.135(21) & 0.12103(1210) & -1.1524(28) \\
7.825  & 1.7388070(42)   &  0.0005625(54)  &  0.0075428(23) & 7.696(51) &  1.154(27) & 0.11751(1175) & -1.1420(33) \\
\hline
\end{tabular}
\caption{The gauge action density, light and strange quark condensates at zero temperatures. Also shown
are the values of $r_1/a$, $\tilde m_s$, $R_{\beta}$ and $R_{\beta} R_m$ used in the calculation of the trace
anomaly along the LCP.}
\label{tab0}
\end{table*}
 \begin{table*}
\begin{tabular}{|l|c|l|l|l|l|l|l|}
\hline
\multicolumn{1}{|c|}{$\beta$} & $T$ [MeV] & 
\multicolumn{1}{c|}{$\langle s_G \rangle $} & 
\multicolumn{1}{c|}{$\langle \bar \psi \psi \rangle_l $} & 
\multicolumn{1}{c|}{$\langle \bar \psi \psi \rangle_s $}  & 
\multicolumn{1}{c|}{$\chi^{disc}_l$} & \multicolumn{1}{c|}{$\chi^{disc}_s$} & 
\multicolumn{1}{c|}{$L_{bare}$} \\
\hline
5.900 &  133.8 & 2.632253(64)  &  0.044948(25)  &  0.105106(15) &  0.577(10) & 0.0644(17) &  0.002599(20) \\
5.950 &  140.3 & 2.596445(58)  &  0.038962(28)  &  0.097108(17) &  0.655(12) & 0.0681(19) &  0.003157(22) \\
6.000 &  147.2 & 2.560827(68)  &  0.032932(30)  &  0.088920(16) &  0.741(11) & 0.0704(14) &  0.004077(20) \\
6.025 &  150.8 & 2.543240(18)  &  0.029995(27)  &  0.085152(10) &  0.810(16) & 0.0711(15) &  0.004600(32) \\
6.050 &  154.5 & 2.525666(19)  &  0.027029(32)  &  0.081458(18) &  0.946(26) & 0.0754(27) &  0.005294(33) \\
6.075 &  158.2 & 2.508305(33)  &  0.024144(50)  &  0.078122(26) &  1.061(19) & 0.0777(21) &  0.006013(28) \\
6.100 &  162.1 & 2.490962(34)  &  0.020983(31)  &  0.074290(16) &  1.148(23) & 0.0798(18) &  0.007027(18) \\
6.125 &  166.0 & 2.473804(31)  &  0.017975(47)  &  0.070757(17) &  1.223(24) & 0.0816(29) &  0.008156(31) \\
6.150 &  170.1 & 2.456840(33)  &  0.014837(53)  &  0.067264(24) &  1.313(28) & 0.0889(19) &  0.009463(53) \\
6.175 &  174.3 & 2.439980(29)  &  0.012008(39)  &  0.063819(18) &  1.148(26) & 0.0906(31) &  0.010942(55) \\
6.195 &  177.8 & 2.426644(30)  &  0.009958(46)  &  0.061004(18) &  1.009(22) & 0.0932(15) &  0.012192(42) \\
6.245 &  186.5 & 2.394214(29)  &  0.006302(37)  &  0.054910(31) &  0.440(12) & 0.0866(23) &  0.015619(71) \\
6.285 &  194.1 & 2.368816(49)  &  0.004572(23)  &  0.050296(23) &  0.1963(71) & 0.0753(28) &  0.018535(49) \\
6.341 &  205.0 & 2.334602(54)  &  0.003229(14)  &  0.044690(24) &  0.0686(34) & 0.0724(20) &  0.022822(75)  \\
6.354 &  207.6 & 2.326948(71)  &  0.0030128(68)  &  0.043460(18) &  0.05033(157) & 0.06293(215) &  0.023805(68) \\ 
6.423 &  222.1 & 2.286667(62)  &  0.0022268(37)  &  0.037597(16) &  0.01467(44) & 0.04245(122) &  0.029564(72)  \\
6.488 &  236.6 & 2.250583(43)  &  0.0018235(37)  &  0.033135(14) &  0.00656(46) & 0.02796(45) &  0.034892(104) \\
6.515 &  243.0 & 2.236043(44)  &  0.0017067(23)  &  0.031671(12) &  0.00394(25) & 0.02268(48) &  0.037358(81) \\
6.550 &  251.4 & 2.217540(36)  &  0.0015783(23)  &  0.029853(10) &  0.00289(28) & 0.01663(45) &  0.040428(102) \\
6.575 &  257.7 & 2.204590(57)  &  0.0014903(19)  &  0.028517(10) &  0.00186(20) & 0.01368(46) &  0.042698(84) \\
6.608 &  266.1 & 2.187842(66)  &  0.0013966(12)  &  0.026962( 8) &  0.00166(13) & 0.01048(32) &  0.045373(81) \\
6.664 &  281.0 & 2.160154(56)  &  0.00127253(92)  &  0.0249128(67) &  0.000646(55) & 0.006665(202) &  0.050566(88) \\ 
6.800 &  320.4 & 2.096708(55)  &  0.00104636(41)  &  0.0207485(33) &  0.000170(48) & 0.002140(77) &  0.062640(66) \\
6.950 &  369.6 & 2.032442(34)  &  0.00086791(28)  &  0.0172826(20) &  0.000055(14) & 0.000795(36) &  0.076417(96)  \\
7.150 &  445.6 & 1.954433(30)  &  0.000696540(50)  &  0.01390563(83) &  0.00000214(44) & 0.0002043(52) &  0.094300(69) \\
7.280 &  502.1 & 1.907706(57)  &  0.000609280(48)  &  0.01216886(50) &  0.00000388(166) & 0.0001201(108) &  0.105861(91) \\
7.373 &  546.3 & 1.875942(38)  &  0.000531641(23)  &  0.01062292(55) &  0.000001297(43) & 0.0002442(32) &  0.113932(98)  \\
7.596 &  666.7 & 1.805019(28)  &  0.000422615(31)  &  0.00844724(40) &  0.000001080(397) & 0.0001474(113) &  0.132910(111) \\
7.825 &  814.8 & 1.738571(16)  &  0.000338793(10)  &  0.00677357(19) &  0.0000003877(22) & 0.00007679(64) &  0.151621(112) \\
\hline
\end{tabular}
\caption{Expectation value of local observables calculated on $N_{\tau}=6$ lattices}
\label{tab6}
\end{table*}
 \begin{table*}
\begin{tabular}{|l|c|l|l|l|l|l|l|}
\hline
\multicolumn{1}{|c|}{$\beta$} & $T$ [MeV] &
\multicolumn{1}{c|}{$\langle s_G \rangle $} &
\multicolumn{1}{c|}{$\langle \bar \psi \psi \rangle_l $} &
\multicolumn{1}{c|}{$\langle \bar \psi \psi \rangle_s $}  &
\multicolumn{1}{c|}{$\chi^{disc}_l$} & \multicolumn{1}{c|}{$\chi^{disc}_s$} &
\multicolumn{1}{c|}{$L_{bare}$} \\
\hline
6.050 &  115.8 & 2.5266237(348)  &  0.0327762(186)  &  0.0828810(115) &  0.4139(69) & 0.05669(150) &  0.0004353(134) \\
6.125 &  124.5 & 2.4753158(262)  &  0.0264181(110)  &  0.0730705(82) &  0.4706(75) & 0.06081(109) &  0.0006204(47)   \\
6.175 &  130.7 & 2.4419268(287)  &  0.0225733(227)  &  0.0669802(122) &  0.4779(81) & 0.05382(134) &  0.0008371(116) \\
6.195 &  133.3 & 2.4288377(376)  &  0.0211403(154)  &  0.0645320(91) &  0.4973(132) & 0.05412(245) &  0.0009414(166) \\
6.245 &  139.9 & 2.3965418(294)  &  0.0177511(150)  &  0.0591829(61) &  0.5346(109) & 0.05172(93) &  0.0012420(117)  \\
6.285 &  145.5 & 2.3713746(283)  &  0.0151984(242)  &  0.0550896(84) &  0.5607(334) & 0.04938(161) &  0.0016083(177) \\
6.341 &  153.7 & 2.3370256(418)  &  0.0118052(185)  &  0.0498389(77) &  0.7507(139) & 0.05228(141) &  0.0022252(131)  \\
6.354 &  155.7 & 2.3292352(137)  &  0.0109211(513)  &  0.0485851(197) &  0.8123(357) & 0.05624(304) &  0.0024708(325) \\
6.390 &  161.3 & 2.3079034(111)  &  0.0088051(453)  &  0.0452635(142) &  0.8915(238) & 0.05593(184) &  0.0031329(387) \\
6.423 &  166.6 & 2.2887805(179)  &  0.0069624(573)  &  0.0426307(216) &  0.9083(238) & 0.05822(342) &  0.0038408(355) \\
6.460 &  172.7 & 2.2678076(214)  &  0.0050971(363)  &  0.0395899(152) &  0.6520(254) & 0.05707(351) &  0.0048401(370) \\
6.488 &  177.5 & 2.2523061(146)  &  0.0040973(324)  &  0.0375676(211) &  0.4791(204) & 0.05586(308) &  0.0055937(523) \\
6.515 &  182.2 & 2.2376344(146)  &  0.0033605(111)  &  0.0358104(81) &  0.3208(65) & 0.05421(204) &  0.0064548(245)   \\
6.550 &  188.6 & 2.2189982(111)  &  0.0026714(70)  &  0.0335964(50) &  0.1676(32) & 0.04600(150) &  0.0076062(174)    \\
6.575 &  193.3 & 2.2059639(144)  &  0.0023153(61)  &  0.0319944(50) &  0.1061(20) & 0.04065(91) &  0.0084632(181)     \\
6.608 &  199.5 & 2.1890447(95)  &  0.0019665(55)  &  0.0300405(69) &  0.0554(18) & 0.03770(106) &  0.0096959(240)     \\
6.664 &  210.7 & 2.1612301(160)  &  0.00161467(242)  &  0.0274505(65) &  0.02910(72) & 0.028704(850) &  0.0118668(196) \\
6.740 &  226.8 & 2.1249421(157)  &  0.00132194(152)  &  0.0243573(47) &  0.01123(37) & 0.017709(487) &  0.0150425(309) \\
6.800 &  240.3 & 2.0975154(118)  &  0.00117166(80)  &  0.0223027(32) &  0.00606(20) & 0.010758(192) &  0.0176361(287) \\
6.880 &  259.3 & 2.0624169(93)  &  0.00102161(56)  &  0.0199079(30) &  0.00282(16) & 0.005774(105) &  0.0213349(344)  \\
6.950 &  277.2 & 2.0329709(115)  &  0.00092680(48)  &  0.0182584(25) &  0.00138(13) & 0.003370(64) &  0.0246964(418) \\
7.030 &  298.9 & 2.0006938(123)  &  0.00083316(37)  &  0.0165135(12) &  0.00080(21) & 0.001763(20) &  0.0286964(278) \\
7.150 &  334.2 &  1.9547446(127)  &  0.00072858(127)  &  0.014507068(795) &  0.0001447(228) & 0.0007811(177) &  0.0349571(493) \\ 
7.280 &  376.6 &  1.9079588(118)  &  0.00063333(87)  &  0.012635689(523) &  0.0000317(121) & 0.0003433(117) &  0.0419502(385) \\
7.373 &  409.7 &  1.8762085(84)  &  0.00055109(22)  &  0.011005082(305) &  0.0000075(24) & 0.0004985(47) &  0.0470148(330) \\
7.596 &  500.0 &  1.8051472(106)  &  0.00043608(17)  &  0.008714430(192) &  0.0000040(28) & 0.0002289(36) &  0.0595132(289) \\
7.825 &  611.1 & 1.73864086(834)  &  0.0003485897(29)  &  0.006968632(67) &  0.000000600( 3) & 0.00011855(51) &  0.0725663(201) \\
\hline
\end{tabular}
\caption{Expectation value of local observables calculated on $N_{\tau}=8$ lattices}
\label{tab8}
\end{table*}

\begin{table*}
\begin{tabular}{|l|c|l|l|l|l|l|l|}
\hline
\multicolumn{1}{|c|}{$\beta$} & $T$ [MeV] &
\multicolumn{1}{c|}{$\langle s_G \rangle $} &
\multicolumn{1}{c|}{$\langle \bar \psi \psi \rangle_l $} &
\multicolumn{1}{c|}{$\langle \bar \psi \psi \rangle_s $}  &
\multicolumn{1}{c|}{$\chi^{disc}_l$} & \multicolumn{1}{c|}{$\chi^{disc}_s$} &
\multicolumn{1}{c|}{$L_{bare}$} \\
\hline
6.488 &  142.0 &  2.2529939(94)  &  0.008717(66)  &  0.0392107(30) &  0.4696(80) & 0.03671(59) &  0.0006332(54)  \\
6.515 &  145.8 &  2.2383351(92)  &  0.007845(53)  &  0.0376092(21) &  0.4983(48) & 0.03601(67) &  0.0007530(46)  \\
6.575 &  154.6 &  2.2066486(87)  &  0.005916(112)  &  0.0339874(45) &  0.6199(111) & 0.03566(87) &  0.0011236(72) \\
6.608 &  159.6 &  2.1897092(97)  &  0.004863(141)  &  0.0320595(57) &  0.6914(140) & 0.03790(54) &  0.0014074(111) \\
6.664 &  168.6 &  2.1618075(67)  &  0.003366(119)  &  0.0293881(48) &  0.5971(104) & 0.03805(76) &  0.0020061(99) \\
6.740 &  181.4 &  2.1254552(49)  &  0.002058(44)  &  0.0260341(30) &  0.2350(27) & 0.03178(63) &  0.0030709(50)   \\
6.800 &  192.2 &  2.0979463(46)  &  0.001551(21)  &  0.0237258(22) &  0.10175(85) & 0.02664(27) &  0.0040471(52)  \\
6.880 &  207.5 &  2.0627661(47)  &  0.001196(11)  &  0.0210031(13) &  0.03805(50) & 0.01794(15) &  0.0055452(50)  \\
6.950 &  221.8 &  2.0332765(52)  &  0.0010256(87)  &  0.0191246(19) &  0.01785(44) & 0.01182(21) &  0.0069963(116) \\
7.030 &  239.1 &  2.0009567(52)  &  0.0008889(70)  &  0.0171719(16) &  0.00732(49) & 0.00655(12) &  0.0088725(130) \\
7.150 &  267.4 &  1.9549570(41)  &  0.0007594(33)  &  0.01496641(85) &  0.002826(227) & 0.002693(65) &  0.0118853(116) \\ 
7.280 &  301.3 &  1.9081112(43)  &  0.0006518(22)  &  0.01296143(67) &  0.000578(120) & 0.001005(28) &  0.0155783(205) \\
7.373 &  327.8 &  1.8763272(51)  &  0.0005648(11)  &  0.01125795(32) &  0.000239(83) & 0.000522(29) &  0.0183751(154)  \\
7.596 &  400.0 &  1.8052323(39)  &  0.00044450(86)  &  0.008876950(228) &  0.0000210(201) & 0.0001413(88) &  0.0256495(265) \\ 
7.825 &  488.9 &  1.7387004(82)  &  0.00035425( 5)  &  0.007080600(105) &  0.00000040(14) & 0.0000461(10) &  0.0337466(261)  \\
\hline
\end{tabular}
\caption{Expectation value of local observables calculated on $N_{\tau}=10$ lattices}
\label{tab10}
\end{table*}
 \begin{table*}
\begin{tabular}{|l|c|l|l|l|l|l|l|}
\hline
\multicolumn{1}{|c|}{$\beta$} & $T$ [MeV] &
\multicolumn{1}{c|}{$\langle s_G \rangle $} &
\multicolumn{1}{c|}{$\langle \bar \psi \psi \rangle_l $} &
\multicolumn{1}{c|}{$\langle \bar \psi \psi \rangle_s $}  &
\multicolumn{1}{c|}{$\chi^{disc}_l$} & \multicolumn{1}{c|}{$\chi^{disc}_s$} &
\multicolumn{1}{c|}{$L_{bare}$} \\
\hline
6.740 &  151.2  & 2.1257086(54)  &  0.004079(97)   &  0.0269239(25)  &  0.4498(119)     & 0.02317(53)    &  0.0004772(73)  \\
6.800 &  160.2  & 2.0981866(76)  &  0.003016(120)  &  0.0246617(30)  &  0.5714(211)     & 0.02569(175)   &  0.0007167(44)  \\
6.880 &  172.9  & 2.0629734(44)  &  0.001882(75)   &  0.0218834(45)  &  0.3838(75)      & 0.02479(102)   &  0.0011690(73)  \\
6.950 &  184.8  & 2.0334633(59)  &  0.001339(52)   &  0.0198904(40)  &  0.1644(30)      & 0.02098(123)   &  0.0016813(86)  \\
7.030 &  199.2  & 2.0011122(72)  &  0.001025(22)   &  0.0177901(23)  &  0.0617(15)      & 0.01489(61)    &  0.0024077(75)  \\
7.150 &  222.8  & 1.9550761(46)  &  0.000808(10)   &  0.0153880(17)  &  0.0183(10)      & 0.00823(55)    &  0.0037030(71)  \\
7.280 &  251.1  & 1.9082143(37)  &  0.000675(13)   &  0.0132415(21)  &  0.0061(12)      & 0.00312(13)    &  0.0054060(154) \\
7.373 &  273.1  & 1.8764082(49)  &  0.0005792(88)  &  0.01146391(89) &  0.0042(12)      & 0.002445(51)   &  0.0068171(125) \\
7.596 &  333.3  & 1.8052889(42)  &  0.00045065(76) &  0.00899511(28) &  0.000093(58)    & 0.000682(11)   &  0.0106861(146) \\
7.825 &  407.4  & 1.7387435(49)  &  0.000358140(90)  &  0.00715650(16) &  0.000000601(50) & 0.0001120(91)  &  0.0153196(362) \\
\hline
\end{tabular}
\caption{Expectation value of local observables calculated on $N_{\tau}=12$ lattices}
\label{tab12}
\end{table*}


\bibliography{HotQCD}

\begin{thebibliography}{61}
\expandafter\ifx\csname natexlab\endcsname\relax\def\natexlab#1{#1}\fi
\expandafter\ifx\csname bibnamefont\endcsname\relax
  \def\bibnamefont#1{#1}\fi
\expandafter\ifx\csname bibfnamefont\endcsname\relax
  \def\bibfnamefont#1{#1}\fi
\expandafter\ifx\csname citenamefont\endcsname\relax
  \def\citenamefont#1{#1}\fi
\expandafter\ifx\csname url\endcsname\relax
  \def\url#1{\texttt{#1}}\fi
\expandafter\ifx\csname urlprefix\endcsname\relax\def\urlprefix{URL }\fi
\providecommand{\bibinfo}[2]{#2}
\providecommand{\eprint}[2][]{\url{#2}}

\bibitem[{\citenamefont{Wilson}(1974)}]{Wilson:1974sk}
\bibinfo{author}{\bibfnamefont{K.~G.} \bibnamefont{Wilson}},
  \bibinfo{journal}{Phys. Rev.} \textbf{\bibinfo{volume}{D10}},
  \bibinfo{pages}{2445} (\bibinfo{year}{1974}).

\bibitem[{\citenamefont{Creutz}(1980)}]{Creutz:1980zw}
\bibinfo{author}{\bibfnamefont{M.}~\bibnamefont{Creutz}},
  \bibinfo{journal}{Phys.Rev.} \textbf{\bibinfo{volume}{D21}},
  \bibinfo{pages}{2308} (\bibinfo{year}{1980}).

\bibitem[{\citenamefont{Engels et~al.}(1981)\citenamefont{Engels, Karsch, Satz,
  and Montvay}}]{Engels:1980ty}
\bibinfo{author}{\bibfnamefont{J.}~\bibnamefont{Engels}},
  \bibinfo{author}{\bibfnamefont{F.}~\bibnamefont{Karsch}},
  \bibinfo{author}{\bibfnamefont{H.}~\bibnamefont{Satz}}, \bibnamefont{and}
  \bibinfo{author}{\bibfnamefont{I.}~\bibnamefont{Montvay}},
  \bibinfo{journal}{Phys. Lett.} \textbf{\bibinfo{volume}{B101}},
  \bibinfo{pages}{89} (\bibinfo{year}{1981}).

\bibitem[{\citenamefont{Aoki et~al.}(2009)\citenamefont{Aoki, Borsanyi, Durr,
  Fodor, Katz et~al.}}]{Aoki:2009sc}
\bibinfo{author}{\bibfnamefont{Y.}~\bibnamefont{Aoki}},
  \bibinfo{author}{\bibfnamefont{S.}~\bibnamefont{Borsanyi}},
  \bibinfo{author}{\bibfnamefont{S.}~\bibnamefont{Durr}},
  \bibinfo{author}{\bibfnamefont{Z.}~\bibnamefont{Fodor}},
  \bibinfo{author}{\bibfnamefont{S.~D.} \bibnamefont{Katz}},
  \bibnamefont{et~al.}, \bibinfo{journal}{JHEP}
  \textbf{\bibinfo{volume}{0906}}, \bibinfo{pages}{088} (\bibinfo{year}{2009}),
  \eprint{0903.4155}.

\bibitem[{\citenamefont{Bazavov
  et~al.}(2012{\natexlab{a}})\citenamefont{Bazavov, Bhattacharya, Cheng, DeTar,
  Ding et~al.}}]{Bazavov:2011nk}
\bibinfo{author}{\bibfnamefont{A.}~\bibnamefont{Bazavov}},
  \bibinfo{author}{\bibfnamefont{T.}~\bibnamefont{Bhattacharya}},
  \bibinfo{author}{\bibfnamefont{M.}~\bibnamefont{Cheng}},
  \bibinfo{author}{\bibfnamefont{C.}~\bibnamefont{DeTar}},
  \bibinfo{author}{\bibfnamefont{H.}~\bibnamefont{Ding}}, \bibnamefont{et~al.},
  \bibinfo{journal}{Phys. Rev.} \textbf{\bibinfo{volume}{D85}},
  \bibinfo{pages}{054503} (\bibinfo{year}{2012}{\natexlab{a}}),
  \eprint{1111.1710}.

\bibitem[{\citenamefont{Borsanyi
  et~al.}(2012{\natexlab{a}})\citenamefont{Borsanyi, Fodor, Katz, Krieg, Ratti
  et~al.}}]{Borsanyi:2011sw}
\bibinfo{author}{\bibfnamefont{S.}~\bibnamefont{Borsanyi}},
  \bibinfo{author}{\bibfnamefont{Z.}~\bibnamefont{Fodor}},
  \bibinfo{author}{\bibfnamefont{S.~D.} \bibnamefont{Katz}},
  \bibinfo{author}{\bibfnamefont{S.}~\bibnamefont{Krieg}},
  \bibinfo{author}{\bibfnamefont{C.}~\bibnamefont{Ratti}},
  \bibnamefont{et~al.}, \bibinfo{journal}{JHEP}
  \textbf{\bibinfo{volume}{1201}}, \bibinfo{pages}{138}
  (\bibinfo{year}{2012}{\natexlab{a}}), \eprint{1112.4416}.

\bibitem[{\citenamefont{Bazavov et~al.}(2012{\natexlab{b}})}]{Bazavov:2012jq}
\bibinfo{author}{\bibfnamefont{A.}~\bibnamefont{Bazavov}} \bibnamefont{et~al.}
  (\bibinfo{collaboration}{HotQCD Collaboration}), \bibinfo{journal}{Phys.Rev.}
  \textbf{\bibinfo{volume}{D86}}, \bibinfo{pages}{034509}
  (\bibinfo{year}{2012}{\natexlab{b}}), \eprint{1203.0784}.

\bibitem[{\citenamefont{Bazavov et~al.}(2014)\citenamefont{Bazavov, Ding,
  Hegde, Kaczmarek, Karsch et~al.}}]{Bazavov:2014xya}
\bibinfo{author}{\bibfnamefont{A.}~\bibnamefont{Bazavov}},
  \bibinfo{author}{\bibfnamefont{H.~T.} \bibnamefont{Ding}},
  \bibinfo{author}{\bibfnamefont{P.}~\bibnamefont{Hegde}},
  \bibinfo{author}{\bibfnamefont{O.}~\bibnamefont{Kaczmarek}},
  \bibinfo{author}{\bibfnamefont{F.}~\bibnamefont{Karsch}},
  \bibnamefont{et~al.} (\bibinfo{year}{2014}), \eprint{1404.6511}.

\bibitem[{\citenamefont{Petreczky}(2012{\natexlab{a}})}]{Petreczky:2012rq}
\bibinfo{author}{\bibfnamefont{P.}~\bibnamefont{Petreczky}},
  \bibinfo{journal}{J. Phys.} \textbf{\bibinfo{volume}{G39}},
  \bibinfo{pages}{093002} (\bibinfo{year}{2012}{\natexlab{a}}),
  \eprint{1203.5320}.

\bibitem[{\citenamefont{Philipsen}(2013)}]{Philipsen:2012nu}
\bibinfo{author}{\bibfnamefont{O.}~\bibnamefont{Philipsen}},
  \bibinfo{journal}{Prog. Part. Nucl. Phys.} \textbf{\bibinfo{volume}{70}},
  \bibinfo{pages}{55} (\bibinfo{year}{2013}), \eprint{1207.5999}.

\bibitem[{\citenamefont{DeTar and Heller}(2009)}]{DeTar:2009ef}
\bibinfo{author}{\bibfnamefont{C.}~\bibnamefont{DeTar}} \bibnamefont{and}
  \bibinfo{author}{\bibfnamefont{U.}~\bibnamefont{Heller}},
  \bibinfo{journal}{Eur.Phys.J.} \textbf{\bibinfo{volume}{A41}},
  \bibinfo{pages}{405} (\bibinfo{year}{2009}), \eprint{0905.2949}.

\bibitem[{\citenamefont{Linde}(1980)}]{Linde:1980ts}
\bibinfo{author}{\bibfnamefont{A.~D.} \bibnamefont{Linde}},
  \bibinfo{journal}{Phys. Lett.} \textbf{\bibinfo{volume}{B96}},
  \bibinfo{pages}{289} (\bibinfo{year}{1980}).

\bibitem[{\citenamefont{Braun-Munzinger
  et~al.}(2003)\citenamefont{Braun-Munzinger, Redlich, and
  Stachel}}]{BraunMunzinger:2003zd}
\bibinfo{author}{\bibfnamefont{P.}~\bibnamefont{Braun-Munzinger}},
  \bibinfo{author}{\bibfnamefont{K.}~\bibnamefont{Redlich}}, \bibnamefont{and}
  \bibinfo{author}{\bibfnamefont{J.}~\bibnamefont{Stachel}}
  (\bibinfo{year}{2003}), \bibinfo{note}{to appear in Quark Gluon Plasma 3,
  eds. R.C. Hwa and Xin-Nian Wang, World Scientific Publishing},
  \eprint{nucl-th/0304013}.

\bibitem[{\citenamefont{Ejiri et~al.}(2006)\citenamefont{Ejiri, Karsch, and
  Redlich}}]{Ejiri:2005wq}
\bibinfo{author}{\bibfnamefont{S.}~\bibnamefont{Ejiri}},
  \bibinfo{author}{\bibfnamefont{F.}~\bibnamefont{Karsch}}, \bibnamefont{and}
  \bibinfo{author}{\bibfnamefont{K.}~\bibnamefont{Redlich}},
  \bibinfo{journal}{Phys.Lett.} \textbf{\bibinfo{volume}{B633}},
  \bibinfo{pages}{275} (\bibinfo{year}{2006}), \eprint{hep-ph/0509051}.

\bibitem[{\citenamefont{Gale et~al.}(2013)\citenamefont{Gale, Jeon, and
  Schenke}}]{Gale:2013da}
\bibinfo{author}{\bibfnamefont{C.}~\bibnamefont{Gale}},
  \bibinfo{author}{\bibfnamefont{S.}~\bibnamefont{Jeon}}, \bibnamefont{and}
  \bibinfo{author}{\bibfnamefont{B.}~\bibnamefont{Schenke}},
  \bibinfo{journal}{Int.J.Mod.Phys.} \textbf{\bibinfo{volume}{A28}},
  \bibinfo{pages}{1340011} (\bibinfo{year}{2013}), \eprint{1301.5893}.

\bibitem[{\citenamefont{Umeda et~al.}(2012)}]{Umeda:2012er}
\bibinfo{author}{\bibfnamefont{T.}~\bibnamefont{Umeda}} \bibnamefont{et~al.}
  (\bibinfo{collaboration}{WHOT-QCD Collaboration}),
  \bibinfo{journal}{Phys.Rev.} \textbf{\bibinfo{volume}{D85}},
  \bibinfo{pages}{094508} (\bibinfo{year}{2012}), \eprint{1202.4719}.

\bibitem[{\citenamefont{Heller et~al.}(1999)\citenamefont{Heller, Karsch, and
  Sturm}}]{Heller:1999xz}
\bibinfo{author}{\bibfnamefont{U.~M.} \bibnamefont{Heller}},
  \bibinfo{author}{\bibfnamefont{F.}~\bibnamefont{Karsch}}, \bibnamefont{and}
  \bibinfo{author}{\bibfnamefont{B.}~\bibnamefont{Sturm}},
  \bibinfo{journal}{Phys. Rev.} \textbf{\bibinfo{volume}{D60}},
  \bibinfo{pages}{114502} (\bibinfo{year}{1999}), \eprint{hep-lat/9901010}.

\bibitem[{\citenamefont{Blum et~al.}(1997)\citenamefont{Blum, Detar, Gottlieb,
  Rummukainen, Heller et~al.}}]{Blum:1996uf}
\bibinfo{author}{\bibfnamefont{T.}~\bibnamefont{Blum}},
  \bibinfo{author}{\bibfnamefont{C.~E.} \bibnamefont{Detar}},
  \bibinfo{author}{\bibfnamefont{S.~A.} \bibnamefont{Gottlieb}},
  \bibinfo{author}{\bibfnamefont{K.}~\bibnamefont{Rummukainen}},
  \bibinfo{author}{\bibfnamefont{U.~M.} \bibnamefont{Heller}},
  \bibnamefont{et~al.}, \bibinfo{journal}{Phys. Rev.}
  \textbf{\bibinfo{volume}{D55}}, \bibinfo{pages}{1133} (\bibinfo{year}{1997}),
  \eprint{hep-lat/9609036}.

\bibitem[{\citenamefont{Orginos et~al.}(1999)\citenamefont{Orginos, Toussaint,
  and Sugar}}]{Orginos:1999cr}
\bibinfo{author}{\bibfnamefont{K.}~\bibnamefont{Orginos}},
  \bibinfo{author}{\bibfnamefont{D.}~\bibnamefont{Toussaint}},
  \bibnamefont{and} \bibinfo{author}{\bibfnamefont{R.}~\bibnamefont{Sugar}}
  (\bibinfo{collaboration}{MILC Collaboration}), \bibinfo{journal}{Phys. Rev.}
  \textbf{\bibinfo{volume}{D60}}, \bibinfo{pages}{054503}
  (\bibinfo{year}{1999}), \eprint{hep-lat/9903032}.

\bibitem[{\citenamefont{Karsch et~al.}(2000)\citenamefont{Karsch, Laermann, and
  Peikert}}]{Karsch:2000ps}
\bibinfo{author}{\bibfnamefont{F.}~\bibnamefont{Karsch}},
  \bibinfo{author}{\bibfnamefont{E.}~\bibnamefont{Laermann}}, \bibnamefont{and}
  \bibinfo{author}{\bibfnamefont{A.}~\bibnamefont{Peikert}},
  \bibinfo{journal}{Phys. Lett.} \textbf{\bibinfo{volume}{B478}},
  \bibinfo{pages}{447} (\bibinfo{year}{2000}), \eprint{hep-lat/0002003}.

\bibitem[{\citenamefont{Bernard et~al.}(2007)\citenamefont{Bernard, Burch,
  DeTar, Gottlieb, Levkova et~al.}}]{Bernard:2006nj}
\bibinfo{author}{\bibfnamefont{C.}~\bibnamefont{Bernard}},
  \bibinfo{author}{\bibfnamefont{T.}~\bibnamefont{Burch}},
  \bibinfo{author}{\bibfnamefont{C.~E.} \bibnamefont{DeTar}},
  \bibinfo{author}{\bibfnamefont{S.}~\bibnamefont{Gottlieb}},
  \bibinfo{author}{\bibfnamefont{L.}~\bibnamefont{Levkova}},
  \bibnamefont{et~al.}, \bibinfo{journal}{Phys. Rev.}
  \textbf{\bibinfo{volume}{D75}}, \bibinfo{pages}{094505}
  (\bibinfo{year}{2007}), \eprint{hep-lat/0611031}.

\bibitem[{\citenamefont{Cheng et~al.}(2008)\citenamefont{Cheng, Christ, Datta,
  van~der Heide, Jung et~al.}}]{Cheng:2007jq}
\bibinfo{author}{\bibfnamefont{M.}~\bibnamefont{Cheng}},
  \bibinfo{author}{\bibfnamefont{N.}~\bibnamefont{Christ}},
  \bibinfo{author}{\bibfnamefont{S.}~\bibnamefont{Datta}},
  \bibinfo{author}{\bibfnamefont{J.}~\bibnamefont{van~der Heide}},
  \bibinfo{author}{\bibfnamefont{C.}~\bibnamefont{Jung}}, \bibnamefont{et~al.},
  \bibinfo{journal}{Phys. Rev.} \textbf{\bibinfo{volume}{D77}},
  \bibinfo{pages}{014511} (\bibinfo{year}{2008}), \eprint{0710.0354}.

\bibitem[{\citenamefont{Bazavov et~al.}(2009)\citenamefont{Bazavov,
  Bhattacharya, Cheng, Christ, DeTar et~al.}}]{Bazavov:2009zn}
\bibinfo{author}{\bibfnamefont{A.}~\bibnamefont{Bazavov}},
  \bibinfo{author}{\bibfnamefont{T.}~\bibnamefont{Bhattacharya}},
  \bibinfo{author}{\bibfnamefont{M.}~\bibnamefont{Cheng}},
  \bibinfo{author}{\bibfnamefont{N.}~\bibnamefont{Christ}},
  \bibinfo{author}{\bibfnamefont{C.}~\bibnamefont{DeTar}},
  \bibnamefont{et~al.}, \bibinfo{journal}{Phys. Rev.}
  \textbf{\bibinfo{volume}{D80}}, \bibinfo{pages}{014504}
  (\bibinfo{year}{2009}), \eprint{0903.4379}.

\bibitem[{\citenamefont{Cheng et~al.}(2010)\citenamefont{Cheng, Ejiri, Hegde,
  Karsch, Kaczmarek et~al.}}]{Cheng:2009zi}
\bibinfo{author}{\bibfnamefont{M.}~\bibnamefont{Cheng}},
  \bibinfo{author}{\bibfnamefont{S.}~\bibnamefont{Ejiri}},
  \bibinfo{author}{\bibfnamefont{P.}~\bibnamefont{Hegde}},
  \bibinfo{author}{\bibfnamefont{F.}~\bibnamefont{Karsch}},
  \bibinfo{author}{\bibfnamefont{O.}~\bibnamefont{Kaczmarek}},
  \bibnamefont{et~al.}, \bibinfo{journal}{Phys. Rev.}
  \textbf{\bibinfo{volume}{D81}}, \bibinfo{pages}{054504}
  (\bibinfo{year}{2010}), \eprint{0911.2215}.

\bibitem[{\citenamefont{Borsanyi
  et~al.}(2010{\natexlab{a}})\citenamefont{Borsanyi, Endrodi, Fodor, Jakovac,
  Katz et~al.}}]{Borsanyi:2010cj}
\bibinfo{author}{\bibfnamefont{S.}~\bibnamefont{Borsanyi}},
  \bibinfo{author}{\bibfnamefont{G.}~\bibnamefont{Endrodi}},
  \bibinfo{author}{\bibfnamefont{Z.}~\bibnamefont{Fodor}},
  \bibinfo{author}{\bibfnamefont{A.}~\bibnamefont{Jakovac}},
  \bibinfo{author}{\bibfnamefont{S.~D.} \bibnamefont{Katz}},
  \bibnamefont{et~al.}, \bibinfo{journal}{JHEP}
  \textbf{\bibinfo{volume}{1011}}, \bibinfo{pages}{077}
  (\bibinfo{year}{2010}{\natexlab{a}}), \eprint{1007.2580}.

\bibitem[{\citenamefont{Borsanyi et~al.}(2014)\citenamefont{Borsanyi, Fodor,
  Hoelbling, Katz, Krieg et~al.}}]{Borsanyi:2013bia}
\bibinfo{author}{\bibfnamefont{S.}~\bibnamefont{Borsanyi}},
  \bibinfo{author}{\bibfnamefont{Z.}~\bibnamefont{Fodor}},
  \bibinfo{author}{\bibfnamefont{C.}~\bibnamefont{Hoelbling}},
  \bibinfo{author}{\bibfnamefont{S.~D.} \bibnamefont{Katz}},
  \bibinfo{author}{\bibfnamefont{S.}~\bibnamefont{Krieg}},
  \bibnamefont{et~al.}, \bibinfo{journal}{Phys.Lett.}
  \textbf{\bibinfo{volume}{B370}}, \bibinfo{pages}{99} (\bibinfo{year}{2014}),
  \eprint{1309.5258}.

\bibitem[{\citenamefont{Follana et~al.}(2007)}]{Follana:2006rc}
\bibinfo{author}{\bibfnamefont{E.}~\bibnamefont{Follana}} \bibnamefont{et~al.}
  (\bibinfo{collaboration}{HPQCD Collaboration, UKQCD Collaboration}),
  \bibinfo{journal}{Phys. Rev.} \textbf{\bibinfo{volume}{D75}},
  \bibinfo{pages}{054502} (\bibinfo{year}{2007}), \eprint{hep-lat/0610092}.

\bibitem[{\citenamefont{Bazavov and
  Petreczky}(2013{\natexlab{a}})}]{Bazavov:2013yv}
\bibinfo{author}{\bibfnamefont{A.}~\bibnamefont{Bazavov}} \bibnamefont{and}
  \bibinfo{author}{\bibfnamefont{P.}~\bibnamefont{Petreczky}},
  \bibinfo{journal}{Phys.Rev.} \textbf{\bibinfo{volume}{D87}},
  \bibinfo{pages}{094505} (\bibinfo{year}{2013}{\natexlab{a}}),
  \eprint{1301.3943}.

\bibitem[{\citenamefont{Cea et~al.}(2014)\citenamefont{Cea, Cosmai, and
  Papa}}]{Cea:2014xva}
\bibinfo{author}{\bibfnamefont{P.}~\bibnamefont{Cea}},
  \bibinfo{author}{\bibfnamefont{L.}~\bibnamefont{Cosmai}}, \bibnamefont{and}
  \bibinfo{author}{\bibfnamefont{A.}~\bibnamefont{Papa}}
  (\bibinfo{year}{2014}), \eprint{1403.0821}.

\bibitem[{\citenamefont{Bazavov
  et~al.}(2012{\natexlab{c}})\citenamefont{Bazavov, Ding, Hegde, Kaczmarek,
  Karsch et~al.}}]{Bazavov:2012vg}
\bibinfo{author}{\bibfnamefont{A.}~\bibnamefont{Bazavov}},
  \bibinfo{author}{\bibfnamefont{H.}~\bibnamefont{Ding}},
  \bibinfo{author}{\bibfnamefont{P.}~\bibnamefont{Hegde}},
  \bibinfo{author}{\bibfnamefont{O.}~\bibnamefont{Kaczmarek}},
  \bibinfo{author}{\bibfnamefont{F.}~\bibnamefont{Karsch}},
  \bibnamefont{et~al.}, \bibinfo{journal}{Phys. Rev. Lett.}
  \textbf{\bibinfo{volume}{109}}, \bibinfo{pages}{192302}
  (\bibinfo{year}{2012}{\natexlab{c}}), \eprint{1208.1220}.

\bibitem[{\citenamefont{Bazavov
  et~al.}(2013{\natexlab{a}})\citenamefont{Bazavov, Ding, Hegde, Kaczmarek,
  Karsch et~al.}}]{Bazavov:2013dta}
\bibinfo{author}{\bibfnamefont{A.}~\bibnamefont{Bazavov}},
  \bibinfo{author}{\bibfnamefont{H.~T.} \bibnamefont{Ding}},
  \bibinfo{author}{\bibfnamefont{P.}~\bibnamefont{Hegde}},
  \bibinfo{author}{\bibfnamefont{O.}~\bibnamefont{Kaczmarek}},
  \bibinfo{author}{\bibfnamefont{F.}~\bibnamefont{Karsch}},
  \bibnamefont{et~al.}, \bibinfo{journal}{Phys. Rev. Lett.}
  \textbf{\bibinfo{volume}{111}}, \bibinfo{pages}{082301}
  (\bibinfo{year}{2013}{\natexlab{a}}), \eprint{1304.7220}.

\bibitem[{\citenamefont{Bazavov
  et~al.}(2013{\natexlab{b}})\citenamefont{Bazavov, Ding, Hegde, Karsch, Miao
  et~al.}}]{Bazavov:2013uja}
\bibinfo{author}{\bibfnamefont{A.}~\bibnamefont{Bazavov}},
  \bibinfo{author}{\bibfnamefont{H.~T.} \bibnamefont{Ding}},
  \bibinfo{author}{\bibfnamefont{P.}~\bibnamefont{Hegde}},
  \bibinfo{author}{\bibfnamefont{F.}~\bibnamefont{Karsch}},
  \bibinfo{author}{\bibfnamefont{C.}~\bibnamefont{Miao}}, \bibnamefont{et~al.}
  (\bibinfo{year}{2013}{\natexlab{b}}), \eprint{1309.2317}.

\bibitem[{\citenamefont{Bazavov and
  Petreczky}(2013{\natexlab{b}})}]{Bazavov:2013zha}
\bibinfo{author}{\bibfnamefont{A.}~\bibnamefont{Bazavov}} \bibnamefont{and}
  \bibinfo{author}{\bibfnamefont{P.}~\bibnamefont{Petreczky}},
  \bibinfo{journal}{Eur.Phys.J.} \textbf{\bibinfo{volume}{A49}},
  \bibinfo{pages}{85} (\bibinfo{year}{2013}{\natexlab{b}}), \eprint{1303.5500}.

\bibitem[{\citenamefont{Kim et~al.}(2013)\citenamefont{Kim, Petreczky, and
  Rothkopf}}]{Kim:2013seh}
\bibinfo{author}{\bibfnamefont{S.}~\bibnamefont{Kim}},
  \bibinfo{author}{\bibfnamefont{P.}~\bibnamefont{Petreczky}},
  \bibnamefont{and} \bibinfo{author}{\bibfnamefont{A.}~\bibnamefont{Rothkopf}}
  (\bibinfo{year}{2013}), \eprint{1310.6461}.

\bibitem[{\citenamefont{Petreczky}(2012{\natexlab{b}})}]{Petreczky:2012gi}
\bibinfo{author}{\bibfnamefont{P.}~\bibnamefont{Petreczky}}
  (\bibinfo{collaboration}{HotQCD Collaboration}), \bibinfo{journal}{PoS}
  \textbf{\bibinfo{volume}{LATTICE2012}}, \bibinfo{pages}{069}
  (\bibinfo{year}{2012}{\natexlab{b}}), \eprint{1211.1678}.

\bibitem[{\citenamefont{Bazavov}(2013)}]{Bazavov:2012bp}
\bibinfo{author}{\bibfnamefont{A.}~\bibnamefont{Bazavov}}
  (\bibinfo{collaboration}{HotQCD Collaboration}),
  \bibinfo{journal}{Nucl.Phys.A904-905} \textbf{\bibinfo{volume}{2013}},
  \bibinfo{pages}{877c} (\bibinfo{year}{2013}), \eprint{1210.6312}.

\bibitem[{\citenamefont{Bazavov and Petreczky}(2010)}]{Bazavov:2010sb}
\bibinfo{author}{\bibfnamefont{A.}~\bibnamefont{Bazavov}} \bibnamefont{and}
  \bibinfo{author}{\bibfnamefont{P.}~\bibnamefont{Petreczky}}
  (\bibinfo{collaboration}{HotQCD collaboration}), \bibinfo{journal}{J. Phys.
  Conf. Ser.} \textbf{\bibinfo{volume}{230}}, \bibinfo{pages}{012014}
  (\bibinfo{year}{2010}), \eprint{1005.1131}.

\bibitem[{\citenamefont{Sommer}(1994)}]{Sommer:1993ce}
\bibinfo{author}{\bibfnamefont{R.}~\bibnamefont{Sommer}},
  \bibinfo{journal}{Nucl. Phys.} \textbf{\bibinfo{volume}{B411}},
  \bibinfo{pages}{839} (\bibinfo{year}{1994}), \eprint{hep-lat/9310022}.

\bibitem[{\citenamefont{Bernard et~al.}(2005)}]{Bernard:2004je}
\bibinfo{author}{\bibfnamefont{C.}~\bibnamefont{Bernard}} \bibnamefont{et~al.}
  (\bibinfo{collaboration}{MILC Collaboration}), \bibinfo{journal}{Phys. Rev.}
  \textbf{\bibinfo{volume}{D71}}, \bibinfo{pages}{034504}
  (\bibinfo{year}{2005}), \eprint{hep-lat/0405029}.

\bibitem[{\citenamefont{Bazavov et~al.}(2010{\natexlab{a}})}]{Bazavov:2010hj}
\bibinfo{author}{\bibfnamefont{A.}~\bibnamefont{Bazavov}} \bibnamefont{et~al.}
  (\bibinfo{collaboration}{MILC Collaboration}), \bibinfo{journal}{PoS}
  \textbf{\bibinfo{volume}{LATTICE2010}}, \bibinfo{pages}{074}
  (\bibinfo{year}{2010}{\natexlab{a}}), \eprint{1012.0868}.

\bibitem[{\citenamefont{Borsanyi
  et~al.}(2012{\natexlab{b}})\citenamefont{Borsanyi, Durr, Fodor, Hoelbling,
  Katz et~al.}}]{Borsanyi:2012zs}
\bibinfo{author}{\bibfnamefont{S.}~\bibnamefont{Borsanyi}},
  \bibinfo{author}{\bibfnamefont{S.}~\bibnamefont{Durr}},
  \bibinfo{author}{\bibfnamefont{Z.}~\bibnamefont{Fodor}},
  \bibinfo{author}{\bibfnamefont{C.}~\bibnamefont{Hoelbling}},
  \bibinfo{author}{\bibfnamefont{S.~D.} \bibnamefont{Katz}},
  \bibnamefont{et~al.}, \bibinfo{journal}{JHEP}
  \textbf{\bibinfo{volume}{1209}}, \bibinfo{pages}{010}
  (\bibinfo{year}{2012}{\natexlab{b}}), \eprint{1203.4469}.

\bibitem[{\citenamefont{Petreczky}(2009)}]{Petreczky:2009at}
\bibinfo{author}{\bibfnamefont{P.}~\bibnamefont{Petreczky}},
  \bibinfo{journal}{Nucl. Phys.} \textbf{\bibinfo{volume}{A830}},
  \bibinfo{pages}{11C} (\bibinfo{year}{2009}), \eprint{0908.1917}.

\bibitem[{\citenamefont{Karsch et~al.}(2003)\citenamefont{Karsch, Redlich, and
  Tawfik}}]{Karsch:2003vd}
\bibinfo{author}{\bibfnamefont{F.}~\bibnamefont{Karsch}},
  \bibinfo{author}{\bibfnamefont{K.}~\bibnamefont{Redlich}}, \bibnamefont{and}
  \bibinfo{author}{\bibfnamefont{A.}~\bibnamefont{Tawfik}},
  \bibinfo{journal}{Eur. Phys. J.} \textbf{\bibinfo{volume}{C29}},
  \bibinfo{pages}{549} (\bibinfo{year}{2003}), \bibinfo{note}{dedicated to Rolf
  Hagedorn}, \eprint{hep-ph/0303108}.

\bibitem[{\citenamefont{Cheng et~al.}(2009)\citenamefont{Cheng, Hendge, Jung,
  Karsch, Kaczmarek et~al.}}]{Cheng:2008zh}
\bibinfo{author}{\bibfnamefont{M.}~\bibnamefont{Cheng}},
  \bibinfo{author}{\bibfnamefont{P.}~\bibnamefont{Hendge}},
  \bibinfo{author}{\bibfnamefont{C.}~\bibnamefont{Jung}},
  \bibinfo{author}{\bibfnamefont{F.}~\bibnamefont{Karsch}},
  \bibinfo{author}{\bibfnamefont{O.}~\bibnamefont{Kaczmarek}},
  \bibnamefont{et~al.}, \bibinfo{journal}{Phys. Rev.}
  \textbf{\bibinfo{volume}{D79}}, \bibinfo{pages}{074505}
  (\bibinfo{year}{2009}), \eprint{0811.1006}.

\bibitem[{\citenamefont{Huovinen and Petreczky}(2010)}]{Huovinen:2009yb}
\bibinfo{author}{\bibfnamefont{P.}~\bibnamefont{Huovinen}} \bibnamefont{and}
  \bibinfo{author}{\bibfnamefont{P.}~\bibnamefont{Petreczky}},
  \bibinfo{journal}{Nucl. Phys.} \textbf{\bibinfo{volume}{A837}},
  \bibinfo{pages}{26} (\bibinfo{year}{2010}), \eprint{0912.2541}.

\bibitem[{\citenamefont{Borsanyi et~al.}(2010{\natexlab{b}})}]{Borsanyi:2010bp}
\bibinfo{author}{\bibfnamefont{S.}~\bibnamefont{Borsanyi}} \bibnamefont{et~al.}
  (\bibinfo{collaboration}{Wuppertal-Budapest Collaboration}),
  \bibinfo{journal}{JHEP} \textbf{\bibinfo{volume}{1009}}, \bibinfo{pages}{073}
  (\bibinfo{year}{2010}{\natexlab{b}}), \eprint{1005.3508}.

\bibitem[{\citenamefont{Beringer et~al.}(2012)}]{Beringer:1900zz}
\bibinfo{author}{\bibfnamefont{J.}~\bibnamefont{Beringer}} \bibnamefont{et~al.}
  (\bibinfo{collaboration}{Particle Data Group}), \bibinfo{journal}{Phys.Rev.}
  \textbf{\bibinfo{volume}{D86}}, \bibinfo{pages}{010001}
  (\bibinfo{year}{2012}).

\bibitem[{\citenamefont{{R Core Team}}(2013)}]{Rpackage}
\bibinfo{author}{\bibnamefont{{R Core Team}}}, \emph{\bibinfo{title}{R: A
  Language and Environment for Statistical Computing}}
  (\bibinfo{address}{Vienna, Austria}, \bibinfo{year}{2013}),
  \urlprefix\url{http://www.R-project.org/}.

\bibitem[{\citenamefont{Venables and Ripley}(2002)}]{RpackageMASS}
\bibinfo{author}{\bibfnamefont{W.~N.} \bibnamefont{Venables}} \bibnamefont{and}
  \bibinfo{author}{\bibfnamefont{B.~D.} \bibnamefont{Ripley}},
  \emph{\bibinfo{title}{Modern Applied Statistics with S}}
  (\bibinfo{publisher}{Springer}, \bibinfo{address}{New York},
  \bibinfo{year}{2002}), \bibinfo{edition}{4th} ed., \bibinfo{note}{iSBN
  0-387-95457-0}, \urlprefix\url{http://www.stats.ox.ac.uk/pub/MASS4}.

\bibitem[{\citenamefont{Jr et~al.}(2014)\citenamefont{Jr, with
  contributions~from Charles~Dupont, and many others.}}]{RpackageHmisc}
\bibinfo{author}{\bibfnamefont{F.~E.~H.} \bibnamefont{Jr}},
  \bibinfo{author}{\bibnamefont{with contributions~from Charles~Dupont}},
  \bibnamefont{and} \bibinfo{author}{\bibnamefont{many others.}},
  \emph{\bibinfo{title}{Hmisc: Harrell Miscellaneous}} (\bibinfo{year}{2014}),
  \bibinfo{note}{r package version 3.14-4},
  \urlprefix\url{http://CRAN.R-project.org/package=Hmisc}.

\bibitem[{\citenamefont{Hung and Shuryak}(1995)}]{Hung:1994eq}
\bibinfo{author}{\bibfnamefont{C.}~\bibnamefont{Hung}} \bibnamefont{and}
  \bibinfo{author}{\bibfnamefont{E.~V.} \bibnamefont{Shuryak}},
  \bibinfo{journal}{Phys.Rev.Lett.} \textbf{\bibinfo{volume}{75}},
  \bibinfo{pages}{4003} (\bibinfo{year}{1995}), \eprint{hep-ph/9412360}.

\bibitem[{\citenamefont{Engels and Karsch}(2012)}]{Engels:2011km}
\bibinfo{author}{\bibfnamefont{J.}~\bibnamefont{Engels}} \bibnamefont{and}
  \bibinfo{author}{\bibfnamefont{F.}~\bibnamefont{Karsch}},
  \bibinfo{journal}{Phys. Rev.} \textbf{\bibinfo{volume}{D85}},
  \bibinfo{pages}{094506} (\bibinfo{year}{2012}), \eprint{1105.0584}.

\bibitem[{\citenamefont{Haque et~al.}(2014)\citenamefont{Haque, Bandyopadhyay,
  Andersen, Mustafa, Strickland et~al.}}]{Haque:2014rua}
\bibinfo{author}{\bibfnamefont{N.}~\bibnamefont{Haque}},
  \bibinfo{author}{\bibfnamefont{A.}~\bibnamefont{Bandyopadhyay}},
  \bibinfo{author}{\bibfnamefont{J.~O.} \bibnamefont{Andersen}},
  \bibinfo{author}{\bibfnamefont{M.~G.} \bibnamefont{Mustafa}},
  \bibinfo{author}{\bibfnamefont{M.}~\bibnamefont{Strickland}},
  \bibnamefont{et~al.}, \bibinfo{journal}{JHEP}
  \textbf{\bibinfo{volume}{1405}}, \bibinfo{pages}{027} (\bibinfo{year}{2014}),
  \eprint{1402.6907}.

\bibitem[{\citenamefont{Laine and Schroder}(2006)}]{Laine:2006cp}
\bibinfo{author}{\bibfnamefont{M.}~\bibnamefont{Laine}} \bibnamefont{and}
  \bibinfo{author}{\bibfnamefont{Y.}~\bibnamefont{Schroder}},
  \bibinfo{journal}{Phys.Rev.} \textbf{\bibinfo{volume}{D73}},
  \bibinfo{pages}{085009} (\bibinfo{year}{2006}), \eprint{hep-ph/0603048}.

\bibitem[{\citenamefont{Clark et~al.}(2005)\citenamefont{Clark, Kennedy, and
  Sroczynski}}]{Clark:2004cp}
\bibinfo{author}{\bibfnamefont{M.}~\bibnamefont{Clark}},
  \bibinfo{author}{\bibfnamefont{A.}~\bibnamefont{Kennedy}}, \bibnamefont{and}
  \bibinfo{author}{\bibfnamefont{Z.}~\bibnamefont{Sroczynski}},
  \bibinfo{journal}{Nucl. Phys. Proc. Suppl.} \textbf{\bibinfo{volume}{140}},
  \bibinfo{pages}{835} (\bibinfo{year}{2005}), \eprint{hep-lat/0409133}.

\bibitem[{\citenamefont{Hasenbusch}(2001)}]{Hasenbusch:2001ne}
\bibinfo{author}{\bibfnamefont{M.}~\bibnamefont{Hasenbusch}},
  \bibinfo{journal}{Phys.Lett.} \textbf{\bibinfo{volume}{B519}},
  \bibinfo{pages}{177} (\bibinfo{year}{2001}), \eprint{hep-lat/0107019}.

\bibitem[{\citenamefont{Bazavov et~al.}(2010{\natexlab{b}})}]{Bazavov:2010ru}
\bibinfo{author}{\bibfnamefont{A.}~\bibnamefont{Bazavov}} \bibnamefont{et~al.}
  (\bibinfo{collaboration}{MILC collaboration}), \bibinfo{journal}{Phys. Rev.}
  \textbf{\bibinfo{volume}{D82}}, \bibinfo{pages}{074501}
  (\bibinfo{year}{2010}{\natexlab{b}}), \eprint{1004.0342}.

\bibitem[{\citenamefont{Allton}(1997)}]{Allton:1996dn}
\bibinfo{author}{\bibfnamefont{C.~R.} \bibnamefont{Allton}},
  \bibinfo{journal}{Nucl.Phys.Proc.Suppl.} \textbf{\bibinfo{volume}{53}},
  \bibinfo{pages}{867} (\bibinfo{year}{1997}), \eprint{hep-lat/9610014}.

\bibitem[{\citenamefont{Bazavov et~al.}(2013{\natexlab{c}})}]{Bazavov:2013gca}
\bibinfo{author}{\bibfnamefont{A.}~\bibnamefont{Bazavov}} \bibnamefont{et~al.}
  (\bibinfo{collaboration}{The MILC Collaboration})
  (\bibinfo{year}{2013}{\natexlab{c}}), \eprint{1311.1474}.

\bibitem[{\citenamefont{Davies et~al.}(2010)\citenamefont{Davies, Follana,
  Kendall, Lepage, and McNeile}}]{Davies:2009tsa}
\bibinfo{author}{\bibfnamefont{C.}~\bibnamefont{Davies}},
  \bibinfo{author}{\bibfnamefont{E.}~\bibnamefont{Follana}},
  \bibinfo{author}{\bibfnamefont{I.}~\bibnamefont{Kendall}},
  \bibinfo{author}{\bibfnamefont{G.}~\bibnamefont{Lepage}}, \bibnamefont{and}
  \bibinfo{author}{\bibfnamefont{C.}~\bibnamefont{McNeile}}
  (\bibinfo{collaboration}{HPQCD Collaboration}), \bibinfo{journal}{Phys. Rev.}
  \textbf{\bibinfo{volume}{D81}}, \bibinfo{pages}{034506}
  (\bibinfo{year}{2010}), \eprint{0910.1229}.

\bibitem[{\citenamefont{Aoki et~al.}(2013)\citenamefont{Aoki, Aoki, Bernard,
  Blum, Colangelo et~al.}}]{Aoki:2013ldr}
\bibinfo{author}{\bibfnamefont{S.}~\bibnamefont{Aoki}},
  \bibinfo{author}{\bibfnamefont{Y.}~\bibnamefont{Aoki}},
  \bibinfo{author}{\bibfnamefont{C.}~\bibnamefont{Bernard}},
  \bibinfo{author}{\bibfnamefont{T.}~\bibnamefont{Blum}},
  \bibinfo{author}{\bibfnamefont{G.}~\bibnamefont{Colangelo}},
  \bibnamefont{et~al.} (\bibinfo{year}{2013}), \eprint{1310.8555}.

\end{thebibliography}

\end{document}